**Theoretical Framework for Phase-space Local-Angle-Domain Waveform Inversion in Anisotropic Elastic Media**

**Z. Koren** zvi.koren@emerson.com

Emerson -AspenTech - Subsurface Science & Engineering
Tel Aviv University - Geophysics Department

## ABSTRACT

This study presents a phase-space formulation of elastic full-waveform inversion (FWI) based on local-angle-domain (LAD) representations of seismic scattering. Instead of minimizing acquisition-domain waveform misfits, the method minimizes image-domain inconsistencies in a phase-space domain parameterized by position, propagation direction, scattering angle, azimuth, and time.

The formulation is motivated by the observation that velocity and impedance perturbations occupy largely distinct regions of phase space. For anisotropic elastic media, we derive a generalized objective function together with the corresponding adjoint-state gradient and Hessian operators. Representing LAD images through directional and opening wavenumbers provides a unified description of propagation and scattering effects.

The approach is independent of the forward-modeling engine and can be implemented with finite-difference, finite-element, spectral-element, ray-based, or beam-based methods. Under Born linearization, ray/beam formulations allow explicit Hessian computation. Owing to the near-orthogonality of directional and opening-angle sensitivities, the Hessian is more diagonally dominant than in conventional FWI, improving parameter separation, reducing velocity-reflectivity crosstalk, and enhancing inversion stability.

By organizing seismic information in a physically meaningful phase-space representation, LAD-FWI supports adaptive, resolution-driven inversion and may reduce cycle skipping through attributes that vary more smoothly than waveform residuals. The framework naturally integrates directional focusing, reflector-consistency analysis, anisotropic parameter estimation, converted-wave imaging, and diffraction-based characterization, providing a unified workflow for elastic FWI, migration-velocity analysis, impedance inversion, and high-resolution subsurface imaging.

## 1. INTRODUCTION

Full-waveform inversion (FWI) has become a fundamental methodology for quantitative subsurface characterization, enabling the estimation of seismic velocities, anisotropic parameters, and impedance distributions from recorded wavefields (Lailly, 1983; Tarantola, 1984; Pratt, 1999; Symes, 2008; Virieux and Operto, 2009). Despite its success, FWI remains a highly nonlinear and often ill-conditioned inverse problem. In anisotropic media, parameter coupling, incomplete illumination, cycle skipping, limited acquisition geometry, and modeling uncertainties frequently hinder convergence and reduce model resolution.

A fundamental limitation of conventional FWI is that inversion objectives are formulated in the acquisition-data domain, where propagation effects and scattering effects are simultaneously encoded within the seismic waveforms. Consequently, velocity-related and reflectivity-related model updates are recovered from strongly coupled sensitivities, often leading to parameter crosstalk, poor Hessian conditioning, and reduced inversion stability. Moreover, conventional multiscale strategies typically use temporal frequency as a surrogate for model resolution, even though subsurface resolution depends jointly on frequency, propagation direction, and scattering geometry.

Angle-domain imaging has demonstrated that seismic wavefields can be decomposed according to their local scattering geometry. In particular, the local-angle-domain (LAD) formulation of Koren and Ravve (2011) provides a generalized phase-space representation in which wavefield components are organized according to propagation direction and scattering angle and azimuth. This representation offers direct access to illumination, directional focusing, reflector orientation, anisotropic behavior, and scattering characteristics within the subsurface image domain.

This paper extends these concepts to formulate a generalized phase-space waveform inversion framework for anisotropic elastic media. Rather than minimizing waveform misfits in the acquisition-data domain, the proposed methodology operates directly on an in-situ phase-space representation of the scattering wavefield parameterized by spatial position, propagation direction, scattering (opening) angle and azimuth, and the temporal support of the scattered energy. The framework therefore treats propagation and scattering phenomena within a common representation and enables inversion objectives to be expressed in terms of physically meaningful wavefield attributes.

The central hypothesis of this work is that velocity-sensitive and reflectivity-sensitive wavefield components occupy predominantly different regions of phase space. In this study, reflectivity is defined as the normalized spatial gradient of the impedance field rather than as a conventional reflection coefficient. Impedance, rather than density, is adopted as the contrast-type parameter because the radiation pattern of a density perturbation is largely confined to small opening angles, where it is nearly indistinguishable from that of an impedance contrast, while its sensitivity vanishes at the wide angles that constrain velocity; density is therefore poorly resolved as an independent parameter, whereas the (velocity, impedance) parameterization approximately aligns each parameter class with a distinct region of phase space (Tarantola, 1986; Forgues and Lambaré, 1997; Operto et al., 2013; see Section 7). Velocity information is expected to concentrate at large opening angles and low directional wavenumbers, whereas reflectivity information is expected to concentrate at smaller opening angles and broader directional bandwidths. Although these regions may overlap in complex geological settings, particularly in the presence of steeply dipping structures and multipathing, their partial separation suggests a mechanism for improving parameter discrimination, reducing velocity-reflectivity coupling, and enhancing inversion conditioning. The analytic and numerical benchmarks of **Appendix F** verify this separation, together with its resolution and conditioning consequences.

Based on this hypothesis, we formulate inversion objectives directly in phase space and derive the corresponding adjoint-state gradients and Hessian operators. In addition to conventional waveform-consistency constraints, the proposed framework incorporates directional focusing, reflector-orientation consistency, scattering-gather behavior, and adaptive phase-space weighting. The resulting formulation naturally accommodates reflections, diffractions, transmitted waves, mode-converted waves recorded by multicomponent acquisition (**Appendix D, Figures D1 and D2**), anisotropic propagation effects, and model-resolution analysis within a unified inversion methodology.

An important consequence of the proposed approach is that inversion scheduling can be guided by model-wavenumber illumination, parameter sensitivity, resolution gain, and Hessian conditioning rather than by temporal frequency alone. This leads to a generalized phase-space continuation strategy in which inversion progresses through increasingly resolved spatial, directional, and scattering wavenumbers. Such a framework is expected to improve parameter separability, reduce crosstalk, and potentially mitigate cycle-skipping by operating on wavefield attributes that vary more smoothly with model perturbations than raw waveform residuals. Each of these effects is demonstrated exactly in the analytic benchmark and approximately in the numerical example of **Appendix F**.

The primary contribution of this paper is the formulation of FWI directly in LAD phase space, where velocity-type and reflectivity-type sensitivities exhibit different wavenumber support and therefore produce a Hessian structure with reduced parameter coupling compared to conventional data-domain FWI.

It is useful to state explicitly the two levels at which this paper operates. The first level is the general LAD-FWI framework: the LAD and LDP representations, the inversion objectives, and the associated adjoint-state gradients are formulated independently of the numerical engine used to construct the decomposed wavefields - any propagation and imaging operator capable of producing local-angle-domain decompositions, whether based on wave-equation extrapolation or on asymptotic ray/beam superposition, may serve as the imaging kernel. The second level is a specific specialization: when the Born approximation is adopted and the operator is implemented with the ray/beam engine, additional capabilities become available that are not accessible in the general setting - semi-analytic Fréchet derivatives, explicit assembly of the Gauss-Newton Hessian blocks and the resulting resolution and conditioning analysis (**Appendix A**), and operator-level discrimination of multiply scattered energy (**Appendix E**). Sections 3–8 belong to the first level; Section 9 and **Appendices A** and **F** develop the second. Statements of efficiency, explicit Hessian access, and built-in multiple suppression should therefore be read as properties of the ray/beam specialization, whereas the phase-space objectives and their gradients hold for the framework at large.

The generalized Local-Angle-Domain (LAD) representation provides a directional-scattering phase-space description of the seismic wavefield. When local waveform information is retained through the Local Directional Projection (LDP) decomposition, the representation becomes a phase-space-time description. Throughout the paper, the term *LAD* is used generically to denote this family of directional-scattering representations.

The paper is organized as follows:

The following appendices provide additional theoretical and computational developments:

Although the present work focuses on the theoretical formulation of LAD-FWI and does not include numerical inversion field examples, it is important to note that the proposed methodology builds upon the LAD imaging framework introduced by Koren et al. (2007, 2008, 2010) and Koren and Ravve (2011). Over the past decade, LAD-based imaging has been applied successfully in a wide range of exploration and reservoir-characterization workflows, including full-azimuth depth imaging, migration-velocity model building, tomography, AVAZ and VVAZ analysis, Q compensation, diffraction imaging, fracture characterization, and reservoir studies in both marine and land environments (see Section 2 for a review of previous work).

Consequently, several of the physical attributes and phase-space quantities utilized in the proposed inversion framework are not purely theoretical constructs. Rather, they are extensions of phase-space observables that have already demonstrated practical value in imaging and interpretation workflows. The present study therefore focuses on establishing the mathematical framework required to elevate these LAD-derived attributes from imaging and analysis products to primary inversion-domain variables. In this formulation, parameter sensitivities, model resolution, and Hessian conditioning can be analyzed and adaptively controlled through recoverable directional and scattering wavenumbers, so that the inversion is organized according to information content and resolving power rather than temporal frequency alone.

## NOTATION

Throughout this paper, scalar quantities are denoted by italic symbols, vectors by bold symbols, and matrices or linear operators by bold uppercase symbols. In particular,

- $\mathbf{x}$ denotes spatial position;
- $\mathbf{k}_{in}$, $\mathbf{k}_{sc}$, $\mathbf{k}_{\nu}$, and $\mathbf{k}_{\gamma}$ denote incident, scattered, directional, and opening wavenumber vectors, respectively;
- $\mathbf{u}$, $\boldsymbol{\lambda}$, and $\mathbf{q}$ denote forward, adjoint, and residual wavefields;
- $\mathbf{R}$ denotes the impedance-gradient reflectivity vector, with magnitude $R = \| \boldsymbol{R} \|$; and $\mathbf{n}_{\mathrm{r}} = \mathbf{R}/R$ denotes the local reflector-normal unit vector.
- The angular variables $\boldsymbol{\nu} = (\nu_1, \nu_2)$ and $\boldsymbol{\gamma} = (\gamma_1, \gamma_2)$ represent directional and scattering-angle coordinate pairs, respectively, and are treated as spherical coordinates rather than vectors. The Jacobian (Fréchet derivative), Hessian, wave-equation, sensitivity-kernel, weighting, and imaging operators are denoted by $\mathbf{J}$, $\mathbf{H}$, $\mathbf{L}$, $\mathbf{K}$, $\mathbf{W}$, and $\mathbf{M}$, respectively.
- Fourth-order elastic properties are represented by indexed tensor notation, such as the stiffness tensor $c_{ijkl}$ and its density-normalized counterpart $a_{ijkl}$.
- $\ln(\cdot)$ denotes the natural logarithm; all logarithmic impedance quantities are defined using the natural-log transform.

## 2. BACKGROUND AND PREVIOUS WORK

Angle-domain imaging and extended-image methodologies have demonstrated that seismic wavefields can be decomposed according to their local scattering geometry, revealing information that is largely hidden in conventional stacked images. Over the past three decades, these developments have progressively transformed angle-domain representations from imaging diagnostics into powerful tools for velocity analysis, model building, and inversion.

The foundations of the generalized Radon transform (GRT) for angle-domain seismic imaging were introduced by Beylkin (1985), Beylkin and Burridge (1990), and de Hoop and Bleistein (1997). Wave-equation-based extended imaging was established by ten Kroode et al. (1994) and Nolan and Symes (1996), who generalized imaging conditions and demonstrated that extended image spaces contain valuable velocity-model information that is largely suppressed in conventional migrated images. Their work established the conceptual basis for image-domain migration velocity analysis (Symes and Carazzone, 1991; Sava and Biondi, 2004) and laid the foundation for subsequent angle-domain methodologies.

Building on these ideas, Rousseau et al. (2000), Xu et al. (2001), Audebert et al. (2002), Rickett and Sava (2002), Koren et al. (2002), Sava and Fomel (2003), Sollid and Ursin (2003), Soubaras (2003), Bleistein et al. (2005a, b), Wu and Chen (2006), and Biondi (2007a, b) developed the mathematical and practical framework for angle-domain common-image gathers (ADCIGs), with extensions to converted waves by Rosales et al. (2008). These methods demonstrated that reflection-angle information provides powerful constraints for velocity analysis, residual moveout estimation, illumination studies, amplitude-versus-angle (AVA) analysis, uncertainty assessment, and inversion.

More recently, Dafni and Symes (2016, 2018) provided a rigorous theoretical treatment of angle-domain imaging and clarified the relationships between subsurface-offset extensions, reflection-angle gathers, and scattering-angle parameterizations. Their work strengthened the mathematical understanding of how different extended-image coordinates encode model perturbations and provided important insights into the interpretation of angle-domain sensitivity kernels.

Despite their success, most conventional angle-domain formulations remain primarily focused on the analysis of specular reflections. Although non-specular wave phenomena such as diffractions, transmitted waves, and more complex scattering interactions can be represented within extended-image frameworks, they are generally not treated as first-class inversion attributes. Furthermore, many angle-domain approaches emphasize scattering (opening-angle) information while providing a more limited description of propagation direction, illumination, and directional wavefield behavior.

Several authors expanded the angle-domain concept toward a more complete description of the local scattering wavefield. Notably, Brandsberg-Dahl et al. (2003) developed formulations that incorporated broader scattering geometries. This line of development culminated in the generalized Local-Angle-Domain (LAD) representation introduced by Koren et al. (2007, 2008, 2010) and Kozlov et al. (2009), and subsequently formalized by Koren and Ravve (2011) and Ravve and Koren (2011). In this framework, the seismic wavefield is decomposed into two orthogonal components: directional wavenumbers, which describe propagation and illumination, and opening wavenumbers, which characterize scattering behaviors. The resulting phase-space representation generalizes conventional reflection-angle gathers by simultaneously describing propagation and scattering phenomena and by accommodating not only specular reflections but also diffractions, transmitted waves, and other components of the in-situ scattering wavefield.

Since its introduction, the LAD methodology has been successfully applied to a wide range of seismic tomography, imaging and reservoir-characterization problems in both marine and land environments. Representative applications include fracture characterization and VVAZ and AVAZ analyses (Inozemtsev et al., 2013; Hosgood et al., 2015; Korkidi et al., 2018), shale and unconventional reservoir imaging (Podolak et al., 2014), full-azimuth depth imaging and velocity model building (Inozemtsev et al., 2015, 2017, 2019; Chase and Koren, 2015; Eskozha et al., 2017; Olneva et al., 2019; Shustak et al., 2022, 2024, 2026), Q-compensation and amplitude-preserving imaging (Dekel et al., 2016, 2017), machine-learning-assisted seismic characterization (Itan et al., 2018), diffraction imaging and fracture detection (De Ribet et al., 2018; Ghosh et al., 2023; Singh et al., 2025), and the recent development of Local Directional Projection (LDP) methodologies for enhanced reflectivity characterization and scattering analysis (Kletenik-Edelman et al., 2025; Shustak et al., 2026).

The present study extends these LAD-based applications by introducing a unified inversion framework, based on the adjoint-state method, for the simultaneous estimation of subsurface velocity and impedance fields. The framework generates high-resolution, model-driven and data-driven reflectivity and diffractivity seismic images, together with their associated attributes. By formulating inversion directly within the LAD, it exploits the complementary sensitivity of propagation and scattering information to separate velocity-related and reflectivity-related updates, thereby integrating waveform inversion, migration-velocity analysis, impedance inversion, diffraction imaging, and advanced subsurface characterization within a single methodology.

### 2.1 From Explicit Fréchet Matrices to Implicit Adjoint-State Gradients

The velocity-model inversion employed in the LAD-based applications cited above relies primarily on nonlinear, ray-based tomography, in which model updates are driven by an explicitly constructed Fréchet derivative (Jacobian) matrix. Each row of this matrix describes the sensitivity of an individual traveltime or attribute residual to perturbations in the model parameters. Because the matrix must be explicitly computed, stored, and inverted, typically within a regularized least-squares framework, the model space is usually parameterized on a relatively coarse grid or represented by smooth basis functions. Moreover, the sensitivities are generally restricted to ray paths or first-order finite-frequency kernels. As a result, the recovered updates are inherently limited to the long-wavelength components of the model, with achievable resolution determined primarily by ray coverage and parameterization rather than by the full information content of the seismic data.

In the present study, we adopt the implicit adjoint-state method implied in the phase-space domain, which eliminates both limitations. The gradient of the objective function is obtained directly without explicitly forming, storing, or inverting the Fréchet matrix. Its computational cost, essentially one forward ("migration") and one adjoint wavefield propagation ("de-migration") per source, is independent of the number of model parameters, allowing the model to be sampled on the native wave-propagation grid. Because the sensitivities are carried by the full wavefields, the resulting gradient naturally contains the complete finite-frequency response of the data, enabling model updates with resolution approaching the diffraction limit of the recorded wavefield.

In conventional FWI methods, this resolution, however, comes at a price: the correlation-based gradient implicitly combines (entangles) the kinematic (long-wavelength velocity) and dynamic (high-resolution reflectivity) sensitivities within a single update - a limitation discussed further below and addressed by the LAD formulation developed in this paper.

The central hypothesis of this work is that velocity-sensitive and reflectivity-sensitive perturbations occupy distinct, although partially overlapping, regions of phase space. The analytical and numerical benchmarks presented in **Appendix F** verify this separation in a controlled setting. Consequently, objective functions formulated directly in the LAD domain provide a natural mechanism for discriminating between propagation-related and scattering-related sensitivities. This separation has the potential to improve the recovery of both velocity and impedance while establishing a unified inversion framework that bridges waveform inversion, migration-velocity analysis, impedance inversion, diffraction imaging, and advanced subsurface characterization.

### 2.2 Extended FWI and Cycle-Skipping Mitigation

A substantial body of work has addressed the cycle-skipping (local-minima) problem of conventional FWI by extending the model or source space - deliberately enlarging the parameterization with non-physical degrees of freedom so that the data can be fit even when the background velocity is far from correct, and then progressively annihilating the non-physical component to recover a physically consistent model. The conceptual foundation is Symes' extended-modeling framework, rooted in differential semblance optimization (Symes and Carazzone, 1991; Symes, 2008), which showed that suitably chosen extensions convexify the objective function with respect to the long-wavelength velocity components.

The first major family operates in the image domain: the model is extended along subsurface space-lag or time-lag axes, and velocity errors manifest as defocusing of the extended image away from zero lag. Differential-semblance and annihilator-based objectives then drive the velocity update (Shen and Symes, 2008; Sava and Vasconcelos, 2011). Biondi and Almomin (2014) unified this wave-equation migration-velocity-analysis sensitivity with conventional FWI through the time-lag extension (tomographic FWI), demonstrating recovery of both low- and high-wavenumber model components without an accurate starting model. Barnier et al. (2023a, b) provided a systematic treatment of FWI by model extension, including its theory, optimization strategies, and practical design choices.

The second family extends the source or wavefield rather than the image. In wavefield reconstruction inversion (van Leeuwen and Herrmann, 2013), the wave equation is imposed only as a quadratic penalty, so the reconstructed wavefield is allowed to violate the physics just enough to fit the data; the wave-equation residual then acts as a secondary (extended) source that vanishes as the model converges. Closely related extended-source formulations recast this as inversion for a distributed source perturbation with a penalty on its spatial or temporal support (Wang et al., 2016; Huang et al., 2018; Symes, 2020). In a related spirit, time-reversal full wavefield inversion (Landa and Fomel, 2026) abandons waveform-residual minimization altogether in favor of a wavefield-focusing criterion: the recorded data are propagated backward in time and the degree of focusing at the true source location and emission-onset time is evaluated, an objective that attains a distinct maximum for the correct velocity model and is thereby inherently robust against cycle skipping.

A third, data-domain family achieves an analogous effect through matching filters: adaptive waveform inversion (Warner and Guasch, 2016) computes Wiener filters between predicted and observed traces and penalizes energy at non-zero lags, while deconvolution- and correlation-based objectives (van Leeuwen and Mulder, 2010; Luo and Sava, 2011) similarly measure kinematic misfit in a way that degrades gracefully beyond half a cycle. These filter-based methods can be interpreted as particular source-extension schemes. Complementary to model and source extension, modified misfit functionals - most prominently optimal-transport distances (Engquist and Froese, 2014; Métivier et al., 2016) and envelope- or phase-based measures - pursue the same convexification goal without enlarging the model space.

Common to all these approaches is a trade-off: robustness to poor starting models is purchased with a substantially enlarged unknown space (and hence memory and computational cost), penalty-weight continuation schedules that must be tuned, and, in the image-domain variants, sensitivity to the same illumination and sampling limitations that affect extended-image gathers. The LAD representation developed in this paper offers a complementary route: rather than introducing non-physical extension axes, it operates on physically meaningful directional and scattering observables, whose smoother dependence on model perturbations provides an alternative mechanism for mitigating cycle skipping while retaining a physically interpretable inversion domain. This limitation is revisited quantitatively in **Figure 9** (Section 8), which contrasts the basins of attraction of the data-domain and phase-space objectives for a representative reflection event.

Finally, as extended formulations and high-performance computing have pushed FWI toward the full usable bandwidth of the recorded data, the method has evolved from a velocity-model-building tool into a direct imaging engine. In the resulting FWI-imaging and FWI-derived reflectivity (FDR) approaches, high-resolution reflectivity volumes are computed directly from the inverted velocity or impedance models rather than through a separate migration step (Kalinicheva et al., 2020; Zhang et al., 2020; He et al., 2021; McLeman et al., 2021, 2022), and their practical value for interpretation and reservoir characterization, including elastic and prestack extensions, has been demonstrated on field data (Kumar and Ali, 2024; Cheng et al., 2025). This development is directly relevant to the present framework: the model-driven reflectivity field produced by LAD-FWI is the phase-space counterpart of FDR, and its interplay with the complementary data-driven reflectivity image is developed in Section 10.

## 3. ELASTIC WAVE PROPAGATION AND MODEL PARAMETERIZATION

We consider the elastic wave equation in a heterogeneous anisotropic medium,

$$\rho(\mathbf{x})\,\frac{\partial^2 u_i}{\partial t^2} = \partial_j\left(c_{ijkl}(\mathbf{x})\frac{\partial u_k}{\partial x_l}\right) + f_i(\mathbf{x},t), \qquad i,j,k,l\ \in 1,2,3 \tag{3.1}$$

where $\mathbf{f}(\mathbf{x},t)$ and $\mathbf{u}(\mathbf{x},t)$ denote the source and corresponding displacement-vector fields, respectively, defined and computed on a fine three-dimensional spatial grid. The medium is characterized by the density field $\rho(\mathbf{x})$ and the fourth-order stiffness tensor $\mathbf{c}(\mathbf{x})$, which fully describe its elastic properties.

A key observation motivating the proposed formulation is the markedly different spatial behavior of velocity-type and impedance-type parameters. Velocity and anisotropic parameters generally vary smoothly throughout the subsurface and are therefore dominated by low spatial wavenumbers. In contrast, impedance variations are concentrated near geological interfaces and are characterized by significantly broader spatial bandwidth. These differing spectral characteristics lead to distinct sensitivities in the local-angle-domain phase space and form the basis for the proposed inversion methodology.

In this study, we adopt the following parameterization:

$$\mathbf{m}(\mathbf{x}) = \{a_{ijkl}(\mathbf{x}), Z_P(\mathbf{x}), Z_S(\mathbf{x})\}, \tag{3.2a}$$

where $a_{ijkl} = \frac{c_{ijkl}}{\rho}$ is the density-normalized fourth-order stiffness tensor, and $Z_P = \rho V_P$ and $Z_S = \rho V_S$ denote the compressional- and shear-wave impedances, respectively. The components of $a_{ijkl}$ have units of velocity squared.
Alternatively, the tensor components $a_{ijkl}$ may be expressed in terms of Thomsen-type, Tsvankin-type, or other anisotropic parameterizations (Thomsen, 1986; Tsvankin, 1997). In this case, the model vector becomes,

$$\mathbf{m}(\mathbf{x}) = \{\mathbf{v}(\mathbf{x}), Z_P(\mathbf{x}), Z_S(\mathbf{x})\}, \tag{3.2b}$$

where $\mathbf{v}$ symbolizes the set of anisotropic parameters associated with the chosen anisotropic elastic symmetry. For example, a sedimentary sequence can be represented by polar anisotropy, also referred to as transverse isotropy with a tilted symmetry axis (TTI), and $\mathbf{v}(\mathbf{x})$ can be represented as,

$$\mathbf{v}(\mathbf{x}) = \{V_P, f_{SP}, \delta, \varepsilon, \gamma_T, \mathbf{n}_{\mathrm{a}}\}(\mathbf{x}), \quad f_{SP} = 1 - \left(V_S/V_p\right)^2 \quad ; \quad \mathbf{n}_{\mathrm{a}} = \{\theta, \phi\}_a \quad , \tag{3.3}$$

where $V_p$ and $V_S$ are the axial compressional and shear velocities, respectively, and $\delta, \varepsilon$ and $\gamma_T$ are the Thomsen parameters ($\gamma_T$ is the shear-wave anisotropy coefficient, conventionally denoted $\gamma$); the subscript distinguishes it from the opening angle used throughout this paper. The direction unit vector $\mathbf{n}_{\mathrm{a}}$ defines the local direction of the tilted axis with the dip and azimuth angles $\{\theta, \phi\}_a$.

The corresponding normal reflectivity vector fields, $\mathbf{R}_P = \{R_P, \theta, \phi\}$ and $\mathbf{R}_S = \{R_S, \theta, \phi\}$, are defined as constrained (curl-free) relative gradients of the impedance fields, or equivalently as one-half the gradients of the logarithmic impedance fields,

$$\mathbf{R}(\mathbf{x}) = \frac{1}{2}\frac{\nabla Z}{Z} = \frac{1}{2}\nabla(\ln Z), \tag{3.4}$$

with $R = |\mathbf{R}|$ and unit reflector-normal direction vector,

$$\mathbf{n}_{\mathrm{r}} = \mathbf{R}/|\boldsymbol{R}| = \begin{bmatrix} \sin\theta\cos\phi \\ \sin\theta\sin\phi \\ \cos\theta \end{bmatrix} , \tag{3.5}$$

where $\theta$ and $\phi$ denote the dip and azimuth angles of the reflector normal, respectively. It is important to emphasize that throughout this paper, reflectivity refers to the relative spatial gradient of impedance (equation 3.4) and not to conventional plane-wave reflection coefficients.

## 4. CONVENTIONAL TIME-DOMAIN FWI

Conventional full-waveform inversion (FWI) is formulated as a nonlinear optimization problem aimed at estimating the subsurface model parameters, $\mathbf{m}(\mathbf{x})$, by minimizing an objective function defined as the misfit between the observed time-domain seismic data, recorded at the acquisition source-receiver locations $(\mathbf{x}_{\mathrm{s}}, \mathbf{x}_{\mathrm{r}})$, and the corresponding calculated (modeled) data,

$$\Phi(\mathbf{m}) = \delta D(\mathbf{x}_{\mathrm{s}}, \mathbf{x}_{\mathrm{r}}, t \,|\, \mathbf{m}) = \left\| D_{\mathrm{cal}}(\mathbf{x}_{\mathrm{s}}, \mathbf{x}_{\mathrm{r}}, t \,|\, \mathbf{m}) - D_{\mathrm{obs}}(\mathbf{x}_{\mathrm{s}}, \mathbf{x}_{\mathrm{r}}, t) \right\| . \tag{4.1}$$

The data residual $\delta D(\mathbf{x_s}, \mathbf{x_r}, t|\mathbf{m})$ is injected at the receiver locations and backpropagated through the current model to form the adjoint wavefield $\lambda(\mathbf{x}, t)$, which satisfies the adjoint wave equation:

$$\mathbf{L}^{\dagger}(\mathbf{m})\boldsymbol{\lambda}(\mathbf{x},t) = \sum_{\mathrm{r}} \delta D(\mathbf{x}_{\mathrm{s}},\mathbf{x}_{\mathrm{r}},t)\,\delta(\mathbf{x}-\mathbf{x}_{\mathrm{r}}). \tag{4.2}$$

The gradient of the misfit functional with respect to model parameters $\mathbf{m}(\mathbf{x})$ is then obtained via the zero-lag correlation between the forward wavefield $U(\mathbf{x}, t)$ and the adjoint wavefield:

$$\nabla\Phi(\mathbf{m})(\mathbf{x}) = \int_0^T \lambda(\mathbf{x},t)\,\frac{\partial\mathbf{L}}{\partial\mathbf{m}} U(\mathbf{x},t)\,dt\ . \tag{4.3}$$

For example, in acoustic velocity inversion:

$$\nabla\Phi(V)(\mathbf{x}) = -2\int_0^T \frac{1}{V^3(\mathbf{x})}\frac{\partial^2 U(\mathbf{x},t)}{\partial t^2}\lambda(\mathbf{x},t)dt\ . \tag{4.4}$$

Thus, the conventional FWI gradient is constructed through the space-time-domain correlation of the forward and adjoint (residual) wavefields, which implicitly entangles kinematic sensitivities related to the long-wavelength velocity model with dynamic sensitivities associated with the high-resolution reflectivity field. Despite its widespread use, this formulation is subject to several well-known limitations, including cycle skipping, parameter crosstalk, limited resolution and decoupling of anisotropic parameters, dependence on accurate source-wavelet estimation, and pronounced sensitivity to noise, especially in land-seismic applications.

The extended-FWI methodologies reviewed in Section 2 were developed largely in response to the first of these limitations. By enlarging the model, source, or data space with additional auxiliary (non-physical) degrees of freedom such as subsurface space-lag and time-lag image extensions, extended sources with relaxed wave-equation constraints, or matching-filter and optimal-transport misfit measures, they restore convexity of the objective function with respect to the long-wavelength velocity components and thereby mitigate cycle skipping.

However, these methods retain the fundamental structure of the conventional data-domain gradient. The extension axes serve as auxiliary mathematical constructs that are intended to collapse as the inversion converges, rather than as physically meaningful observables. Consequently, the entanglement of kinematic and dynamic sensitivities within the correlation-based gradient remains largely unchanged.

The LAD formulation developed in the following sections pursues a complementary strategy. Rather than introducing auxiliary extension axes into the model or source space, it maps the recorded wavefield directly into a physically interpretable phase space defined by orthogonal directional and scattering wavenumbers. Within this representation, velocity-related and reflectivity-related sensitivities occupy distinguishable, although partially overlapping, regions, and the resulting inversion attributes vary more smoothly with model perturbations than do raw waveform residuals.

## 5. LOCAL-ANGLE-DOMAIN (LAD) REPRESENTATION

### 5.1 The Phase-Space-Local Angle-Domain

The proposed phase-space (image) data is obtained by applying a migration-type operator, $\mathfrak{M}(\mathbf{m})$, and binning the resulting image contributions into a local-angle-domain (LAD) multidimensional grid, following the workflow:

$$D_{\mathrm{obs}}(\mathbf{x}_s,\mathbf{x}_r,t)\xrightarrow{\mathfrak{M}(\mathbf{m})} I(\mathbf{x},\mathbf{k}_{\mathrm{in}},\mathbf{k}_{\mathrm{sc}})\xrightarrow[\mathbf{k}_\gamma=\mathbf{k}_{\mathrm{in}}-\mathbf{k}_{\mathrm{sc}}]{\mathbf{k}_\nu=\mathbf{k}_{\mathrm{in}}+\mathbf{k}_{\mathrm{sc}}} I(\mathbf{x},\mathbf{k}_\nu,\mathbf{k}_\gamma)\rightarrow I(\mathbf{x},\boldsymbol{\nu},\boldsymbol{\gamma}) \ . \tag{5.1}$$

Here, $\mathbf{k}_{\mathrm{in}}(\mathbf{x})$ and $\mathbf{k}_{\mathrm{sc}}(\mathbf{x})$ denote the incident and scattered wavenumber vectors, respectively; in our notation convention, both point toward the scattering point $\mathbf{x}$. These vectors encode both the propagation directions (to point $\mathbf{x}$ from the sources $\mathbf{x}_s$ and receivers $\mathbf{x}_r$, respectively) and local wavelengths, through their magnitudes,

$$k = |\mathbf{k}| = \omega|\mathbf{p}| = \frac{\omega}{v_{\mathrm{phs}}[\mathbf{c}(\mathbf{x}),\hat{\mathbf{k}}]} \ , \tag{5.2}$$

where $\omega$ is the angular frequency, $\mathbf{p}$ is the slowness vector and $v_{\mathrm{phs}}[\mathbf{c}(\mathbf{x}),\hat{\mathbf{k}}]$ is the phase velocity magnitude in direction $\hat{\mathbf{k}}$, which can be obtained from the Christoffel equation.

Following, for example, Koren and Ravve (2011), we adopt an orthogonal wavenumber-based phase-space parameterization $(\mathbf{x},\mathbf{k}_\nu,\mathbf{k}_\gamma)$, or equivalently, an angle-domain representation $(\mathbf{x},\boldsymbol{\nu},\boldsymbol{\gamma})$,

$$\left.\begin{array}{l}\mathbf{k}_\nu=\mathbf{k}_{\mathrm{in}}+\mathbf{k}_{\mathrm{sc}} \quad \text{and} \quad \mathbf{k}_\gamma=\mathbf{k}_{\mathrm{in}}-\mathbf{k}_{\mathrm{sc}} \\ \mathbf{k}_{\mathrm{in}}=\dfrac{\mathbf{k}_\nu-\mathbf{k}_\gamma}{2} \quad \text{and} \quad \mathbf{k}_{\mathrm{sc}}=\dfrac{\mathbf{k}_\nu+\mathbf{k}_\gamma}{2}\end{array}\right\} \ , \ (\mathbf{k}_\nu\perp\mathbf{k}_\gamma: \ \ \mathbf{k}_\nu\cdot\mathbf{k}_\gamma=0) \quad . \tag{5.3}$$

The directional wavenumber $\mathbf{k}_\nu$ is the vector sum of the incident and scattered wavenumbers. Its direction $\hat{\mathbf{k}}_\nu$ characterizes the local propagation direction and illumination geometry. It therefore controls dip sensitivity, directional illumination, and acquisition-related effects.

In contrast, the opening wavenumber $\mathbf{k}_\gamma$ quantifies the angular separation between the incident and scattered wavefields. Its magnitude is directly related to scattering angle and determines the recoverable spatial frequencies of the model. Consequently, $\mathbf{k}_\gamma$ governs image resolution, reflectivity response, and angle-dependent amplitude behavior.

Since $\mathbf{k}_\nu$ and $\mathbf{k}_\gamma$ are orthogonal, they are expected to provide a physically meaningful decomposition of the seismic wavefield into propagation and scattering components.

### 5.2 Angle-Domain Representation

The wavenumber vectors, $\mathbf{k}_\nu$ and $\mathbf{k}_\gamma$ can be mapped into directional (illumination) phase angles $\boldsymbol{\nu}=\{\nu_1,\nu_2\}$ and scattering phase angles $\boldsymbol{\gamma}=\{\gamma_1,\gamma_2\}$ through (e.g., Ravve and Koren, 2011, Part 2):

- **directional dip**: $\nu_1=\arccos\left(\dfrac{k_{\nu z}}{\|\mathbf{k}_\nu\|}\right)$ and **directional azimuth**: $\nu_2=\operatorname{atan2}(k_{\nu y},k_{\nu x})$ , (5.4)

and,

- **opening angle**: $\gamma_1=\arccos\left(\dfrac{\mathbf{k}_{\mathrm{in}}\cdot\mathbf{k}_{\mathrm{sc}}}{\|\mathbf{k}_{\mathrm{in}}\|\|\mathbf{k}_{\mathrm{sc}}\|}\right)$ , and
  **opening azimuth**: $\gamma_2=\operatorname{atan2}\left[\mathbf{k}_\gamma\cdot\left(\hat{\mathbf{k}}_\nu\times\mathbf{e}_1\right),\mathbf{k}_\gamma\cdot\mathbf{e}_1\right]$, with $\mathbf{e}_1=\dfrac{\hat{\mathbf{N}}-\left(\hat{\mathbf{N}}\cdot\hat{\mathbf{k}}_\nu\right)\hat{\mathbf{k}}_\nu}{\left\|\hat{\mathbf{N}}-\left(\hat{\mathbf{N}}\cdot\hat{\mathbf{k}}_\nu\right)\hat{\mathbf{k}}_\nu\right\|}$ . (5.5)

The opening (scattering) azimuth $\gamma_2$ is defined in the reflection plane associated with each incident–scattered wavenumber pair, i.e. the plane perpendicular to $\hat{\mathbf{k}}_\nu$. It is measured with respect to a reference direction $\mathbf{e}_1$, defined as the projection of the North direction $\hat{\mathbf{N}}$ onto this plane.

**Figure 1** illustrates the incident and scattered wavenumbers and their mapping into the local-angle-domain (LAD) representation.

**The LAD Scattering system | Local Directional Projection**

**Figure 1.** The local-angle-domain (LAD) scattering-angle system at a subsurface scattering point $(x, y, z)$. An incident wave from the source S and the scattered wave toward the receiver R are decomposed into two complementary pairs of local angles. In this study we consider both waves pointing inward - toward the scattering point. The directional (propagation) angles - In/Sc directional dip $\nu_1$ and azimuth $\nu_2$ - describe the orientation of the local propagation direction, characterized by the unit vector $\mathbf{n}_\nu$. The opening (scattering) angles - In/Sc opening angle $\gamma_1$ and azimuth $\gamma_2$ - describe the angular separation between the incident and scattered waves/rays (wavenumber/slowness vectors). The associated local-directional-projection (LDP) signal (see Section 6) is shown along the directional slowness vector. This decomposition into directional and opening components forms the basis of the LAD representation used throughout the paper.

### 5.3 LAD Image Formation

In this study, the term phase-space domain is used in a broader sense than the conventional angle-domain representation employed in reflection imaging. Unlike conventional angle-domain representations, which focus primarily on reflection and opening angles associated with specular events, the phase-space domain is expected to provide a generalized decomposition of the complete in-situ scattered wavefield at each subsurface point (scattering element).

The wavefield is separated into two complementary components: directional wavenumber vectors (or their equivalent angular representation) and opening wavenumber vectors (or opening angles). The directional components describe the local propagation and scattering directions of the wavefield, whereas the opening components characterize the angular relationship between the incident and scattered wavefields. This extended phase-space formulation provides a unified framework for analyzing both specular and non-specular energy, including reflections, diffractions, transmitted waves corresponding to opening angles approaching 180°, and other scattering modes, thereby enabling a richer characterization of the subsurface wavefield and its associated rock properties.

A generalized phase-space representation can be expressed as,

$$I(\mathbf{x},\mathbf{k}_\nu,\mathbf{k}_\gamma) = \iint \overbrace{U(\mathbf{x}+\frac{\mathbf{r}}{2},\omega)U^*(\mathbf{x}-\frac{\mathbf{r}}{2},\omega)\exp(-i\mathbf{k}_\nu \cdot \mathbf{x})}^{(\mathbf{a})} \; \overbrace{\exp(-i\mathbf{k}_\gamma \cdot \mathbf{r})}^{(\mathbf{b})} \, d\mathbf{r}\, d\omega \; . \qquad (5.6)$$

In its most general form, equation 5.6 represents a bilinear phase-space kernel**,** where the image is constructed from local correlations of the wavefield through a Wigner-type distribution. Specifically, the integrand consists of the product of wavefields evaluated at two nearby points, $\mathbf{x} \pm \frac{\mathbf{r}}{2}$ , modulated by a complex exponential kernel of the form $e^{-i\mathbf{k}_\gamma \cdot \mathbf{r}}$. This kernel performs a localized Fourier transform with respect to the spatial lag vector $\mathbf{r}$, mapping spatial correlations into scattering (wavenumber) components. As a result, equation 5.6 provides a joint representation of the wavefield in both space and local wavenumber, capturing both amplitude and directional information.

In equation 5.6, Term (a) represents the spatial Fourier transform of two-point correlations of the wavefield in the vicinity of location $\mathbf{x}$, comparing wavefields that are slightly shifted in space. This term captures the local scattering behavior, rather than merely the energy at a single point. The separation vector $\mathbf{r}$ defines the spatial lag, i.e. a displacement relative to the scattering point $\mathbf{x}$, within a small neighborhood centered at $\mathbf{x}$.

Term (b) projects the local wavefield correlation onto specific scattering directions by mapping spatial lags onto opening-angle components within the reflection plane (orthogonal to the propagation direction $\hat{\mathbf{k}}_\nu$). This operation effectively decomposes the signal into angle-resolved scattering contributions.

It is important to note that a direct implementation of equation 5.6 using numerical wave-equation solvers (e.g. elastic or viscoelastic formulations) is computationally demanding in terms of both runtime and memory. Specifically, at each FWI iteration, the phase-space decomposition must be evaluated at every time step and for all spatial grid points. To render the proposed LAD-FWI formulation computationally feasible - while preserving its generality - practical approximations can be applied.

**5.4 Born Approximation**

The Born approximation considers small model perturbations $\delta\mathbf{m}(\mathbf{x})$ and neglects multiple scattering (e.g., Beylkin, 1985; Beylkin and Burridge, 1990), which makes it possible to convert the nonlinear equation 5.6 to a directly solvable equation in which the relation between the model perturbations and the scattered data becomes linear,

$$U_{\text{sc}}(\mathbf{x},\omega) = \int G(\mathbf{x},\mathbf{x}',\omega)\,\delta\mathbf{m}(\mathbf{x}')\,U_{\text{in}}(\mathbf{x}',\omega)\,d\mathbf{x}', \tag{5.7}$$

where $G$ is the Green's function connecting the subsurface scatterer point $\mathbf{x}'$ to the image point $\mathbf{x}$ , $\delta\mathbf{m}(\mathbf{x}')$ is a small model perturbation, and $U_{\text{in}}$ is the assumed known incident field at $\mathbf{x}'$. Substituting (5.7) into (5.6), the phase-space image becomes linear in $\delta\mathbf{m}$, so that,

$$I(\mathbf{x},\mathbf{k}_\nu,\mathbf{k}_\gamma) \propto \int \ \mathbf{K}(\mathbf{x},\mathbf{x}',\mathbf{k}_\nu,\mathbf{k}_\gamma)\,\delta\mathbf{m}(\mathbf{x}')\,d\mathbf{x}', \tag{5.8}$$

where $\mathbf{K}$ denotes the phase-space sensitivity kernel,

$$\mathbf{K}(\mathbf{x},\mathbf{x}',\mathbf{k}_\nu,\mathbf{k}_\gamma) = \int d\omega \int d\mathbf{r}\, G(\mathbf{x}+\frac{\mathbf{r}}{2},\mathbf{x}',\omega)\ \ G(x-\frac{\mathbf{r}}{2},\mathbf{x}',\omega) \left|U_{\text{in}}(\mathbf{x}',\omega)\right|^2 \exp\left(-i\mathbf{k}_\gamma \cdot \mathbf{r}\right) \quad , \tag{5.9}$$

associated with the Born-linearized imaging operator. Equivalently, $\mathbf{K}$ is the integral-kernel representation of the Jacobian of the phase-space image with respect to the model perturbation, evaluated about the reference model $\mathbf{m}_0$, $\mathbf{J}^{(I)} = \partial I/\partial \boldsymbol{m}$. In other words, the $\mathbf{K}$ kernel or the $\mathbf{J}$ operator defines the sensitivity of the multidimensional phase-space image to (small) changes in the model parameters (e.g., velocity-type and reflectivity-type perturbations).

The corresponding Gauss-Newton Hessian is obtained from the normal operator $\mathbf{H} \approx \mathbf{K}^{*}\mathbf{K}$, which acts as the model-resolution (or blurring) operator. Individual columns of $\mathbf{H}$ define the point-spread functions (PSFs) that characterize the spatial and directional resolution attainable in the inversion.

The scattered wavefield is now linearly related to small perturbations in the model parameters. Under this assumption, the recorded wavefield can be expressed as a superposition of singly scattered contributions, and although equation 5.6 is bilinear in the wavefields, substitution of the Born-scattered field renders the resulting image to be linear with respect to the model perturbation $\delta\mathbf{m}$.

As mentioned, this linearization suppresses (i.e., neglects) multiple scattering effects and enables the interpretation of the phase-space image as a superposition of independent scattering events. Consequently, each phase-space bin $(\mathbf{k}_{\nu}, \mathbf{k}_{\gamma})$ can be associated with a unique local interaction between the incident field and the perturbation, which significantly simplifies the inversion framework.

The phase-space image data $I(\mathbf{x}, \mathbf{k}_{\nu}, \mathbf{k}_{\gamma})$ is discretized on a seven-dimensional (7D) grid comprising three spatial coordinates $\mathbf{x} = \{x, y, z\}$, and two pairs of spherical angular parameters $\boldsymbol{\nu} = \{\nu_1, \nu_2\}$ and $\boldsymbol{\gamma} = \{\gamma_1, \gamma_2\}$ (4D in total) representing the direction angles of the orthogonal propagation and scattering wavenumber vectors, respectively. The discretization of this multidimensional coordinate system should be designed to follow the local frequency bandwidth and dominant wavenumber content of the in-situ scattering wavefield, thereby adapting the sampling density to the expected spatial and angular variations of the data.

**Figure 2** summarizes the general workflow of LAD imaging applied to real data. The first column on the left shows the input data, consisting of the reference subsurface model (bottom) and the recorded seismic data represented by common shot gathers (top). The second column illustrates the LAD migration engine, which maps the seismic wavefield into subsurface image points and bins the resulting image contributions into multidimensional 5D LAD gathers. An example of a single 5D LAD gather is displayed in the third column. The vertical axis represents subsurface depth, while the remaining dimensions correspond to the two LAD coordinate systems: (1) the illumination (propagation) direction, parameterized by dip and azimuth angles $\boldsymbol{\nu}$, and (2) the opening (scattering) angle and azimuth, parameterized by $\boldsymbol{\gamma}$.

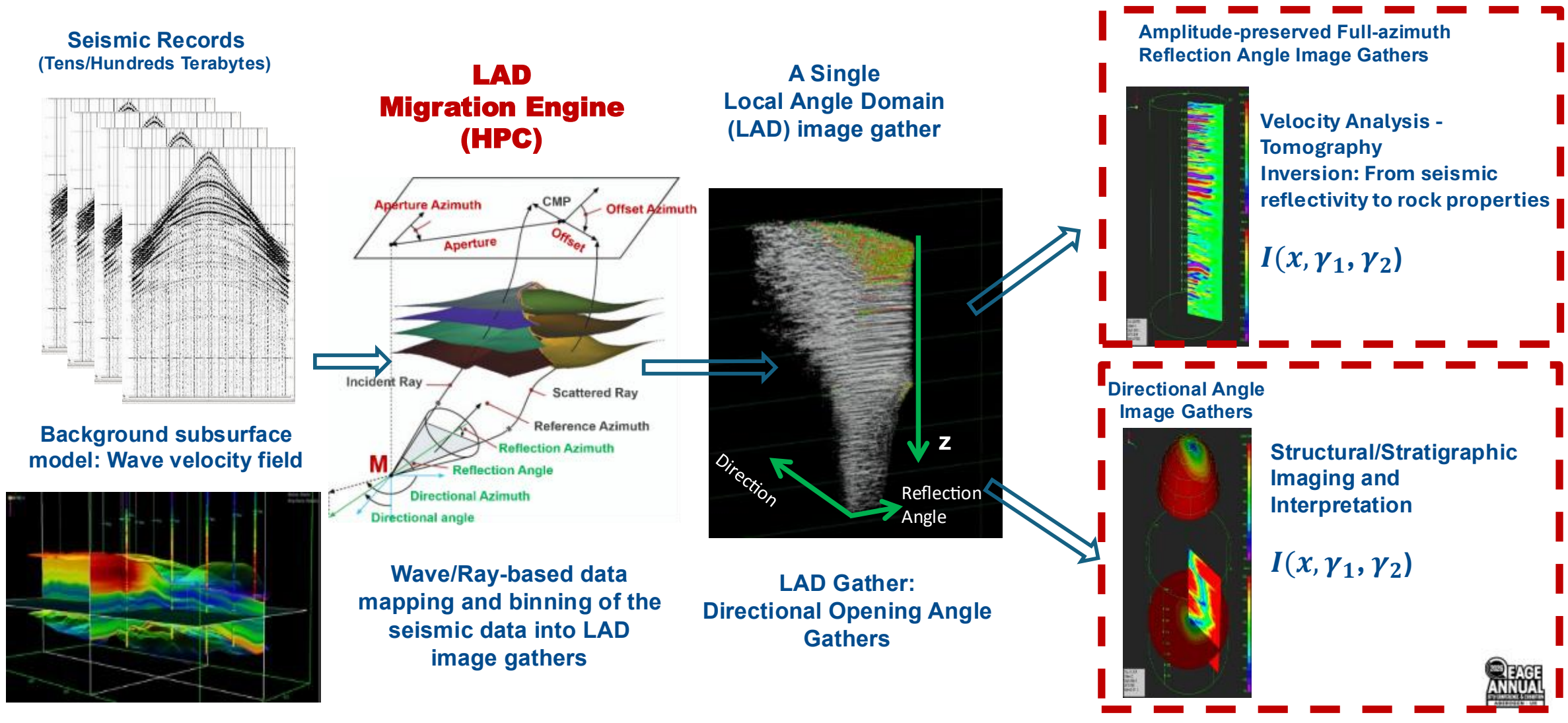

**Figure 2.** General workflow of LAD imaging applied to real data. **Left column:** input data - the reference (background) subsurface velocity-model (bottom) and the recorded seismic data shown as common-shot gathers (top). **Second column:** the LAD migration engine, which maps the recorded wavefield to subsurface image points and bins the image contributions into multidimensional 5D LAD gathers. **Third column:** an example of a single 5D LAD image gather; the vertical axis is subsurface depth, and the remaining dimensions are the two LAD angular coordinate systems - the illumination (propagation) direction, parameterized by dip and azimuth $(\nu_1, \nu_2)$, and the opening (scattering) angle and azimuth, parameterized by $(\gamma_1, \gamma_2)$. **Right column:** two complementary 3D full-azimuth angle gathers (cylindrical display) obtained by partial integration of the 5D gather - an amplitude-preserved, azimuthally varying reflection-angle image gather $I(\mathbf{x}, \gamma_1, \gamma_2)$, formed by weighted integration over the directional coordinates and used for velocity-analysis tomography (VVAZ and AVAZ) and rock-property inversion (top), and a directional-angle (directional-energy) gather, formed by weighted integration over the opening-angle coordinates and used for structural/stratigraphic imaging and interpretation (bottom). Together these projections separate the local scattering and propagation characteristics of the wavefield and form the basis for the proposed LAD-FWI. The last column presents two types of 3D full-azimuth angle gathers obtained by partial integration of the 5D LAD gather. The upper panel shows an azimuthally varying reflection-angle image gather, generated by weighted integration of the LAD image $I$ over the directional coordinates $\mathbf{\Omega_\nu}$,

$$I_\gamma(\mathbf{x}, \boldsymbol{\gamma}) = \int \mathbf{W}_\nu\,(\mathbf{x}, \boldsymbol{\nu}, \boldsymbol{\gamma})\, I(\mathbf{x}, \boldsymbol{\nu}, \boldsymbol{\gamma})\, d\boldsymbol{\nu}.$$

The lower panel displays the directional-energy gather, obtained by weighted integration of the LAD energy $I^2$over the opening-angle coordinates $\mathbf{\Omega_\gamma}$,

$$E_\nu(\mathbf{x}, \boldsymbol{\nu}) = \int \mathbf{W}_\gamma\,(\mathbf{x}, \boldsymbol{\nu}, \boldsymbol{\gamma})\, I^2(\mathbf{x}, \boldsymbol{\nu}, \boldsymbol{\gamma})\, d\boldsymbol{\gamma}.$$

Together, these complementary projections provide separate views of the local scattering and propagation characteristics of the wavefield and form the basis for the LAD analysis employed throughout the proposed framework.

***Sensitivity of velocity-type and reflectivity-type parameters in phase space***

Assuming a locally stationary scattering event with equal incident and scattered wavenumber magnitudes, $|\mathbf{k}_{in}| = |\mathbf{k}_{sc}| = k$, the magnitude of the directional wavenumber can be written as (e.g., Beylkin, 1985; Bleistein et al., 2001),

$$k_\nu = |\mathbf{k}_\nu| = 2k\cos\left(\frac{\gamma}{2}\right),\; k = \frac{\omega}{v_{\text{phs}}\left[\mathbf{c}(\mathbf{x}), \hat{\mathbf{k}}_\nu\right]}, \tag{5.10}$$

where $\gamma$ denotes the scattering (opening) angle and $v_{\text{phs}}$ is the local phase velocity.
Equation (5.10) establishes a direct relationship between scattering angle and directional bandwidth through the obliquity factor, $\frac{k_\nu}{2k} = \cos\left(\frac{\gamma}{2}\right)$. This relationship forms the foundation for separating velocity and reflectivity sensitivities in phase space.

**Figure 3** illustrates the behavior of $k_\nu$ for two limiting cases. For large opening angles ($\gamma \to 180°$), the directional wavenumber approaches zero ($k_\nu \to 0$). This regime corresponds to wide-angle propagation and low spatial wavenumbers. The corresponding image-space components primarily constrain the long-wavelength background velocity and anisotropic model parameters. In contrast, for small opening angles ($\gamma \to 0°$), the directional wavenumber reaches its maximum value ($k_\nu \to 2k$), providing the highest spatial resolution and the greatest sensitivity to directional scattering perturbations such as reflectivity and diffractivity. In this limit, specular reflections are concentrated near reflector-normal directions, whereas diffraction energy is distributed over a wider range of propagation directions.

## The LAD System : Representing direct (diving) waves as a limiting case of scattering waves – with zero reflectivity

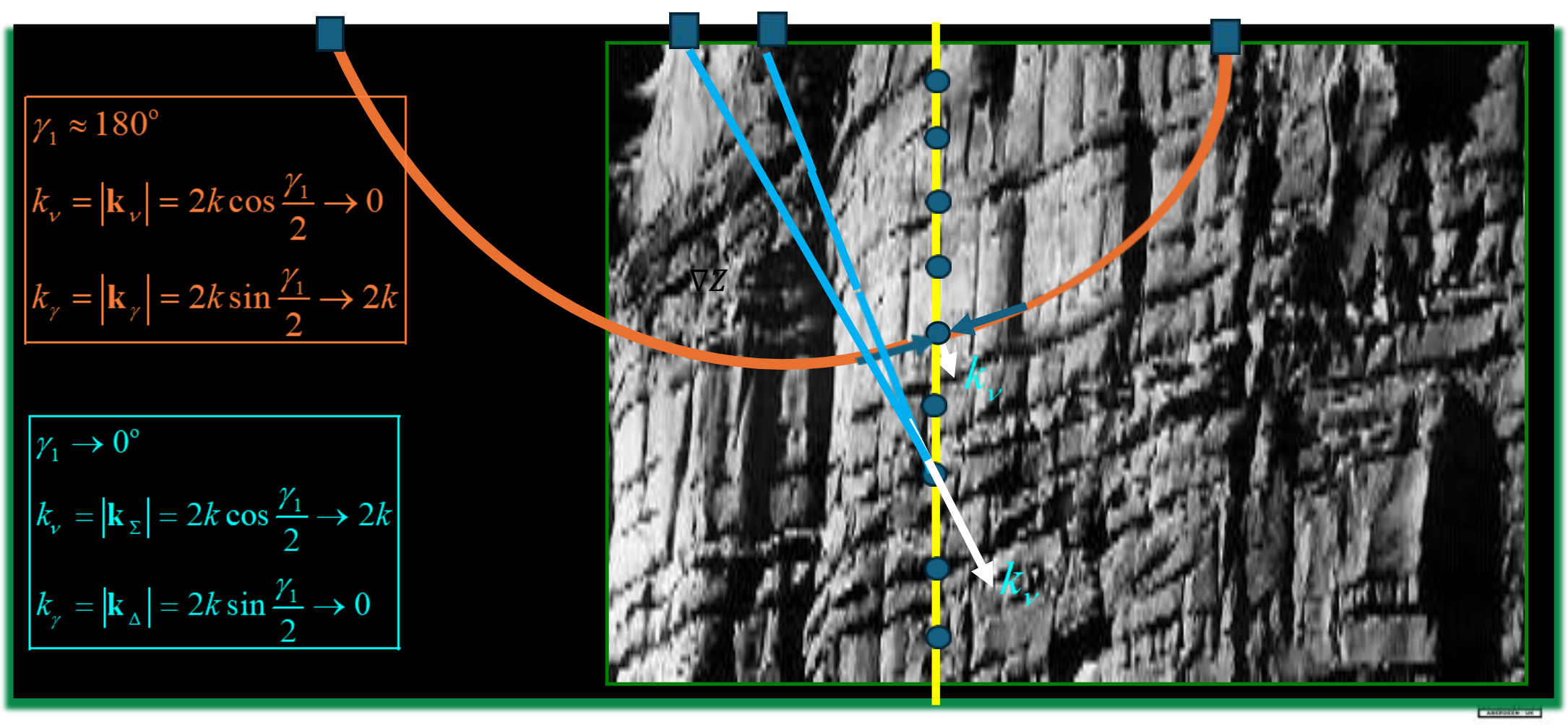

**Figure 3.** Two limiting scattering configurations. **Blue:** small opening angle ($\gamma \approx 0°$) and large directional wavenumber ($|\mathbf{k}_\nu| \approx 2k$). **Orange:** wide opening angle ($\gamma \to 180°$) and small directional wavenumber ($|\mathbf{k}_\nu| \to 0$), corresponding to a transmitted-wave configuration through the scattering point. The right panel shows the obliquity factor of equation (5.10), which links the opening angle to the magnitude of the directional wavenumber; the two limiting configurations are marked on the curve.

The two projected domains emphasize complementary physical properties: the reflection-angle gathers are primarily sensitive to background kinematics and anisotropy, whereas the directional-energy gathers emphasize propagation directivity, scattering diversity, and non-specular wavefield components.

**Figure 4** presents the normalized LAD distribution of reflectivity and velocity sensitivities for three representative reflector dips ($\theta = 0°, 30°$, and $60°$) in a 2D medium. The horizontal axis represents the normalized directional wavenumber, $k_\nu = \sin(\nu)$, with $k_\nu \in [-1, 1]$, while the vertical axis corresponds to the half-opening angle, $\gamma/2 \in [0°, 90°]$. The sensitivity functions shown in the figure are computed using the expressions given below.

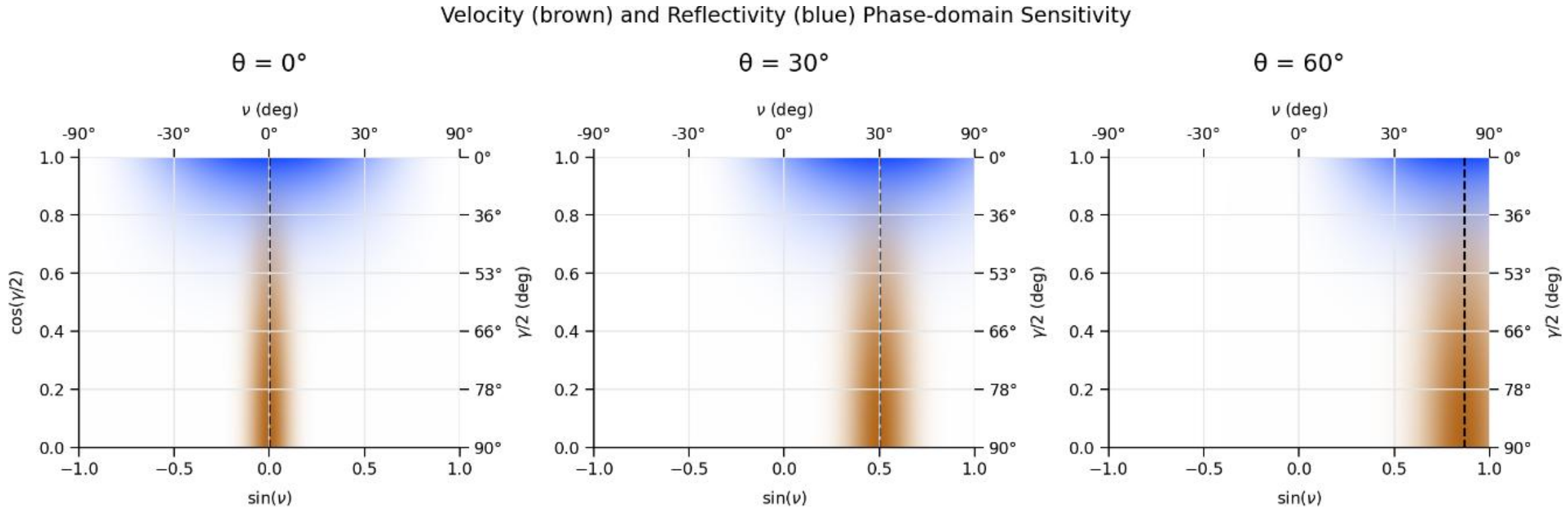

**Figure 4.** Velocity (brown) and reflectivity (blue) phase-domain sensitivity kernels in $(\sin\nu, \cos(\gamma/2))$ space for reflector dips $\theta = 0°, 30°$ and $60°$. The top and right axes show the corresponding angles $\nu$ and $\gamma/2$. Dashed vertical lines mark the specular direction $\sin(\nu_0) = \sin(\theta)$. Both sensitivities shift and broaden with dip, reflecting their kinematic nature and indicating increased parameter coupling at higher dips.

Reflectivity sensitivity remains strongly localized around the dip-dependent specular direction, whereas velocity sensitivity occupies a substantially broader region of phase space. As reflector dip increases, the velocity sensitivity spreads further away from the specular direction, reflecting its predominantly kinematic origin and resulting in stronger parameter coupling.

In summary, the LAD representation reveals a fundamental distinction between velocity-type and reflectivity-type sensitivities:

- ***Reflectivity sensitivity*** is concentrated at small opening angles and around the dip-dependent specular direction. Under the Born approximation, it remains highly localized in phase space because it originates from interface scattering. Consequently, reflectivity updates are associated with broadband directional wavenumbers and high-resolution spatial information.
- ***Velocity sensitivity***, in contrast, exhibits a predominantly kinematic character. It vanishes at zero opening angle and increases with scattering angle, approximately following a $\sin^2(\gamma/2)$ dependence. As a result, velocity sensitivity occupies a broader region of phase space and reaches its maximum at large opening angles, where it primarily constrains the low-wavenumber components of the background model.

These observations indicate that, under the Born approximation, velocity-type and reflectivity-type perturbations occupy distinct, albeit partially overlapping, regions of phase space. Reflectivity sensitivity remains focused near specular directions, whereas velocity sensitivity broadens with increasing reflector dip and scattering angle. The overlap between the two sensitivity patterns becomes more pronounced for steeply dipping structures, resulting in stronger parameter coupling and increased potential for inversion crosstalk.

This separation provides the theoretical foundation for the proposed LAD-FWI formulation. By defining objective functions within carefully selected regions of phase space, velocity and reflectivity updates can be estimated with reduced mutual interference, thereby improving parameter discrimination, conditioning, and inversion stability.

**Appendix A (Figures A1** and **A2)** provides a more detailed analysis of the sensitivity kernels, resolution characteristics, and parameter conditioning associated with the LAD-FWI framework.

**5.5 Local and Global Coordinate Systems**

The LAD representation may be defined in either a global Cartesian coordinate system or a local, structure-oriented coordinate system. In the global system, directional angles are measured with respect to a fixed vertical axis. In the local system, the coordinate frame is rotated such that its vertical axis coincides with the local reflector-normal direction, $\mathbf{n_r}$. We adopt the latter because it conforms naturally to the structural geometry of the reference model. In this framework, small directional angles correspond to propagation directions close to the local reflector normal, enabling physically meaningful migration apertures that are symmetrically defined around $\mathbf{n_r}$ and thereby improving amplitude preservation in angle-domain imaging. Although the local parameterization introduces additional nonlinearity, since the coordinate system evolves with the reference model, its structural consistency provides significant advantages for inversion. Accordingly, all directional wavenumbers $\mathbf{k}_\nu$ and directional angles $\nu$are defined in the local coordinate system, with $\nu_1 = \nu_2 = 0$ representing propagation aligned with $\mathbf{n_r}$.

**5.6 Inward and Outward Scattering Wavefields**

At each image location, the scattering wavefield is further separated into **inward** and **outward** components according to the sign of $\mathbf{k}_\nu \cdot \mathbf{n_r}$. Inward scattering ($\mathbf{k}_\nu \cdot \mathbf{n_r} > 0$) corresponds to propagation directions aligned with the reflector normal, whereas outward scattering ($\mathbf{k}_\nu \cdot \mathbf{n_r} < 0$) represents wavefields arriving from opposing directions. This decomposition improves the analysis of illumination and directional bias and is particularly important for velocity updating because inward and outward wavefields generally sample different propagation paths, especially in structurally complex regions. Furthermore, the two families may exhibit different illumination strengths and opposite reflectivity polarities, reflecting mirrored sampling of the local impedance gradient. Treating them separately therefore provides a more reliable characterization of subsurface illumination and contributes to more stable and physically consistent model updates.

## 6. LOCAL DIRECTIONAL PROJECTION (LDP) SIGNALS

A fundamental limitation of seismic imaging (migration), in which the image is represented by uniformly discretized vertical traces, is the well-known signal-stretching (deformation) effect. This phenomenon results from projecting time-domain signals, sampled at constant intervals $dt$ and associated with scattering angles $\theta$ (representing both $\gamma_1$ and $\nu_1$) onto the vertical axis, which is discretized at constant vertical-time intervals $dt_v$. Since $\frac{dt}{dt_v} \approx \frac{1}{\cos\theta}$, the amount of stretching increases with the increasing scattering angle. Consequently, signals associated with large $\theta$ values become progressively elongated, resulting in waveform distortion and reduced vertical resolution.

To mitigate this limitation, Kletenik-Edelman et al. (2025) proposed a post-imaging procedure for generating local directional projection (LDP) signals from existing directional angle image gathers, $I(\mathbf{x}, \boldsymbol{\nu})$, using the Fourier slice theorem. **Figure 5** schematically illustrates the LDP construction process: (a) back-projection of a local 3D spatial window to a specified directional plane (out of the many others), (b) the corresponding LDP vector and projected signal (brown), and (c) the complete ensemble of LDP vectors and signals spanning all illumination directions.

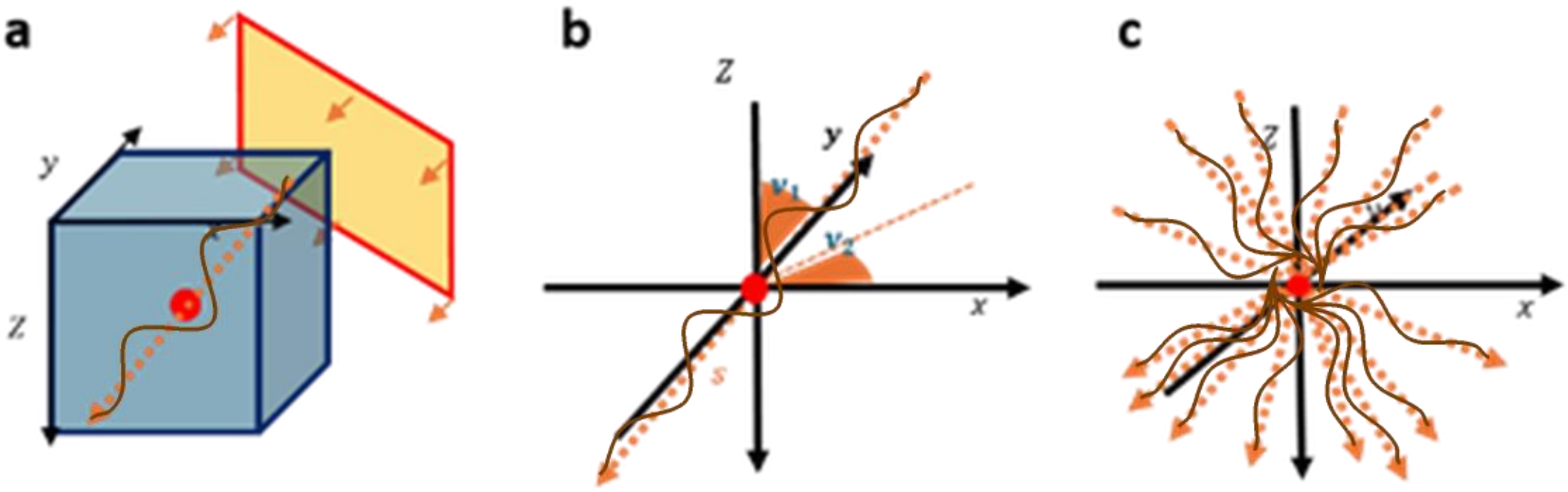

**Figure 5**. Illustration of the construction of local-directional-projection (LDP) signals. (a) back-projection of a local 3D spatial window for generating a given direction plane, (b) the corresponding LDP vector and signal (brown), and (c) the set of all LDP vectors. Representative continuous low-frequency sinusoidal signals (brown) are drawn along the orange dashed propagation directions, Gaussian-modulated so that the amplitudes are largest at the central node and decay toward the ends; these are the locally projected (LDP) signals associated with each direction.

In the present study, we extend this concept by generating the LDP signals directly during the imaging process. Rather than mapping isolated data samples onto the vertical image axis during the migration process, we project localized wavefield segments of duration $T$, centered about the stationary propagation time $T_{sr}$. Introducing the local time-lag coordinate, $\tau \in \left[-\frac{T}{2}, \frac{T}{2}\right]$ the wavefield time is expressed as $t = T_{sr} + \tau$.

The resulting image is no longer represented as a collection of vertical traces. Instead, it is defined on a spatially varying grid whose nodes contain local directional projection (LDP) time signals (see the synthetic example on **Appendix F**). Accordingly, the imaging process described in Section 5 is reformulated for the LDP workflow as,

$$D_{\mathrm{obs}}(\mathbf{x}_{\mathrm{s}},\mathbf{x}_{\mathrm{r}},t)\xrightarrow{\mathfrak{M}(\mathbf{m})} I(\mathbf{x}_{\mathrm{p}},\mathbf{k}_{\mathrm{in}},\mathbf{k}_{\mathrm{sc}},\tau)\xrightarrow[\mathbf{k}_{\gamma}=\mathbf{k}_{\mathrm{in}}-\mathbf{k}_{\mathrm{sc}}]{\mathbf{k}_{\nu}=\mathbf{k}_{\mathrm{in}}+\mathbf{k}_{\mathrm{sc}}} I(\mathbf{x}_{\mathrm{p}},\mathbf{k}_{\nu},\mathbf{k}_{\gamma},\tau)\rightarrow I(\mathbf{x}_{\mathrm{p}},\boldsymbol{\nu},\boldsymbol{\gamma},\tau)\,. \tag{6.1}$$

The resulting dataset, $I(\mathbf{x}_{\mathrm{p}},\boldsymbol{\nu},\boldsymbol{\gamma},\tau)$, constitutes the primary input for the proposed LAD-FWI workflow and for high-resolution imaging applications. The dataset is eight-dimensional (8D), comprising three spatial coordinates, $\boldsymbol{x}_{\mathrm{p}}$, two pairs of spherical angular parameters, $\boldsymbol{\nu} = \{\nu_1, \nu_2\}$ and $\boldsymbol{\gamma} = \{\gamma_1, \gamma_2\}$, and a local time-lag coordinate $\tau$. The discretization of this multidimensional coordinate system should be designed according to the local frequency bandwidth and dominant wavenumber content of the scattered wavefield, thereby adapting the spatial, angular, and temporal sampling densities to the expected variability of the data.

It is important to emphasize that the 8D representation does not necessarily require a proportional increase in data volume. The same seismic information used to construct the 7D image volume $I(\mathbf{x},\mathbf{k}_{\nu},\mathbf{k}_{\gamma})$ is reorganized to form $I(\mathbf{x}_{\mathrm{p}},\mathbf{k}_{\nu},\mathbf{k}_{\gamma},\tau)$, where the resolution of the space grid points $\mathbf{x}_{\mathrm{p}}$ is lower than the resolution of $\mathbf{x}$: the additional $\tau$ dimension primarily replaces part of the densely sampled vertical axis rather than introducing entirely new information. **Appendix B (Figure B1)** provides a detailed discussion of the discretization criteria for each coordinate axis.

The projection length $T$ is chosen to be proportional to the local dominant period and therefore inversely proportional to the dominant frequency. This wave-packet-based projection preserves the intrinsic temporal structure of the scattered signal, including its phase and spectral characteristics. As a result, the LDP representation provides a more faithful description of the in-situ scattering wavefield while remaining fully consistent with the underlying LAD formulation.

For obtaining high-resolution images defined with fine spatial grid points $\mathbf{x}_{\mathrm{h}}$, the directional projection maps a signal sample associated with lag $\tau$ to the spatial location,

$$\mathbf{x}_{\mathrm{h}} = \mathbf{x}_{\mathrm{p}} + \boldsymbol{v}_{\mathrm{ray}}\,[\boldsymbol{c}(\mathbf{x}), \boldsymbol{n}_{\boldsymbol{\nu}}]\,\tau\,, \qquad \mathbf{n}_{\nu} = \hat{\mathbf{k}}_{\nu} = \frac{\mathbf{k}_{\nu}}{\|\,\mathbf{k}_{\nu}\,\|}, \tag{6.2}$$

where $\boldsymbol{v}_{\mathrm{ray}}$ denotes the local group-velocity corresponding to the phase-propagation direction $\mathbf{n}_{\nu}$, computed from the elastic model parameters through the Christoffel equation. The projection operation can therefore be expressed as,

$$I(\mathbf{x}_{\mathrm{p}},\mathbf{k}_{\nu},\mathbf{k}_{\gamma},\tau)\rightarrow I(\mathbf{x}_{\mathrm{h}},\mathbf{k}_{\nu},\mathbf{k}_{\gamma})\;. \tag{6.3}$$

The principal advantage of the LDP representation is that it preserves the original temporal waveform while distributing the signal energy along physically meaningful propagation directions. Multiple independently illuminated wave packets contribute to each image location from different dips, azimuths, and opening angles. Their collective response provides enhanced focusing of localized reflectivity features and a richer description of the underlying scattering process. Within the LAD framework, this multidirectional accumulation is expected to yield higher effective resolution and improved discrimination of scattering mechanisms than conventional stacked-image representations.

The LDP-projected signals can be viewed as time signals illuminating a specific point-like subsurface region from all available directions and opening angles. Naturally, the degree of localization depends on the accuracy of the background model. In this sense, the proposed LAD-FWI can be regarded as a Time-LAD-FWI.

In the next section we will investigate the sensitivity of the LDP image representation $I(\mathbf{x}_{\mathrm{p}},\mathbf{k}_{\nu},\mathbf{k}_{\gamma},\tau)$ to changes of the model parameters $\delta\mathbf{m} = \{\delta V, \delta\mathbf{R}\}$ - velocity-type and reflectivity-type perturbations, respectively-under the Born approximation.

### 6.1 Separation into Two Phase-Domain Projected Image Data: Directional and Opening Angle

Note that for a *true* model and any fixed scattering wavenumber $\mathbf{k}_{\boldsymbol{\gamma}}$ - or equivalently, fixed scattering angle $\boldsymbol{\gamma}$ - the integral,

$$I(\mathbf{x},\gamma,\tau)=\int_{\Omega_{\mathbf{v}}} W_{\nu}(\mathbf{x}_{\mathrm{p}},\mathbf{v},\gamma)\ I(\mathbf{x}_{\mathrm{p}},\mathbf{v},\gamma,\tau)\ d\mathbf{v}\ , \tag{6.4}$$

yields physically consistent subsurface image data. This type of image data provides the optimal domain for reflection-angle analysis, including residual moveout estimation for velocity updating and amplitude analysis through VVA(Z) and AVA(Z).

Moreover, to obtain amplitude-preserved angle-domain reflectivities, it is preferable to restrict the integration in equation 6.4 to the vicinity of the specular direction, where the integration range should include only the Fresnel-based aperture (a small range of directional angles).

In contrast, the integral,

$$I(\mathbf{x},\mathbf{v},\tau)=\int_{\Omega_{\gamma}} W_{\gamma}(\mathbf{x}_{\mathrm{p}},\mathbf{v},\boldsymbol{\gamma})\ I(\mathbf{x}_{\mathrm{p}},\mathbf{v},\boldsymbol{\gamma},\tau)\, d\boldsymbol{\gamma}\ , \tag{6.5}$$

yields a directional image corresponding to a single propagation direction $\nu$. Owing to the strong amplitude and phase variations across directional angles, these images enable the separation of specular reflection energy from non-specular components, such as diffractions and other scattered-wave events.

As previously discussed, the signals associated with small opening angles provide higher sensitivity and higher resolution than those associated with wide opening angles. Hence, integral 6.5 can be performed with relatively small opening angles - a trade-off between high resolution (small opening angles) and stability (larger opening angles) should be considered.

It is important to mention at this stage that the size of the full LAD projected data $I(\mathbf{x}_{\mathrm{p}},\mathbf{v},\boldsymbol{\gamma},\tau)$ can be prohibitively large. Separating this huge multidimensional volume into (1) specular scattering-angle projection data of feasible size (equation 6.4) and (2) directional-angle projection data with small opening angles (equation 6.5) makes the generation of the LAD data considerably more efficient, with much smaller data sizes.

Here, $\mathbf{W}_{\nu}(\mathbf{x}_{\mathbf{p}},\mathbf{v},\boldsymbol{\gamma})$ and $\mathbf{W}_{\gamma}(\mathbf{x}_{\mathbf{p}},\mathbf{v},\boldsymbol{\gamma})$, are weighting functions that primarily account for non-uniform illumination in phase space. So, from this point forward, the LAD image data consists of two types of projected image data:

$I_{\gamma}=I(\mathbf{x}_{\mathrm{p}},\gamma,\tau)$ - specular-directed, omni-scattering angle/azimuth projected signals, and
$I_{\mathrm{v}}=I(\mathbf{x}_{\mathrm{p}},\mathbf{v},\tau)$ - omni-directional dip/azimuth generated with small opening angles.

For multicomponent elastic data, the LDP construction generalizes naturally to mode-decomposed projected signals: the wavefield at the image point is first decomposed into its quasi-compressional and quasi-shear constituents by projection onto the polarization eigenvectors of the local Christoffel matrix, and the directional projections are then formed per wave mode. This extension, which underlies converted-wave imaging and inversion within the proposed framework, is developed in **Appendix D**.

## 7. CHOICE OF INVERSION PARAMETER: IMPEDANCE VS. REFLECTIVITY

The sensitivity kernels derived in the preceding sections are expressed naturally in terms of the reflectivity vector fields **R**, because seismic scattering responds to the local impedance contrast rather than to the impedance itself. The reflectivity is not, however, an independent field: by construction (equation 3.4) each reflectivity vector is the relative gradient of a scalar impedance field, $\mathbf{R}_c = ½\nabla \ln Z_c$, for $c \in \{P, S\}$. Its three components are therefore constrained by the curl-free condition $\nabla \times \mathbf{R}_c = 0$, so a direct inversion for $\mathbf{R}$ as an unconstrained vector field is over-parameterized, carrying three degrees of freedom per point where the underlying physics has one.

For this reason, we retain the reflectivity vectors as the natural variables of the sensitivity operator but adopt the scalar impedance fields Z (equivalently the log-impedance fields) as the inversion unknowns. The two are linked, per parameter class $c \in \{P, S\}$, by the linear gradient operator G,

$$\mathbf{R}_c(\mathbf{x}) = \frac{1}{2}\nabla \ln Z_c(\mathbf{x}) = \mathbf{G}_c[Z_c] \quad , \tag{7.1}$$

where $\mathbf{G}$ denotes the (curl-free) relative-gradient operator $\mathbf{G}[\bullet] = ½\nabla\ln(\bullet)$. The phase-space Jacobian and Gauss-Newton Hessian with respect to impedance then follow directly from those with respect to reflectivity by the chain rule,

$$\mathbf{J}_Z = \mathbf{J}_R\,\mathbf{G} \quad , \tag{7.2}$$

and, for the reflectivity-sensitive Hessian block and its velocity coupling,

$$\mathbf{H}_{ZZ} = \mathbf{G}^T\,\mathbf{H}_{RR}\,\mathbf{G},\;\; \mathbf{H}_{VZ} = \mathbf{H}_{VR}\,\mathbf{G} \quad . \tag{7.3}$$

This parameterization offers three advantages that reinforce the central hypothesis of this work. First, it is minimal and integrable: solving for Z guarantees that the recovered reflectivity is a valid gradient field, automatically enforcing the curl-free constraint and ensuring the model integrates back to a consistent impedance, which a free-$\mathbf{R}$ inversion does not. Second, it is better conditioned: one scalar unknown per point replaces three coupled components, reducing the null space and the crosstalk among the reflectivity components themselves. Third, because the operator $\mathbf{G}$ is a spatial gradient, and hence high-pass in wavenumber, the mapping $\mathbf{H}_{ZZ} = \mathbf{G}^T\mathbf{H}_{RR}\mathbf{G}$ further concentrates the impedance sensitivity at high model wavenumbers, while the velocity sensitivity remains concentrated at low wavenumbers. The off-diagonal coupling block $\mathbf{H}_{VZ} = \mathbf{H}_{VR}\,\mathbf{G}$ is thereby suppressed relative to the raw $\mathbf{H}_{VR}$ block, strengthening the block-diagonal structure of the LAD Hessian discussed in **Appendix A**.

### 7.1 Impedance versus Density

A complementary question concerns the choice of impedance, rather than density, as the contrast-type inversion parameter. Under the Born approximation, the scattered field produced by a localized density perturbation exhibits a radiation pattern concentrated at small opening angles with weak angular dependence, closely mimicking that of an impedance perturbation at near-normal incidence; conversely, at the large opening angles at which velocity perturbations dominate the scattering response, the sensitivity to density is negligible. For realistic surface-acquisition apertures, density therefore possesses no angular band in which it is the dominant observable: its estimation relies on subtle amplitude-versus-angle variations that are easily contaminated by velocity errors, attenuation, and elastic effects. The multiparameter Hessian associated with a (velocity, density) parameterization is accordingly ill-conditioned, with strong off-diagonal coupling, and density updates tend to absorb residuals originating from velocity inaccuracies (Wu and Aki, 1985; Tarantola, 1986; Forgues and Lambaré, 1997; Operto et al., 2013; Prieux et al., 2013). By contrast, the (velocity, impedance) parameterization approximately diagonalizes the radiation patterns: impedance is the parameter to which small-opening-angle reflectivity is directly and stably sensitive, while velocity governs the kinematics of wide-angle and transmitted arrivals. Within the LAD representation, this parameterization therefore maximizes the separation of parameter sensitivities across the opening-angle and directional-wavenumber support, which is precisely the property that the proposed inversion framework is designed to exploit. Should density itself be required, e.g., for reservoir characterization or pore-pressure applications, it is more robustly derived a posteriori from the inverted impedance and velocity fields, ρ = Z/V, than estimated as a primary inversion variable.

**Note**: We emphasize that the relation $\mathbf{R}_c$ = ½∇ln$Z_c$ represents an isotropic-impedance idealization. In general, anisotropic elastic media, the impedance is direction-dependent, and the scalar fields $Z$ should be understood as the compressional and shear impedance magnitudes associated with the $P$ and $S$ reflectivity classes, respectively; additional anisotropic parameters enter through the velocity term (Gholami et al., 2013; Alkhalifah and Plessix, 2014). The vector-reflectivity derivations of the preceding sections remain the more general object, from which the scalar-impedance inversion is obtained as the practical, well-posed specialization used in the implementation.

## 8. LAD-FWI GENERAL FORMULATION

The generic implicit-adjoint LAD-FWI workflow may be written symbolically as

$$\mathbf{m}^{k+1} = \mathbf{m}^k - \alpha\, \mathbf{H}_m^{-1}\mathbf{g}_m, \tag{8.1}$$

$\mathbf{g}_m$ is the gradient of the objective function $\Phi\left[I\left(\mathbf{x}_\mathrm{p}, \mathbf{v}, \boldsymbol{\gamma}, \tau \mid \mathbf{m}_\mathrm{o}\right)\right]$ with respect to model parameter $\mathbf{m}$, $\mathbf{H}_m^{-1}$ is an approximation of the inverse Hessian, and $\alpha$ is the step-length parameter.

The Hessian plays a central role in the inversion because it quantifies:

- the sensitivity of the objective function to model perturbations,
- the regions of the model constrained by the data,
- underdetermined portions of the model space,
- parameter trade-offs,
- expected resolution and uncertainty, and
- the degree of nonlinearity near the solution.

Under the Born approximation, the forward operator can be locally linearized and the Hessian approximated by the normal operator,

$$\mathbf{H}_m \approx \left(\frac{\partial\Phi}{\partial\mathbf{m}}\right)^* \left(\frac{\partial\Phi}{\partial\mathbf{m}}\right), \tag{8.2}$$

which becomes more diagonally dominant in phase space, $\mathbf{H} \equiv \mathbf{H}(\mathbf{k}_\mathrm{x}, \mathbf{k}_\mathrm{v}, \mathbf{k}_\gamma, f)$ .

### 8.1 Resolution Analysis

The inverse Hessian provides an estimate of model uncertainty and resolution. When the inverse exists,

$$\mathbf{Cov(m)} \propto \mathbf{H}_m^{-1}.$$

Consequently, large diagonal Hessian values correspond to well-constrained model parameters and low uncertainty, whereas small diagonal values indicate poor resolution and elevated uncertainty.

**Appendix A** presents a detailed analysis of Jacobian-based parameter sensitivity and Hessian-based resolution for both velocity-type and reflectivity-type parameters in LAD image data.

### 8.2 Objective Function for LAD-FWI

For simplicity, consider an isotropic Earth model parameterized by compressional velocity and impedance,

$$\mathbf{m}(\mathbf{x}) = \{V(\mathbf{x}), \mathrm{Z}(\mathbf{x})\}. \tag{8.3}$$

At each iteration, a background reflectivity vector $\mathbf{R}(\mathbf{x})$, characterized by amplitude $R$ and orientation $\mathbf{n_r}$, is recomputed from the updated impedance field using equation (3.4). Here, $\mathbf{R}(\mathbf{x})$ denotes an impedance-gradient reflectivity vector and should not be interpreted as a conventional interface reflection coefficient. This is the isotropic specialization of the parameterization discussed in Section 7; the scalar impedance field $Z$ is the inversion unknown, and the reflectivity vector **R** is the derived relative gradient $\mathbf{R} = ½\nabla \ln \mathrm{Z}$ of equation (3.4).

The parameter-sensitivity analysis presented in the preceding sections and **Appendix A** demonstrates that different LAD attributes exhibit complementary sensitivities to velocity and impedance perturbations. This motivates the following composite objective functional:

$$\Phi_{\mathrm{LAD-FWI}} = \sum_i \alpha_i \, \Phi_i, \tag{8.4}$$

where $\Phi_i$ denotes an individual physical constraint or data-domain attribute, and $\alpha_i = \alpha_i(\mathbf{k}_{\mathrm{x}}, \mathbf{k}_{\mathbf{v}}, \mathbf{k}_{\boldsymbol{\gamma}}, f)$ controls its relative contribution during a given inversion stage.
A representative formulation is,

$$\Phi_{\text{LAD-FWI}} = \alpha_1\Phi_1 + \alpha_2\Phi_2 + \alpha_3\Phi_3 + \alpha_4\Phi_4 + \alpha_5\Phi_5. \tag{8.5}$$

Each term is described below. **Figure 6** summarizes the overall iteration workflow and the role of each objective term: every iteration maps the observed data into the LAD image, evaluates the five objective terms, and assembles the composite gradient and Gauss-Newton Hessian that drive the model update of equation (8.1). After each update, the reflectivity vector is recomputed from the impedance field through equation (3.4), maintaining consistency between the inverted parameters, while an outer multiscale loop expands the recoverable wavenumber content and reschedules the weights, as discussed in the Inversion Strategy section below. **Figure 7** details the adjoint-state gradient computation performed within each such iteration.

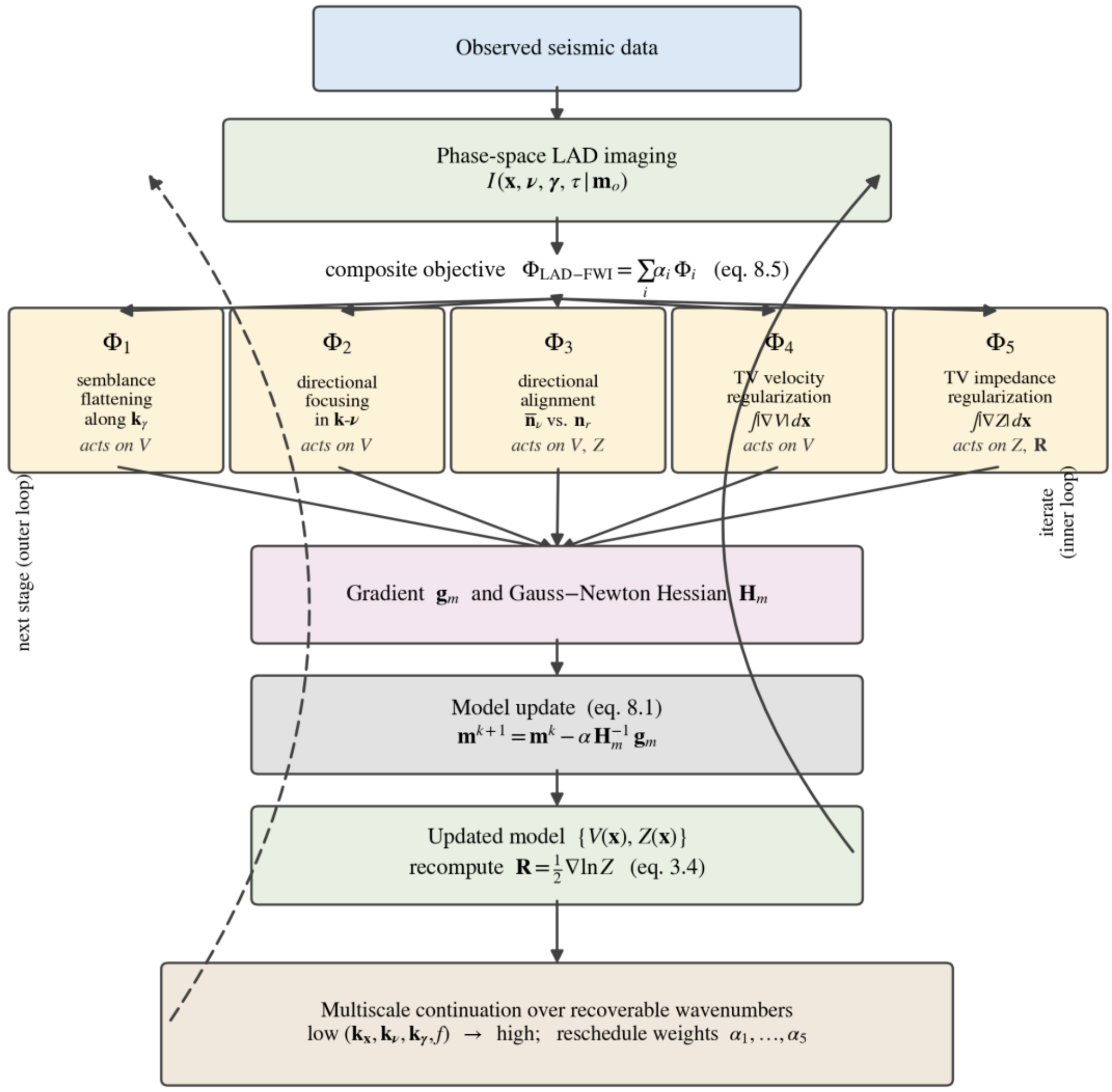

**Figure 6.** Iterative workflow of LAD-FWI. Each iteration maps the observed data into the LAD image $I(\mathbf{x}_p, \boldsymbol{\nu}, \boldsymbol{\gamma}, \tau \mid \mathbf{m}_o)$, evaluates the composite objective of equation (8.5) through the five terms $\Phi_1$ - $\Phi_5$, computes the gradient and Gauss-Newton Hessian with the adjoint-state method and assembles all the components, and updates the model via equation (8.1). The reflectivity vector $\mathbf{R} = \frac{1}{2}\nabla \ln Z$ is recomputed from the updated impedance (equation 3.4) before the next iteration. An outer multiscale loop expands the recoverable wavenumber content and reschedules the weights $\alpha_1, \ldots, \alpha_5$.

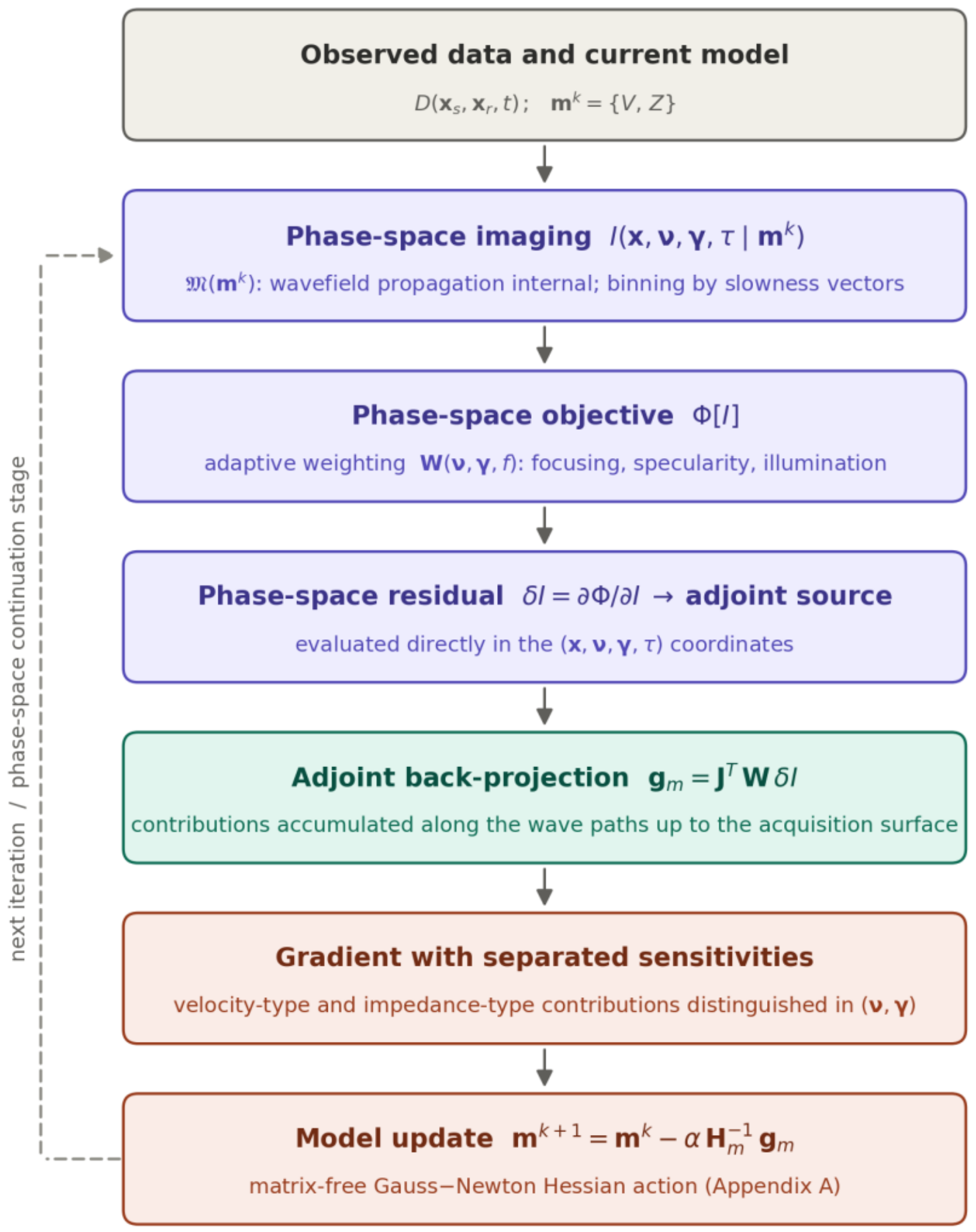

**Figure 7.** Adjoint-state gradient computation within a single LAD-FWI iteration. Each iteration begins by migrating the observed data into the LAD image through the LAD imaging operator evaluated in the current model; the propagation of the source and receiver wavefields is internal to this operator, not a separate synthetic-data modeling step. The objective and its adaptive weights are evaluated directly in phase space, and the resulting residual acts as the adjoint source. The gradient is assembled by back-projection through the adjoint imaging (Jacobian) operator, which distributes and accumulates the weighted residual contributions along the wave paths up to the acquisition surface, with velocity-type and impedance-type sensitivities separated in phase space. The model is updated using the matrix-free Gauss-Newton Hessian action (**Appendix A**); the dashed path indicates iteration and progression through the LAD continuation stages (**Appendix C, Figure C1**).

**Term (1): Semblance-Like Flattening along $\mathbf{k}_\gamma$**

The first term enforces kinematic consistency of reflection events in opening-angle common-image gathers. Ideally, image components associated with different opening angles should focus at the same subsurface location. This requirement may be expressed either as a semblance maximization problem or equivalently as a semblance-misfit minimization problem.

The objective functional is,

$$\Phi_1 = -\int S\,(\mathbf{x})\,d\mathbf{x}, \tag{8.6}$$

where $S$ denotes the semblance measure,

$$S = \frac{A^2}{B}\,, \quad A = \sum\nolimits_{\gamma} I(\mathbf{x},\boldsymbol{\gamma}) \quad , \quad B = N\sum\nolimits_{\gamma} I^2(\mathbf{x},\boldsymbol{\gamma}) \;. \tag{8.7}$$

The first variation yields,

$$\delta S = \frac{2A}{B}\delta A - \frac{A^2}{B^2}\delta B. \tag{8.8}$$

Comparison with the generic variational form identifies the image-space residual field. The corresponding adjoint source is proportional to the mismatch between the local image amplitude $I$ and the semblance-weighted stack amplitude $A/B$. The residual vanishes when all events are perfectly aligned and increases in the presence of residual moveout or imperfect focusing.

The velocity gradient follows directly from the adjoint-state relation.
The second variation is,

$$\delta^2 S = \frac{2}{B}(\delta A)^2 - \frac{4A}{B^2}\delta A\,\delta B + \frac{2A^2}{B^3}(\delta B)^2 + \frac{2A}{B}\delta^2 A - \frac{A^2}{B^2}\delta^2 B. \tag{8.9}$$

The individual contributions have clear physical interpretations:

- $\frac{2}{B}(\delta A)^2$ measures sensitivity to coherent stack-energy variations.
- $-\frac{4A}{B^2}\delta A\delta B$ represents coupling between coherent and total energy perturbations and therefore reflects parameter trade-offs.
- $\frac{2A^2}{B^3}(\delta B)^2$ measures sensitivity to illumination and amplitude variations.
- $-\frac{A^2}{B^2}\delta^2 B$ contains second-order nonlinear effects associated with the imaging operator.

The Hessian of the semblance objective is therefore,

$$\delta^2\Phi_1 = -\int \delta^2\,S\,d\mathbf{x}. \tag{8.10}$$

As noted above, semblance maximization may equivalently be recast as semblance-misfit minimization. Defining the flattening residual $r(\mathbf{x},\gamma) = I(\mathbf{x},\gamma) - \bar{I}(\mathbf{x})$, where $\bar{I} = A/N$ is the opening-angle-averaged image, the identity $1 - S = (N/B)\sum_{\gamma} r^2$ expresses the objective, up to an additive constant, as a normalized residual-energy functional. In the Gauss-Newton approximation, the normalization $1/B$ is frozen and the second-order imaging term is neglected; since $\delta r = \delta I - \delta\bar{I}$, with $\delta\bar{I} = N^{-1}\sum_{\gamma}\delta\,I$, a fixed orthogonal projection of $\delta I$ across opening angles, the image-space Hessian takes the manifestly positive semi-definite form,

$$\delta^2\Phi_1^{GN} = 2N\int \frac{1}{B}\sum_{\gamma}(\delta I - \delta\bar{I})^2\,d\mathbf{x}\,, \quad \mathbf{H}_I = \frac{2N}{B}\left(\mathbf{I}_N - \frac{1}{N}\mathbf{1}\mathbf{1}^T\right)\,, \tag{8.11}$$

where $\mathbf{I}_N$ denotes the identity over the $N$ opening-angle samples and $\mathbf{1}$ the corresponding vector of ones. This operator annihilates $\gamma$-uniform (flat) perturbations and penalizes precisely the residual-moveout components of $\delta I$.

For model parameters $\mathbf{m}$, and $\delta I = \mathbf{L}\,\delta\mathbf{m}$, $\mathbf{L} = \frac{\partial I}{\partial \mathbf{m}}$, and the model-space Hessian becomes,

$$\mathbf{H}_m = \mathbf{L}^T \mathbf{H}_I \mathbf{L} + \left(\frac{\partial \mathbf{J}^{(i)}}{\partial I}\right)\left(\frac{\partial^2 I}{\partial \mathbf{m}^2}\right). \tag{8.12}$$

The second term is neglected in the Gauss-Newton approximation and is generally small near a local minimum.

The weighting coefficient $\alpha_1$ should be largest during the low-frequency, low-wavenumber stages of the inversion, where velocity estimation dominates the optimization.

**Differential semblance**: The flattening term is closely related to differential semblance optimization (DSO; Symes and Carazzone, 1991; Symes, 2008). Both functionals operate on an extended image endowed with a redundant coordinate - the subsurface offset or scattering angle in classical DSO, the local opening angle $\gamma$ here - and both quantify velocity error through the residual dependence of the image on that coordinate, rather than through a direct waveform mismatch.

The deviation-from-stack operator, $\mathbf{I}_N - N^{-1}\mathbf{1}\mathbf{1}^T$, of equation (8.11) is a (non-differential) annihilator whose null space - the $\gamma$-invariant, flat gathers - coincides with that of the differential annihilators of DSO. In the Gauss-Newton approximation, with the normalization 1/B frozen, $\Phi_1$ reduces to a quadratic form in the annihilated image, structurally identical to the DSO objective, and its adjoint-state gradient - the back-projection of the zero-mean residual along the known phase-space paths - is the LAD counterpart of the extended-migration velocity gradients of Shen and Symes (2008).

Term (1) thereby inherits the property that motivated DSO: residual moveout varies smoothly and quasi-monotonically with background-velocity error, so the functional remains informative far from the true model and convexifies the long-wavelength velocity update, in contrast to oscillatory waveform misfits prone to cycle skipping.

Two distinctions are noteworthy: the semblance normalization suppresses the illumination and amplitude imprint to which the unnormalized DSO objective is sensitive, and the opening-angle coordinate is furnished directly by the LAD imaging operator $\mathfrak{M}(m^k)$ applied to the observed data at each iteration, rather than by a subsequent angle transform of a subsurface-offset extension (cf. Dafni and Symes, 2016, 2018).

**Term (2): Directional Focusing in $\mathbf{k}_\nu$**

This objective measures the degree of focusing of a Directional Common-Image Gather. In addition to Term (1), it is designed primarily to update (anisotropic) velocity-related model parameters, since the correct velocity model focuses reflector energy into its specular reflection direction, while incorrect anisotropic model parameters cause directional spreading.

The specular direction is estimated as the energy-weighted directional centroid:

$$\bar{\mathbf{v}}(\mathbf{x}) = \frac{\int \mathbf{v}\; I^2(\mathbf{x},\boldsymbol{\nu})\, d\mathbf{v}}{\int I^2\,(\mathbf{x},\boldsymbol{\nu})\; d\mathbf{v}} \quad . \tag{8.13}$$

The objective function is simply the variance of image energy with respect to direction,

$$\Phi_2 = \frac{1}{2}\int (\mathbf{v} - \bar{\mathbf{v}})^2 I^2(\mathbf{x},\boldsymbol{\nu})\, d\mathbf{v}\, d\boldsymbol{x} \quad . \tag{8.14}$$

Thus, minimizing $\Phi_2$ promotes directional focusing.

The residual associated with the objective is,

$$\mathbf{q}_2(\mathbf{x},\boldsymbol{\nu}) = (\boldsymbol{\nu} - \bar{\boldsymbol{\nu}})I(\mathbf{x},\boldsymbol{\nu}) \quad . \tag{8.15}$$

The residual emphasizes energy occurring away from the specular direction.

The gradient with respect to model parameter $m_i(\mathbf{x}) = \{V\}$ is,

$$g_i(\mathbf{x}) = \int \mathbf{q_2}\,(\mathbf{x},\boldsymbol{\nu})\frac{\partial I(\mathbf{x},\mathbf{v}')}{\partial m_i}\,d\mathbf{v}\,dx \quad , \tag{8.16}$$

where $\frac{\partial I}{\partial m_i}$ is the image sensitivity (Born kernel) with respect to parameter $V$. The gradient measures how changing the model affects directional spreading. Velocity perturbations that reduce the variance contribute negatively to the objective and are favored during inversion.

The Hessian is $\mathbf{H}_{ij} = \frac{\partial \mathbf{g}_i}{\partial \mathbf{m_j}}$ , and describes the curvature of the objective function. Taking the second derivative produces two components.

1. **Gauss-Newton (Local) Term**: $\mathbf{H}_{ij}^{GN} = \int \frac{\partial \mathbf{q_2}}{\partial m_i}\frac{\partial \mathbf{q_2}}{\partial m_j}\,d\mathbf{v}\,dx$ **,** which measures parameter sensitivity. It is positive semi-definite, usually dominates near convergence. It is often the only term retained in practical implementations.
2. **Centroid-Coupling Term**: Because $\bar{\mathbf{v}} = \frac{\int \mathbf{v}I^2 d\mathbf{v}}{\int I^2 d\mathbf{v}}$ depends on the energy at all directions, differentiating $\bar{\mathbf{v}}$ introduces a second directional coordinate, denoted by $\mathbf{v}'$.

The Hessian therefore contains terms of the form,

$$\mathbf{H}_{ij}^{coup} \propto \iint \frac{\partial I(\boldsymbol{\nu})}{\partial \mathbf{m}_i}\,\mathbf{W}(\mathbf{v}'\,\frac{\partial I(\mathbf{v},\mathbf{v}')}{\partial m_j}\,d\mathbf{v}\,d\mathbf{v}' \quad , \tag{8.17}$$

where $\mathbf{W}(\mathbf{v},\mathbf{v}')$ is a weighting function arising from the centroid derivative, and $\mathbf{v}'$ is a secondary (dummy) integration variable introduced when differentiating the centroid, representing another direction within the same image gather.

The physical meaning of the Hessian coupling: The appearance of both $\mathbf{v}$ and $\mathbf{v}'$ means that the objective is nonlocal in directional space. Consequently, the Hessian contains cross-terms linking different directional bins: $(\mathbf{v},\mathbf{v}')$ and describes how changes in energy at one direction influence all other directions through the centroid calculation. The key insight is that the Hessian is not diagonal in $\mathbf{v}$ because the centroid $\bar{\nu}$ depends on energy from the entire directional gather. Therefore, $\mathbf{v}$ and $\mathbf{v}'$ represent interactions between different directions, allowing the inversion to focus energy toward a single specular direction.

**Term (3): Directional Alignment**

This objective enforces consistency between two model-dependent direction vectors:

- the specular propagation direction inferred from the data, $\bar{\mathbf{n}}_{\nu}$, and
- the reflector-normal direction predicted by the current model, $\mathbf{n}_{\mathbf{r}}$.

Such alignment becomes particularly important during intermediate and late stages of inversion, where null-space effects increase and data sensitivity decreases. By encouraging agreement between data-derived propagation directions and model-derived reflector orientations, the inversion is guided toward geologically plausible updates.
The objective function is defined as the alignment misfit as,

$$\Phi_3 = \frac{1}{2}\int (\bar{\mathbf{n}}_\nu(\mathbf{x}) - \mathbf{n}_\mathbf{r}(\mathbf{x}))^2 \, d\boldsymbol{x} \quad , \tag{8.18}$$

where $\bar{\mathbf{n}}_\nu$ is the predicted specular direction vector associated with $\bar{\mathbf{v}}$. The objective function $\Phi_3$ is minimized when $\bar{\mathbf{n}}_\nu = \mathbf{n}_\mathbf{r}$.

First Variation: Taking the first variation yields,

$$\delta\Phi_3 = \int (\mathbf{n}_\nu - \mathbf{n}_\mathrm{r}) \cdot (\delta\mathbf{n}_\nu - \delta\mathbf{n}_\mathrm{r}) \, d\boldsymbol{x} \quad , \tag{8.19}$$

which identifies the model perturbations that reduce directional mismatch.
The corresponding residual is,

$$\mathbf{q_3}(\boldsymbol{x}) = \bar{\mathbf{n}}_\nu(\mathbf{x}) - \mathbf{n}_\mathbf{r}(\mathbf{x}) \quad . \tag{8.20}$$

This term affects both velocity-type and reflectivity-type parameters, hence $\delta\mathbf{m}_i = \{\delta V_i, \delta\mathbf{R}_i\}$.
The gradient,

$$\mathbf{g}_i = \frac{\partial\Phi_3}{\partial\mathbf{m}_i} = \int \boldsymbol{q}_3\,(\boldsymbol{x}) \left(\frac{\partial\bar{\mathbf{n}}_\nu}{\partial\mathbf{m}_i} - \frac{\partial\mathbf{n}_\mathbf{r}}{\partial\mathbf{m}_i}\right) d\mathbf{x} \quad . \tag{8.21}$$

describes how the reflector normal changes with model perturbations.
The Hessian is obtained by differentiating the gradient: $\mathbf{H}_{ij} = \frac{\partial\mathbf{g}_\mathbf{i}}{\partial\mathbf{m}_j}$. Substituting equation 8.21,

$$\mathbf{H}_{ij} = \int \frac{\partial\boldsymbol{q}_3}{\partial\mathbf{m}_j}\left(\frac{\partial\bar{\mathbf{n}}_\nu}{\partial\mathbf{m}_i} - \frac{\partial\mathbf{n}_\mathbf{r}}{\partial\mathbf{m}_i}\right) d\mathbf{x} + \int \mathbf{q}_3 \frac{\partial}{\partial\mathbf{m}_j}\left(\frac{\partial\bar{\mathbf{n}}_\nu}{\partial\mathbf{m}_i} - \frac{\partial\mathbf{n}_\mathbf{r}}{\partial\mathbf{m}_i}\right) d\mathbf{x}.$$

Since $\mathbf{q_3} = \bar{\mathbf{n}}_\nu - \mathbf{n}_\mathbf{r}$, we have $\frac{\partial\mathbf{q_3}}{\partial\mathbf{m}_j} = \frac{\partial\bar{\mathbf{n}}_\nu}{\partial\mathbf{m}_j} - \frac{\partial\mathrm{n}_\mathrm{r}}{\partial\mathbf{m}_j}$. Therefore,

$$\mathbf{H}_{ij} = \int \left(\frac{\partial\mathbf{n}_\nu}{\partial\mathbf{m}_j} - \frac{\partial\mathbf{n}_r}{\partial\mathbf{m}_j}\right) \cdot \left(\frac{\partial\mathbf{n}_\nu}{\partial\mathbf{m}_i} - \frac{\partial\mathbf{n}_\mathrm{r}}{\partial\mathbf{m}_i}\right) d\boldsymbol{x} + \int \mathbf{q}_3 \cdot \left(\frac{\partial^2\mathbf{n}_\nu}{\partial\mathbf{m}_i\partial\mathbf{m}_j} - \frac{\partial^2\mathbf{n}_r}{\partial\mathbf{m}_i\partial\mathbf{m}_j}\right) d\boldsymbol{x} \; . \tag{8.22}$$

Using the Gauss-Newton approximation near convergence, $q_3 \approx 0$, so the second term becomes negligible. The Hessian reduces to,

$$\mathbf{H}_{ij}^{GN} = \int \left(\frac{\partial\bar{\mathbf{n}}_\nu}{\partial\mathbf{m}_i} - \frac{\partial\mathbf{n}_\mathbf{r}}{\partial\mathbf{m}_\mathrm{i}}\right)\left(\frac{\partial\bar{\mathbf{n}}_\nu}{\partial\mathbf{m}_j} - \frac{\partial\mathbf{n}_\mathbf{r}}{\partial\mathbf{m}_j}\right) d\mathbf{x} \tag{8.23}$$

which is positive semi-definite matrix and is recommended to be used in practice.

Finally, the weight $\alpha_3$ is usually kept small during early iterations and increased progressively as inversion advances to higher frequencies and finer spatial scales, where reflector orientation estimates become more reliable.

**Figure 8** illustrates the geometric meaning of the three image-domain terms: term (1) penalizes residual moveout of events in the opening-angle gathers, term (2) penalizes directional spreading of image energy about the specular direction, and term (3) penalizes the misalignment between the data-inferred specular direction and the model-predicted reflector normal. Together, the three terms impose complementary kinematic, focusing, and structural constraints on the model update.

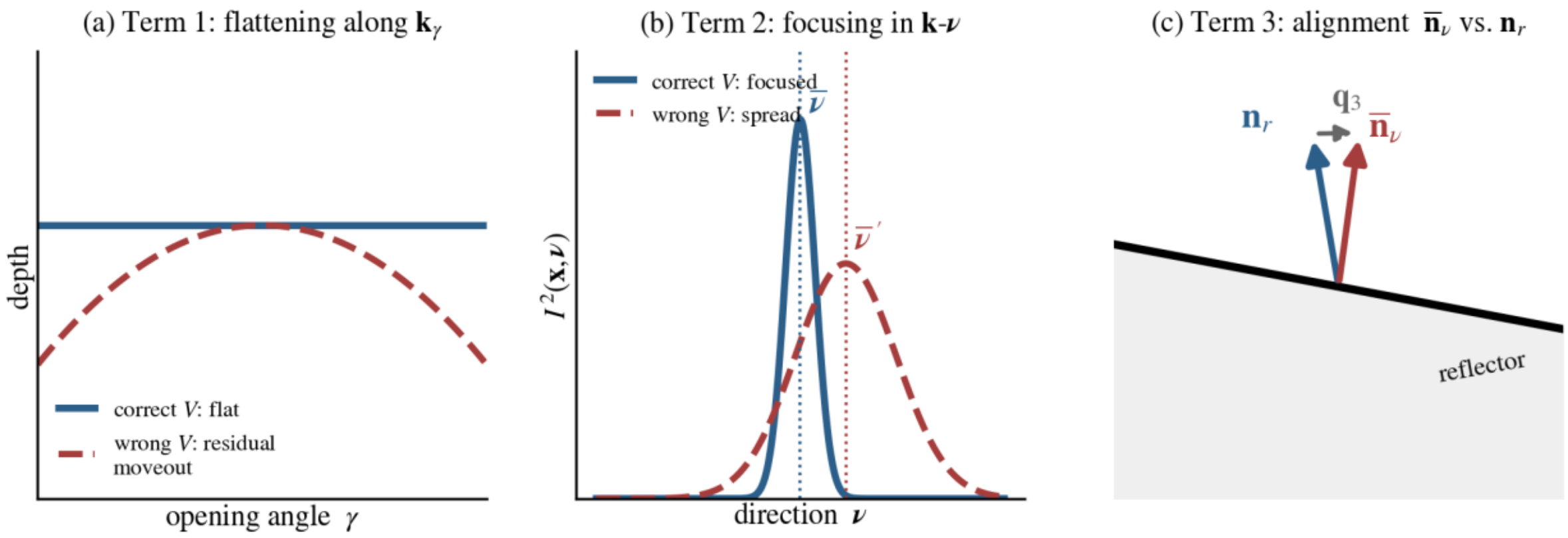

**Figure 8.** Geometric interpretation of the image-domain objective terms. **(a)** Term (1): with the correct velocity, events in the opening-angle gather are flat; velocity errors produce residual moveout, reducing the semblance measure of equation (8.7). **(b)** Term (2): the correct model focuses image energy $I^2$ into a narrow lobe about the specular direction, whereas model errors spread the energy and bias the directional centroid $\bar{\mathbf{v}}$ of equation (8.13). **(c)** Term (3): the alignment residual $\mathbf{q}_3 = \bar{\mathbf{n}}_\nu - \mathbf{n}_\mathbf{r}$ of equation (8.20) measures the mismatch between the data-inferred specular direction and the model-predicted reflector normal.

The practical consequence of formulating Terms (1) - (3) in the image domain is robustness to cycle skipping. **Figure 9** compares, for a single reflection event, the data-domain waveform misfit with the LAD flatness objective $\Phi_1$ as functions of the background-velocity error.

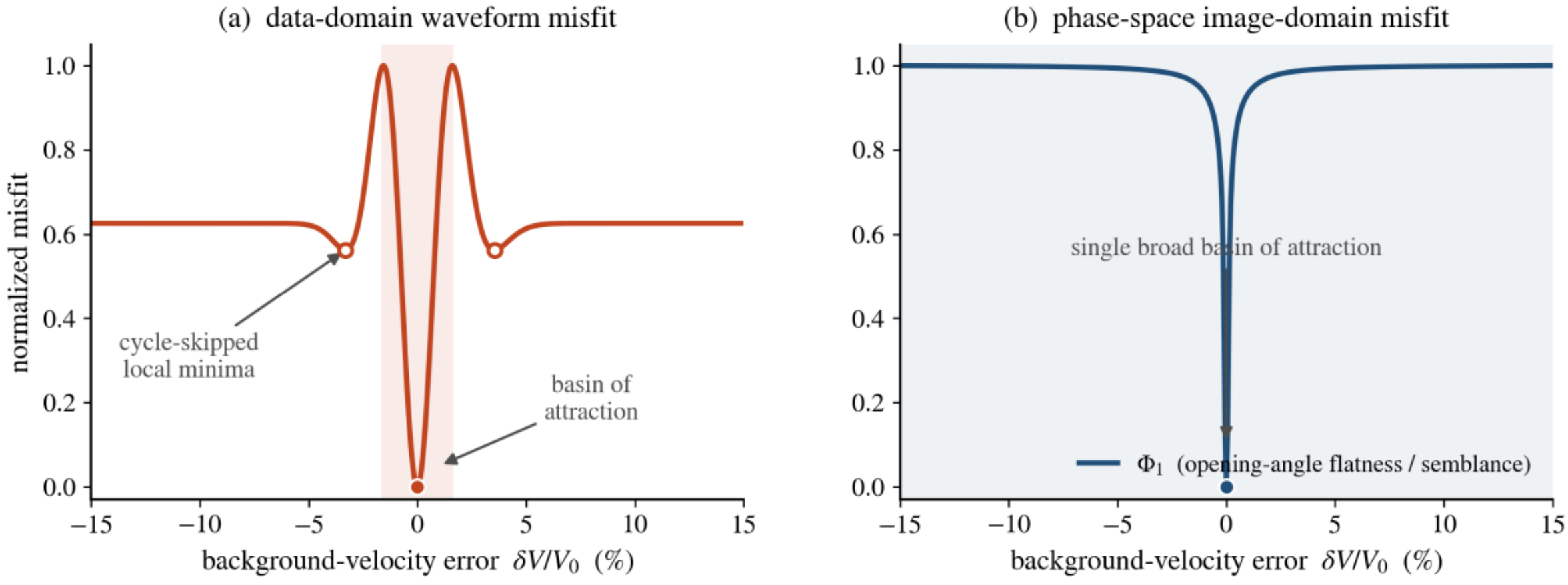

**Figure 9.** Objective-function behavior versus background-velocity error for a single reflection event (depth 2 km, $V_0$ = 2500 m/s, Ricker wavelet with $f_0$ = 15 Hz). (**a**) The data-domain waveform misfit, evaluated over offsets of 0.5–3.5 km, exhibits the classical cycle-skipping pattern: a narrow basin of attraction, of width governed by the half-period criterion, flanked by secondary local minima (open circles). (**b**) The LAD flatness objective $\Phi_1$ of equation (8.7), evaluated on the opening-angle gather of the same event, varies smoothly and remains quasi-convex over the full ±15% range: residual moveout degrades the semblance monotonically, extending the basin of attraction by an order of magnitude.

**Term (4): Velocity Regularization**

Velocity regularization is independent of the migrated image data and should be selected according to the expected geological setting. One possible choice is total-variation (TV) regularization,

$$\Phi_V^{TV} = \int |\nabla V| \; d\mathbf{x}, \tag{8.24}$$

with associated gradient $\mathbf{g}_V^{TV}$ and Hessian $\mathbf{H}_V^{TV}$.

To ensure differentiability where the velocity gradient vanishes, the TV norm is evaluated in its smoothed form,

$$|\nabla V|_\varepsilon = \sqrt{\nabla V \cdot \nabla V + \varepsilon^2}\,, \tag{8.25}$$

where $\varepsilon$ is a small positive constant chosen well below the typical gradient magnitude. The first variation of functional (8.24) yields,

$$\delta\Phi_V^{TV} = \int \frac{\nabla V \cdot \nabla \delta V}{|\nabla V|_\varepsilon} \, d\mathbf{x} = -\int \nabla \cdot \left(\frac{\nabla V}{|\nabla V|_\varepsilon}\right) \delta V \; d\mathbf{x}\,, \tag{8.26}$$

where integration by parts with natural (zero-flux) boundary conditions has been applied. Comparison with the generic variational form identifies the (scalar) TV gradient,

$$\mathrm{g}_V^{TV}(\mathbf{x}) = -\nabla \cdot \left(\frac{\nabla V}{|\nabla V|_\varepsilon}\right). \tag{8.27}$$

The quantity $\mathrm{g}_V^{TV}$ is (minus) the mean curvature of the velocity level sets: it vanishes for locally planar velocity structures regardless of the contrast magnitude and penalizes oscillatory or strongly curved variations. TV updates therefore drive the model toward piecewise-smooth, blocky configurations while preserving sharp planar contrasts.

The Hessian follows by linearizing equation (8.27) with respect to a velocity perturbation $\delta V$. Differentiating the normalized gradient $\nabla V/|\nabla V|_\varepsilon$ with respect to $\nabla V$ yields a projection-type tensor. The resulting Hessian is therefore a linear anisotropic diffusion operator that acts on a velocity perturbation $\delta V$ according to

$$\mathbf{H}_V^{TV}\,\delta V = -\nabla \cdot (\mathbf{D}_\varepsilon \nabla \delta V)\,, \qquad \mathbf{D}_\varepsilon = \frac{1}{|\nabla V|_\varepsilon}\left(\mathbf{I} - \frac{\nabla V \otimes \nabla V}{|\nabla V|_\varepsilon^2}\right). \tag{8.28}$$

The symmetric diffusion tensor $\mathbf{D}_\varepsilon$ has eigenvalue $\varepsilon^2/|\nabla V|_\varepsilon^3$ along the gradient direction and eigenvalue $1/|\nabla V|_\varepsilon$ in the orthogonal plane; it is therefore positive definite, and the Hessian operator is symmetric positive semi-definite (its null space contains the constant models under zero-flux boundary conditions). Smoothing is thus applied predominantly along structural interfaces, perpendicular to $\nabla V$, whereas diffusion across sharp contrasts is suppressed by the factor $\varepsilon^2/|\nabla V|_\varepsilon^3$, which preserves edges. In practical Gauss-Newton implementations, $\mathbf{D}_\varepsilon$ is frozen at the current iterate (lagged-diffusivity fixed-point iteration), yielding a linear, symmetric positive semi-definite Hessian that can be applied efficiently within the model update of equation (8.1).

TV regularization can be beneficial in:

- salt environments with sharp velocity contrasts,
- limited-aperture acquisitions,
- poorly constrained regions,
- inversions requiring structural coupling between velocity and impedance.

However, excessive TV regularization may introduce artificial blockiness and suppress physically meaningful smooth velocity variations.

Alternative regularization operators, including Tikhonov, Laplacian smoothing, structure-oriented smoothing, or anisotropic diffusion, may be more appropriate depending on the geological setting.

The weight $\alpha_4$ is generally largest during early-stage inversion and decreases as data constraints become stronger.

**Term (5): Reflectivity or Impedance Regularization**

To preserve sharp reflectors and maintain edge fidelity, a TV penalty may also be applied to impedance $Z$ (or equivalently to reflectivity $\mathbf{R}$):

$$\Phi_Z^{TV} = \int | \nabla Z | \, d\mathbf{x}. \tag{8.29}$$

The derivation parallels that of Term 4. With the smoothed norm $|\nabla Z|_\varepsilon$ defined as in equation (8.25), the first and second variations of functional (8.29) yield the (scalar) gradient and the anisotropic-diffusion Hessian,

$$\mathrm{g}_Z^{TV}(\mathbf{x}) = -\nabla \cdot \left( \frac{\nabla Z}{|\nabla Z|_\varepsilon} \right), \qquad \mathbf{H}_Z^{TV} \, \delta Z = -\nabla \cdot (\mathbf{D}_\varepsilon(\nabla Z) \, \nabla \delta Z), \tag{8.30}$$

with $\mathbf{D}_\varepsilon$ given by equation (8.28), evaluated for $\nabla Z$. As for the velocity term, the Hessian is symmetric positive semi-definite and is applied in lagged-diffusivity form.

The connection to the reflectivity vector follows from equation (3.4). Since $\mathbf{R} = \frac{1}{2} \nabla \ln Z$ implies $\nabla Z = 2Z\mathbf{R}$, the impedance TV functional may be written as,

$$\Phi_Z^{TV} = 2 \int Z |\mathbf{R}| \, d\mathbf{x}, \tag{8.31}$$

i.e., an amplitude-weighted $L_1$ penalty on the reflectivity magnitude. Alternatively, applying the TV penalty to the logarithmic impedance removes the amplitude weighting,

$$\Phi_{\ln Z}^{TV} = \int |\nabla \ln Z| \, d\mathbf{x} = 2 \int |\mathbf{R}| \, d\mathbf{x}, \tag{8.32}$$

which is a pure sparsity-promoting penalty on the reflectivity: it favors a small number of well-defined reflectors separating blocky impedance intervals, consistent with the layered-medium representation underlying the LAD framework. Its (scalar) gradient with respect to the impedance takes the compact form,

$$g_Z = -\frac{1}{Z} \, \nabla \cdot \left( \frac{\mathbf{R}}{|\mathbf{R}|_\varepsilon} \right) = -\frac{1}{Z} \, \nabla \cdot \mathbf{n_r}, \tag{8.33}$$

i.e., minus the divergence (curvature) of the unit reflector-normal field $\mathbf{n_r}$, scaled by the inverse impedance. The penalty is therefore stationary wherever reflectors are locally planar and acts most strongly on curved or laterally incoherent reflectivity structures. The corresponding Gauss-Newton Hessian retains the anisotropic-diffusion form,

$$\mathbf{H}_Z \approx \frac{1}{Z} \, \mathbf{H}_{\ln Z} \, \frac{1}{Z}, \qquad \mathbf{H}_{\ln Z} \, \delta u = -\nabla \cdot (\mathbf{D}_\varepsilon(\nabla \ln Z) \, \nabla \delta u), \tag{8.34}$$

where $u = \ln Z$. The approximation neglects the first-order term proportional to $g_{\ln Z}/Z^2$, which arises from differentiating the variable transformation $Z = e^u$. This term vanishes at the solution and is therefore omitted within the Gauss-Newton framework.

Unlike velocity regularization, this term becomes increasingly important during the later stages of inversion when higher frequencies are available, and reflector resolution improves.

Accordingly, the weight $\alpha_5$ typically increases as the inversion proceeds toward higher wavenumbers.

The mechanics of the two TV regularization terms are illustrated in **Figure 10**: the TV penalty leaves sharp, blocky profiles unpenalized while suppressing oscillatory variations, and its Gauss-Newton Hessian acts as an anisotropic diffusion operator whose diffusion tensor is elongated along material interfaces and nearly vanishes across them. This anisotropy allows the regularization to smooth along geological structure while preserving sharp velocity and impedance contrasts.

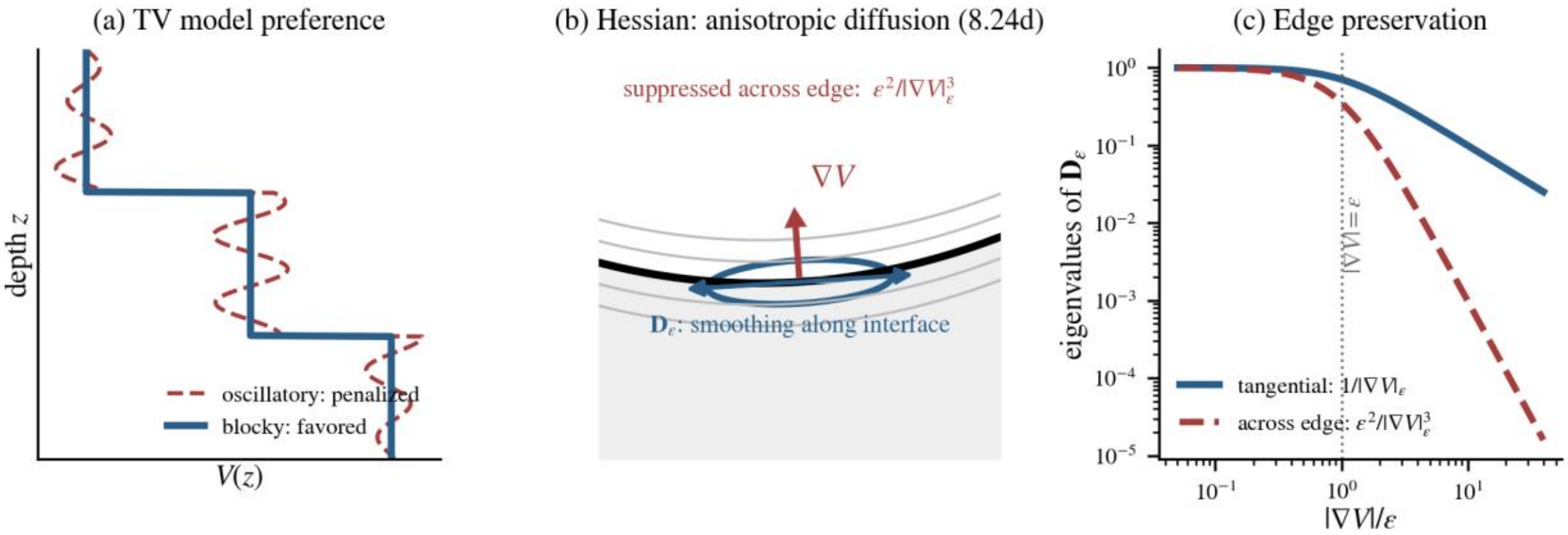

**Figure 10.** Total-variation regularization of Terms (4) and (5). **(a)** TV penalizes oscillatory model variations while leaving sharp, blocky profiles unpenalized. **(b)** The Gauss-Newton Hessian acts as an anisotropic diffusion operator (equation 8.28): the diffusion tensor $\mathbf{D}_\varepsilon$ is elongated along the interface (the level sets of $V$) and nearly vanishes across it. **(c)** Eigenvalues of $\mathbf{D}_\varepsilon$ versus normalized gradient magnitude: tangential smoothing decays slowly as $1/|\nabla V|_\varepsilon$, whereas diffusion across an edge decay as $\varepsilon^2/|\nabla V|_\varepsilon^3$, preserving sharp contrasts.

**Additional Constraints:** The framework readily accommodates additional objective terms whenever justified by the geological setting or reservoir objectives. Examples include fault-enhancement constraints, geological priors, or amplitude-versus-angle (AVA) consistency terms requiring agreement between model-predicted reflection coefficients and observed LAD AVA attributes.

### 8.3 Inversion Strategy

Consistent with conventional multiscale waveform inversion (e.g., Bunks et al. 1995; Sirgue and Pratt, 2004), the optimization begins in the low-wavenumber region of phase space, where the objective is to reconstruct the long-wavelength background velocity model. This stage remains the most nonlinear part of the inversion because only low-resolution information is available and the overlap between velocity and reflectivity sensitivities is strongest.

As the inversion progresses, the accessible LAD expands toward higher wavenumbers. According to the local illumination, scattering angle distribution, and acquisition geometry, progressively finer spatial scales become resolvable. The inversion therefore evolves from recovering the large-scale kinematic structure toward resolving medium- and high-wavenumber components associated with impedance variations, reflector geometry, and structural details.

Within the LAD-FWI framework, the weighting coefficients $\alpha_i$ are adapted according to the locally recoverable wavenumber content and the associated resolution criteria rather than being prescribed solely as a function of frequency. Components primarily associated with low-wavenumber updates, $\alpha_1, \alpha_2, \alpha_4$, receive greater emphasis when the LAD analysis indicates limited resolution and dominant transmission-wave sensitivities. As higher wavenumbers become reliably accessible, $\alpha_3, \alpha_5$, progressively gain importance, reflecting the increasing ability of the data to constrain reflector positioning, impedance contrasts, and structural attributes. **Figure 11** schematically illustrates this weight scheduling: the weights of the kinematic components dominate the early stages of the inversion, when only low wavenumbers are reliably recoverable, and decay as the data constraints strengthen, while the alignment and impedance-regularization weights grow correspondingly as the accessible LAD expands toward higher wavenumbers.

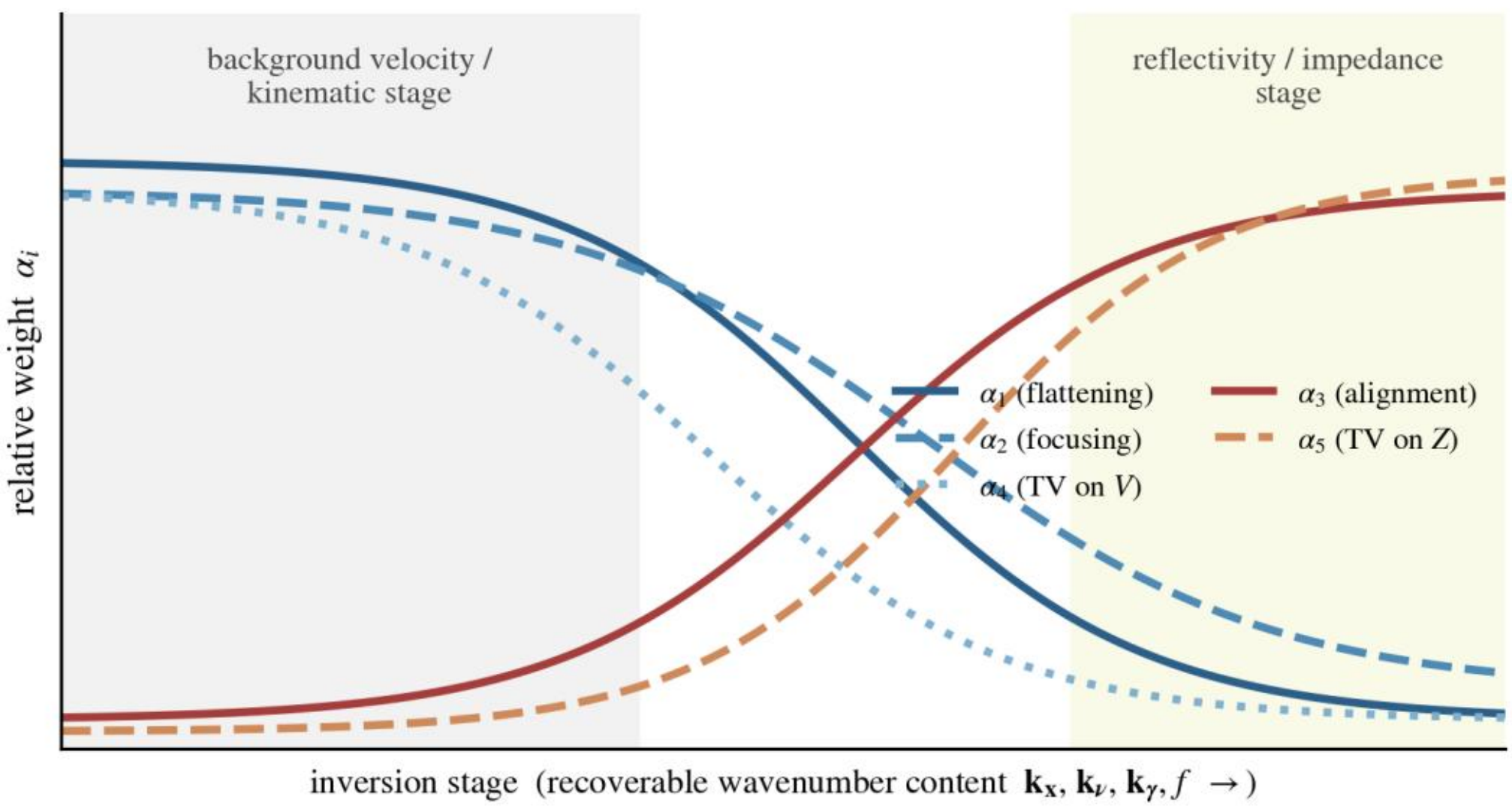

**Figure 11.** Schematic weight scheduling of the composite objective across inversion stages. Components associated with low-wavenumber velocity updates, $\alpha_1, \alpha_2, \alpha_4$, dominate the early, kinematic stages, while the alignment and impedance terms, $\alpha_3, \alpha_5$, progressively gain importance as higher wavenumbers become reliably recoverable. The curves are illustrative; in practice the weights are adapted according to the locally recoverable wavenumber content and the associated resolution criteria.

Consequently, the early stages of the inversion emphasize background velocity reconstruction, gather flattening, wavefield focusing, and inversion stabilization. The transition toward higher-resolution regions of phase space gradually shifts the objective toward reflector imaging, impedance recovery, and structural consistency. This evolution is driven by the estimated local resolving power inferred from the phase-space wavenumber distribution, allowing the inversion objectives to remain consistent with the information content of the data.

**Appendix C** provides a detailed description of the proposed adaptive wavenumber / frequency-based LAD-FWI workflow, where we derive a frequency-dependent Hessian, $\mathbf{H}(f) = \mathbf{J}^T(f)\mathbf{J}(f)$, that provides a velocity–reflectivity coupling coefficient,

$$\rho_{VR}(f) = \frac{\mathbf{J}_V^T \mathbf{J}_R}{\sqrt{(\mathbf{J}_V^T \mathbf{J}_V)(\mathbf{J}_R^T \mathbf{J}_R)}}, \quad \text{where } 0 \leq \rho_{VR} \leq 1. \tag{8.35}$$

(equation (8.35) is also explicitly written in **Appendix C**). Large values indicate strong parameter coupling and a higher likelihood of inversion crosstalk, whereas small values imply increasing independence of the parameter sensitivities and improved recoverability. Accordingly, new frequencies are introduced only when they increase model resolution or reduce parameter coupling. Equation (8.35) thus serves as a quantitative measure of parameter separability and represents a key component of the LAD-FWI framework.

Although the separation between velocity and reflectivity sensitivities improves substantially as the recoverable wavenumber bandwidth increases, the coupling does not disappear completely. Therefore, background velocity updates remain active throughout the inversion, subject to the local resolution criteria. Maintaining these updates ensures consistency between the velocity and impedance models, preserves the physical interpretation of the recovered parameters, and mitigates residual parameter crosstalk within the overlapping regions of phase space.

## 8.4 Advanced LAD Attributes

At mature stages of the inversion workflow, when the structural velocity model, impedance distribution, and vector reflectivity field have been jointly resolved within the adopted modeling assumptions, the incorporation of non-specular energy - including diffractions and other scattered-wave components - enables the extraction of additional high-resolution subsurface attributes.

These LAD attributes may include structural indicators such as fault likelihood, fracture networks, cavities, karst features, and channel systems, as well as material and production-related properties including pressure variations, fluid saturation, and other reservoir-characterization parameters. The ability to analyze both specular and non-specular components within a unified LAD framework provides a powerful mechanism for extending seismic inversion beyond conventional velocity and reflectivity imaging toward a more comprehensive subsurface characterization.

## 8.5 Anisotropy

Anisotropy causes seismic velocity to depend on propagation direction. As a result, events that appear focused in an isotropic model exhibit residual moveout, directional defocusing, and inconsistencies between observed and predicted propagation directions.

Because LAD-FWI decomposes the wavefield into the LAD coordinates $\left(\mathbf{x}_{\mathrm{p}},\mathbf{k}_{\mathrm{v}},\mathbf{k}_{\gamma},\tau\right)$, anisotropic effects appear naturally as direction-dependent distortions of the local-angle-domain gathers. This allows anisotropic parameter estimation to be integrated directly into the inversion framework.

For a model characterized with polar anisotropy (i.e., transverse isotropy with tilted axis of symmetry - TTI), the model parameter vector can be represented with Thomsen parameters $\left[V_P, f_{SP}, \delta, \varepsilon, \gamma_T, \mathbf{n}_{\mathrm{a}}\right]$, and the LAD Jacobian becomes, $\left[\mathbf{J}_{V_P}, \mathbf{J}_{V_S}, \mathbf{J}_{\delta}, \mathbf{J}_{\varepsilon}, \mathbf{J}_{\gamma_T}, \mathbf{J}_{(\theta,\phi)_{\mathrm{a}}}\right]$, where, $V_P$ is the (quasi) compressional velocity, $f_{SP} = 1-\left(V_S / V_P\right)^2$ with $V_S$ the (quasi) shear velocity, $\delta,\varepsilon$ and $\gamma_T$ are the Thomsen polar anisotropy parameters and $\mathbf{n}_{\mathrm{a}} = \left\{\theta,\phi\right\}_{\mathrm{a}}$ is the direction of the tilted symmetry-axis.

Since anisotropy affects wave propagation differently for different directions, its sensitivity is concentrated within specific regions of $k_\nu$ and in wide scattering angles $\gamma$.
The LAD Hessian can be partitioned as,

$$\mathbf{H} = \begin{bmatrix} \mathbf{H}_{VV} & \mathbf{H}_{VA} \\ \mathbf{H}_{AV} & \mathbf{H}_{AA} \end{bmatrix}, \qquad \text{where } \boldsymbol{A} = \{\delta, \epsilon, \gamma_T, \mathbf{n_a}\}. \tag{8.36}$$

The off-diagonal terms, $\mathbf{H}_{VA}$, measure velocity-anisotropy coupling. Because LAD separates energy according to direction and opening angle, the inversion can identify LAD regions where velocity and anisotropic sensitivities become more orthogonal, reducing parameter trade-offs.

In conventional anisotropic FWI, anisotropy is estimated indirectly through travel-time and waveform fitting. In LAD-FWI, anisotropic parameters produce explicit signatures in directional LAD through flattening, focusing, directional centroids, reflector-normal alignment, and Hessian coupling. Consequently, anisotropy estimation becomes a directional-focusing problem in phase space, rather than solely a waveform-matching problem, providing improved parameter separation, stronger geological consistency, and greater inversion stability.

## 9. ON THE CHOICE OF THE WAVEFIELD PROPAGATION ENGINE

The LAD-FWI formulation developed in the preceding sections is agnostic to the numerical engine used to propagate the forward and adjoint wavefields: it may be implemented with direct numerical solvers of the anisotropic elastic wave equation - finite-difference (FD), finite-element (FEM), or spectral-element (SEM) methods - or with ray-based and beam-based (finite-frequency) asymptotic methods. The two families, however, offer complementary advantages, and the choice has practical consequences for the behavior of the inversion.

Direct numerical solvers compute, in principle, the complete wavefield, including all orders of scattering, converted waves, head waves, and diffractions. This completeness makes them the natural choice for simulating realistic synthetic data and for final-stage validation of inverted models.

It comes, however, at a substantial cost: adequate spatial meshing (and the associated temporal sampling) must be maintained throughout the model to control numerical dispersion and stability, rendering three-dimensional anisotropic elastic simulations extremely expensive.

More fundamentally for inversion, the completeness of the wavefield is a mixed blessing. The FWI gradient is formed by correlating two wavefields - the incident (source-generated) wavefield and the back-propagated adjoint residual wavefield - and with a full numerical solution this correlation is inherently uncontrolled: every constituent of the forward wavefield correlates with every constituent of the adjoint wavefield, and the resulting gradient entangles contributions from wave types that carry very different, and sometimes conflicting, model information.

To converge to the correct model, the inversion benefits from knowing which types of fields are being correlated - for example, in order to impose the appropriate constraints on transmission-type versus reflection-type sensitivities, or to restrict the correlation to the LAD regions targeted by the continuation strategy of **Appendix C.** Extracting this control from a full numerical wavefield requires additional wavefield-decomposition machinery applied after the fact.

Finite-frequency ray- and beam-based methods invert this trade-off. Rather than computing everything and selecting afterwards, they construct reasonable approximations to selected, controlled wave/ray paths, so that the identity of each contribution to the gradient - its event type, propagation branch, and scattering geometry - is known by construction. The recently developed EigenRay method (Koren and Ravve, 2021; Ravve and Koren, 2021) solves the two-point boundary-value ray-tracing problem directly from Fermat's variational principle, with high computational efficiency, and delivers substantially richer path coverage and deeper penetration - including multipathing branches - than conventional shooting or bending schemes. Its ongoing extension (referred to the EigenWave method), building on the trajectory-mechanics formulation of finite-frequency wave propagation in anisotropic elastic media (Vasco and Nihei, 2019), replaces each stationary eigenray by a superposition of finite-frequency solutions concentrated around it, thereby endowing the asymptotic solution with realistic finite-frequency (banana-doughnut) sensitivity while retaining the path-level control of the ray framework.

The beam-based approach is particularly well matched to the LAD formulation. Every eigenray/beam intrinsically carries its directivity information - the incident and scattered slowness vectors at each subsurface point - so the mapping of wavefield contributions into the directional and opening-angle coordinates of the LAD phase space is immediate and exact, requiring no auxiliary local wavefield decomposition. For the same reason, the implementation of the adjoint-state method becomes considerably simpler: the adjoint field is assembled from the same controlled set of beams, the sensitivity kernels attach to individual, physically identified paths, and the LAD weighting and constraint operators of Section 8 act directly on per-beam quantities.

The beam framework potentially enables explicit computation of Hessian blocks under the assumptions of the adopted asymptotic approximation. In conventional grid-based FWI, the Jacobian is never explicitly available, and every Hessian-vector product entails additional wave-equation solves; explicit assembly of the Hessian is therefore infeasible, and practical implementations resort to quasi-Newton approximations such as L-BFGS, which estimate the inverse Hessian from a short history of gradient and model differences and capture curvature only along the directions already visited by the iteration.

In the EigenRay/EigenWave framework, by contrast, the Fréchet derivatives are semi-analytic byproducts of the two-point boundary value solution: each eigenray carries the traveltime and amplitude derivatives with respect to the model parameters along its stationary path - the traveltime Hessian along the ray is already delivered by the dynamic eigenray solution (Ravve and Koren, 2021) - and the proposed EigenWave superposition endows each path with its finite-frequency (Fresnel-volume) sensitivity kernel. Rows of the Jacobian are thus computable directly, without additional numerical propagation; in the same spirit, the adjoint application reduces to a fast accumulation of the weighted LAD residuals along the known wave paths, rather than a wavefield correlation. Since each kernel has support only within its Fresnel volume, the resulting Gauss-Newton Hessian is sparse and structured, and its phase-space blocks - the velocity, reflectivity/impedance, and coupling blocks of **Appendix A** - can be assembled, monitored, or applied explicitly rather than merely approximated, as detailed in the beam-based Hessian discussion of **Appendix A**.

In practice, therefore, the two families are best viewed as complementary components of a single workflow: direct numerical solvers for generating full-physics synthetic data and validating final models, and controlled finite-frequency beam propagation for computing the phase-space gradients and Hessian actions that drive the inversion, particularly during the early and intermediate stages of the LAD continuation strategy.

## 10. DATA-DRIVEN AND MODEL-DRIVEN REFLECTIVITY FIELDS

A distinctive feature of the proposed LAD-FWI framework is its ability to produce two complementary representations of subsurface reflectivity:

- **data-driven reflectivity image**, obtained directly from high-resolution wave-equation imaging such as RTM or LAD imaging, representing the scattering behavior required by the recorded seismic wavefield.
- **model-driven reflectivity field**, computed from the high-resolution recovered impedance model and therefore representing the interfaces implied by the inverted subsurface properties.

Together, these products provide both an observational and a model-based description of the subsurface.

### 10.1 Data-Driven Reflectivity

The data-driven reflectivity image is generated through high-resolution seismic imaging of the recorded wavefield. It preserves much of the information present in the measured data, including fine-scale reflectors, diffraction responses, localized scattering anomalies, and other wavefield phenomena.

Because it is derived directly from the observations, it provides the highest-fidelity representation of recorded seismic scattering and is particularly valuable for structural interpretation and the detection of small-scale geological features.

However, data-driven reflectivity remains subject to the limitations of seismic imaging. The image may contain migration artifacts, acquisition-footprint effects, uneven illumination, wavelet sidelobes, wavelet stretching, aliasing, residual multiple contamination, and amplitude distortions associated with finite-aperture acquisition and wavefield reconstruction. Consequently, not all imaged reflectivity necessarily corresponds to physically meaningful impedance contrasts.

### 10.2 Model-Driven Reflectivity

**Background**: The concept of model-driven reflectivity is closely related to the FWI-derived reflectivity (FDR) and FWI-imaging methodologies reviewed at the end of Section 2, in which high-resolution reflectivity volumes are computed directly from the inverted velocity or impedance models rather than from conventional migration, and whose practical value for interpretation and reservoir characterization has been demonstrated on field data.

In the present work, the concept is extended beyond conventional FDR imaging by jointly generating a model-driven reflectivity field from the recovered impedance model and a data-driven reflectivity image from the LAD imaging operator, while explicitly enforcing consistency between the two through Term (3) of the LAD-FWI objective function.

The model-driven reflectivity field is derived directly from the recovered impedance model. For example, it may be represented by the logarithmic impedance-gradient field $\boldsymbol{r}_m(\boldsymbol{x}) = \nabla \ln Z\,(\boldsymbol{x})$, which coincides, up to a factor of two, with the reflectivity vector of equation (3.4), and whose magnitude and orientation describe the strength and geometry of local impedance contrasts. Unlike the data-driven image, this representation is independent of migration amplitudes and wavelet effects and therefore reflects the structure implied by the inversion model itself.

Because it is computed from reconstructed model parameters rather than from the migrated wavefield, the model-driven reflectivity is largely free from migration artifacts, acquisition footprint, imaging aliasing, wavelet-stretching distortions, and other amplitude artifacts associated with wavefield imaging. Residual multiples affect the model-driven reflectivity only indirectly through their influence on the inversion, and their imprint is therefore generally less pronounced than in data-driven imaging products.

Its limitations arise from the inversion itself. The model-driven reflectivity inherits the resolution limits, regularization assumptions, parameterization choices, and possible biases of the recovered impedance model. Consequently, small-scale reflectors, diffraction-generating objects, and other sub-resolution heterogeneities visible in the data-driven image may not be present in the model-driven representation.

### 10.3 Joint Interpretation and Consistency Constraint

The greatest value of the framework arises from the simultaneous availability of both reflectivity representations. The data-driven reflectivity image describes the scattering behavior required by the recorded seismic data, whereas the model-driven reflectivity field describes the interfaces predicted by the recovered impedance model. In conventional workflows, these products are often analyzed independently, and their agreement is assessed qualitatively. In the proposed LAD-FWI framework, however, the relationship between them is incorporated directly into the inversion objective.

Specifically, Term (3) of the LAD-FWI objective function enforces consistency between the reflector geometry implied by the impedance model and the scattering geometry observed in the seismic data. The model-driven reflectivity field provides the local reflector-normal direction through the gradient of the recovered impedance model, while the data-driven reflectivity provides the specular scattering direction extracted from the directional angle gathers. Minimizing the discrepancy between these two directions encourages the recovered impedance interfaces to align with the scattering geometry required by the recorded wavefield.

This objective term establishes a direct coupling between model-space and data-space descriptions of reflectivity. The data-driven image therefore acts not only as an observational product but also as a structural constraint on the inversion, while the model-driven reflectivity provides a physically consistent interpretation of the observed scattering. Agreement between the two indicates consistency between the recovered earth model and the measured wavefield, whereas persistent discrepancies may indicate unresolved velocity or impedance errors, illumination limitations, residual multiples, non-Born scattering effects, or deficiencies in the adopted model parameterization. In this sense, the framework exploits the complementary strengths of both representations and transforms their comparison from a post-inversion diagnostic into an integral component of the inversion process itself.

## 11. DISCUSSION

The central contribution of this study is the formulation of waveform inversion directly within the LAD representation. In this domain, parameter sensitivities, model resolution, and Hessian conditioning can be analyzed and controlled through recoverable directional and scattering wavenumbers. This shift transforms phase space from an imaging and analysis domain into the primary inversion domain, providing a framework for organizing inversion according to information content and parameter separability rather than temporal frequency alone.

A central hypothesis of the methodology is that velocity and reflectivity perturbations occupy predominantly different, although partially overlapping, regions of phase space - a separation demonstrated exactly in the analytic benchmark of **Appendix F**. It is however important to note that separation is not exact. Steep dips, multipathing, shadow zones, anisotropic scattering, finite-frequency effects and imperfect aperture coverage can substantially increase overlap between parameter classes. The LAD representation reduces coupling but does not eliminate it. Under the Born approximation, velocity sensitivity is expected to be concentrated at large opening angles and low directional wavenumbers, whereas reflectivity sensitivity is concentrated at small opening angles and broader directional bandwidths. This organization may reduce off-diagonal Hessian coupling terms and improve parameter separability relative to conventional data-domain FWI, particularly during the early stages of model reconstruction.

An important consequence of the LAD representation is that inversion scheduling can be formulated in terms of recoverable model wavenumbers, local resolution, and parameter conditioning rather than temporal frequency alone. The proposed resolution-driven continuation strategy therefore generalizes classical multiscale FWI by allowing inversion objectives to evolve according to the information content of the data and the estimated resolving power of the LAD Hessian.

The framework also provides a natural mechanism for integrating multiple classes of seismic information within a common inversion environment. Reflections, diffractions, transmitted waves, anisotropic propagation effects, and local directional attributes can all be analyzed within the same LAD coordinates. Consequently, LAD-FWI may provide a unified foundation for velocity model building, impedance inversion, migration-velocity analysis, diffraction-based characterization, and advanced reservoir studies.

A further practical advantage of the framework, when implemented with the beam (finite frequency approximation) propagation engine, is explicit access to Hessian information. Whereas conventional data-domain FWI must rely on quasi-Newton (L-BFGS) approximations because the Jacobian is never explicitly available, the semi-analytic per-beam Fréchet derivatives render the Gauss-Newton Hessian sparse, structured, and directly computable (Section 9; **Appendix A**). This enables explicit assembly of the velocity block, illumination-based approximations of the reflectivity/impedance block, and continuous monitoring of the velocity-impedance coupling throughout the inversion - turning the sensitivity, resolution, and conditioning analyses of **Appendix A** from diagnostic idealizations into operational components of the inversion and providing a direct route to resolution matrices and uncertainty estimates that conventional FWI cannot offer.

An important extension of the framework concerns converted-wave imaging and inversion with multicomponent data. Mode conversion is inherently a local scattering phenomenon, governed by the incident and scattered slowness directions at the image point, and the LAD phase space is therefore its natural representation domain: converted-wave specularity, polarity, registration, and parameter sensitivity are all expressed directly in the in-situ LAD coordinates rather than through model-dependent acquisition-domain constructs such as common-conversion-point binning. This extension, including mode-decomposed LDP signals and the structure of the joint multi-mode Hessian, is developed in **Appendix D**.

Several challenges remain. The proposed methodology introduces additional computational and storage requirements associated with the multidimensional LAD representation. Furthermore, the degree of velocity–reflectivity separation, Hessian diagonal dominance, and resolution improvement remains to be validated quantitatively for realistic three-dimensional anisotropic models, complex salt geometries, and field datasets. These topics are the focus of ongoing research.

### 11.1 Treatment of Multiply Scattered Energy (multiples)

A limitation that deserves explicit discussion concerns the treatment of multiply scattered energy under the Born linearization. The imaging operator $\mathcal{M}(\mathbf{m})$, applied to the observed data at each iteration, is constructed under the Born approximation (Section 5) and therefore maps coherent arrivals only through single-scattering paths. Unlike wave-equation FWI, whose forward-modeling engine can partially account for free-surface and internal multiples, the present framework, in its Born-based formulation, does not explicitly model such events. Consequently, residual multiple energy enters the inversion as coherent noise that cannot be fully explained by the model.

The impact of multiples is objective-term dependent. Surface multiples migrated with the primary-wave operator exhibit residual moveout in the opening-angle gathers that resembles a slow-velocity error and does not flatten for any admissible velocity model. As a result, they can bias the flattening term (Term 1) toward persistently slow background updates. Interbed and peg-leg multiples populate directional wavenumbers that are inconsistent with any physically admissible single-scattering geometry and therefore contaminate the directional-focusing term (Term 2). At the same time, because multiples are often locally specular, the amplitude terms (Terms 3-5) may invert them into spurious depth-periodic reflectivity and impedance interfaces, which total-variation (TV) regularization may inadvertently sharpen rather than suppress.

Moreover, the diagonal-dominance argument developed in **Appendix A** relies on the approximate orthogonality of directional and opening-angle sensitivities under the single-scattering assumption. Multiple-scattering events violate this assumption and can therefore reintroduce velocity-impedance crosstalk beyond the scope of that analysis. Effective multiple attenuation is thus a more stringent prerequisite for LAD-FWI than for full-wavefield inversion methods.

The LAD representation, however, provides an important compensating advantage. Following LAD decomposition, primaries and multiples can be discriminated simultaneously by spatial position, directional wavenumber, and opening-angle wavenumber, rather than by offset and time alone. Multiples separate naturally through their characteristic residual-moveout behavior in $\gamma$, their directional inconsistency in $\nu$, and their tendency to form depth-periodic patterns. In addition, the beam-stack construction of the imaging operator (Koren, 2009; Koren and Ravve, 2011) provides an inherent rejection mechanism: local slant-stack operators with Fresnel-zone apertures strongly attenuate arrivals whose free-surface directivity is inconsistent with the modeled primary-ray pair. Full-azimuth angle-domain implementations of the same principle have also been demonstrated successfully on field data (Inozemtsev et al., 2015).

These observations motivate a LAD gating strategy, implemented through the weighting functions already present in the objective function (Section 8), in which regions of $(\mathbf{x},\boldsymbol{\nu},\boldsymbol{\gamma})$ that consistently fail flattening-consistency and explainability criteria across successive iterations are progressively down-weighted or muted. A practical realization of this strategy, based on five complementary LAD discrimination criteria and a composite weighting function, is presented in **Appendix E**.

### 11.2 Concluding Remarks

Overall, LAD-FWI does not increase the intrinsic information content of the seismic data. Rather, it reorganizes existing information into a physically interpretable LAD representation that is expected to improve parameter separability, model resolution, and inversion conditioning. This principle constitutes the theoretical foundation of the proposed methodology and motivates future validation using both synthetic and field datasets.

Finally, upon convergence of the iterative LAD-FWI process, the inverted model is expected to be consistent with the observed seismic data and, under an appropriate objective function and norm, satisfy the following complementary conditions:

**• Reflection-Angle-Domain Consistency**
Model-predicted azimuthally varying reflection coefficients, computed from the elastic properties of the inverted model (e.g., using the Zoeppritz equations or suitable approximations), should match the corresponding azimuthally varying amplitude-versus-angle (AVA) observations extracted from the reflection-angle image gathers. The match should be achieved in both relative amplitude and phase.

Under these conditions, the inversion can provide quantitative estimates of elastic rock properties and fluid-related parameters, particularly within reservoir intervals.

- **Required conditions:**
- Wide opening-angle coverage to adequately sample the reflection response.
- Broad azimuthal illumination (wide-azimuth acquisition) to resolve azimuth-dependent reflectivity effects.
- Sufficient long-offset recordings to constrain the AVA behavior over a large scattering-angle range.

**• Directional-Domain Consistency**
Model-predicted directional scattering coefficients, computed from the spatial derivatives of the P-wave and S-wave impedance fields in all dip and azimuth directions, should correlate with the corresponding directional energy image gathers $E(\mathbf{x},\boldsymbol{\nu},\boldsymbol{\gamma})$, $E = I^2$, particularly for small opening angles where the directional representation is most sensitive to local reflector geometry.

In the converged solution, the observed directional energy should be maximally focused and aligned with the reflector-normal directions predicted by the inverted model. Accordingly, the inversion seeks not only to match traveltimes and amplitudes, but also to maximize consistency between the directional organization of the data and the directional scattering behavior implied by the model.

- **Required conditions:**
- Illumination from a broad range of propagation directions.
- Large lateral acquisition extent to provide sufficient directional diversity.
- Adequate angular sampling to resolve reflector dip and azimuth variations.

**Integrating Both Conditions**

The two consistency conditions represent complementary aspects of the inversion problem. The reflection-angle domain primarily constrains the magnitude and phase of the local scattering response as a function of opening angle, whereas the directional domain constrains the orientation and focusing of the scattering process as a function of dip and azimuth. Together they provide independent yet mutually consistent constraints on the subsurface model: the reflection-angle gathers validate the local elastic reflectivity, while the directional gathers validate the geometric organization of the reflectors and the associated propagation paths.

Consequently, a fully converged LAD-FWI solution should simultaneously satisfy:

1. **AVA consistency**, whereby the model reproduces the observed angle-dependent reflection amplitudes and phases.
2. **Directional focusing consistency**, whereby the observed directional energy is concentrated along the reflector-normal directions predicted by the model.
3. **Kinematic consistency**, whereby the projected traveltimes are correctly aligned throughout both image domains.

The combined criterion proposed in this study is considerably more stringent than conventional waveform matching alone, as it requires simultaneous agreement between the observed and model-predicted kinematic, directional, and reflection-angle-domain characteristics of the seismic wavefield. Consequently, it provides a stronger physical basis for the joint recovery of velocity, elastic, and fluid-sensitive rock properties from seismic data.

Although the inversion framework proposed here remains to be validated through dedicated numerical and field-data studies, many of its constituent components have already been implemented and successfully applied within the broader LAD imaging ecosystem (see Section 2). In particular, local-angle-domain gathers have been used for velocity model building, tomography, anisotropy analysis, VVAZ/AVAZ studies, Q-compensated imaging, diffraction characterization, and structural interpretation. These applications provide indirect evidence that the LAD attributes exploited by LAD-FWI contain meaningful and recoverable subsurface information. The present work extends these established imaging-domain concepts into a unified inversion framework and provides the theoretical basis for using them as quantitative constraints in waveform inversion.

## 12. CONCLUSIONS

We have presented a phase-space formulation of elastic full-waveform inversion based on Local-Angle-Domain (LAD) representations of the scattered seismic wavefield. Unlike conventional FWI, which operates on acquisition-domain waveform residuals, the proposed methodology performs inversion directly within an in-situ LAD image domain parameterized by spatial location, propagation direction, scattering geometry, and temporal support. This transformation recasts the inversion problem in terms of physically interpretable wavefield attributes that are more directly linked to the underlying subsurface parameters.

A central premise of the approach is that velocity-sensitive and reflectivity-sensitive perturbations occupy predominantly distinct, although partially overlapping, regions of phase space. By exploiting this separation, LAD-FWI provides a natural mechanism for improving parameter discrimination, reducing velocity-reflectivity crosstalk, and enhancing the conditioning of multiparameter inversion. The resulting LAD objective functions, gradients, and Hessian operators establish a unified framework for analyzing both propagation and scattering phenomena in anisotropic elastic media.

Beyond parameter separation, the LAD representation provides direct access to illumination, directional focusing, scattering behavior, reflector orientation, and local wavefield attributes, enabling inversion strategies guided by parameter sensitivity, model resolution, and Hessian conditioning rather than by temporal frequency alone. This perspective motivates a generalized continuation strategy in which inversion progresses through increasingly resolved regions of phase space while adaptively targeting the model components that are most observable.

Although the inversion methodology presented here remains to be validated through dedicated synthetic and field-data studies, it builds upon the LAD imaging technology of Koren and Ravve (2011), which has undergone extensive application in seismic imaging, velocity-model building, tomography, VVAZ/AVAZ analysis, Q-compensated imaging, diffraction characterization, and reservoir studies. Consequently, many of the LAD attributes and data representations employed by LAD-FWI are extensions of concepts that have already demonstrated practical value in real-world applications.

This study establishes the theoretical foundation for a new class of phase-space inversion methods that unifies waveform inversion, migration-velocity analysis, impedance inversion, anisotropic parameter estimation, and diffraction-based subsurface characterization within a common framework. Rather than relying primarily on frequency continuation, the proposed approach organizes inversion according to recoverable model wavenumbers, local resolution, and parameter conditioning. If the underlying hypotheses are confirmed through future synthetic and field-data investigations, LAD-FWI may provide a more robust and physically interpretable pathway for recovering geologically meaningful subsurface models, reducing parameter coupling, improving inversion convergence, and enabling more quantitative integration of imaging and inversion.

## ACKNOWLEDGMENTS

I would like to sincerely acknowledge the outstanding support and dedicated efforts of the AspenTech - SSE P&I - EarthStudy 360 team. In particular, I wish to recognize the R&D team members: Ronit Levy, - Shustak, Eyal Cohen, Daniel Lerner, Anne-Laure Tertois, Yevgeny Ragoza, and Orhan Yilmaz, the Product Management members: Elana Mandelman and Sandra Allwork, and the Geo Science service team: Duane Dopkin and Shiv Singh. Their commitment and contributions have been invaluable.
I would like to thank Emerson - AspenTech for supporting the research on this theme.

## REFERENCES

Alkhalifah, T., and R.-É. Plessix, 2014, A recipe for practical full-waveform inversion in anisotropic media: An analytical parameter resolution study: Geophysics, 79(3), R91–R101, doi: 10.1190/geo2013-0366.1.

Audebert, F., P. Froidevaux, H. Rakotoarisoa, and J. Svay-Lucas, 2002, Insights into migration in the angle domain: 72nd Annual International, Meeting, SEG, Expanded Abstracts, 1188–1191.

Barnier, G., E. Biondi, R. G. Clapp, and B. Biondi, 2023a, Full-waveform inversion by model extension: Theory, design, and optimization: Geophysics, 88, no. 5, R579–R607, doi: 10.1190/geo2022-0350.1.

Barnier, G., E. Biondi, R. G. Clapp, and B. Biondi, 2023b, Full-waveform inversion by model extension: Practical applications: Geophysics, 88, no. 5, R609–R643, doi: 10.1190/geo2022-0382.1.

Beylkin, G., 1985, Imaging of discontinuities in the inverse scattering problem by inversion of a causal generalized Radon transform: Journal of Mathematical Physics, 26, 99–108, doi: 10.1063/1.526755.

Beylkin, G., and R. Burridge, 1990, Linearized inverse scattering problems in acoustics and elasticity: Wave Motion, 12, 15–52, doi: 10.1016/0165-2125(90)90017-X.

Biondi, B., 2007a, Angle-domain common-image gathers from anisotropic migration: Geophysics, 72, no. 2, S81–S91, doi: 10.1190/1.2430561.

Biondi, B., 2007b, Residual moveout in anisotropic angle-domain common-image gathers: Geophysics, 72, no. 2, S93–S103, doi: 10.1190/1.2430562.

Biondi, B., and A. Almomin, 2014, Simultaneous inversion of full data bandwidth by tomographic full-waveform inversion: Geophysics, 79, no. 3, WA129–WA140.

Bleistein, N., J. K. Cohen, and J. W. Stockwell Jr., 2001, Mathematics of multidimensional seismic imaging, migration, and inversion: Springer.

Bleistein, N., Y. Zhang, S. Xu, G. Zhang, and S. H. Gray, 2005a, Kirchhoff inversion in image point coordinates recast as source/receiver point processing: 75th Annual International Meeting, SEG, Expanded Abstracts, 1697–1700.

Bleistein, N., Y. Zhang, S. Xu, G. Zhang, and S. H. Gray, 2005b, Migration/inversion: think image point coordinates, process in acquisition surface coordinates: Inverse Problems, 21, 1715–1744, doi: 10.1088/0266-5611/21/5/013

Brandsberg-Dahl, S., M. V. de Hoop, and B. Ursin, 2003, Focusing in dip and AVA compensation on scattering-angle/azimuth common image gathers: Geophysics, 68, 232–254, doi: 10.1190/1.1543210.

Bunks, C., Saleck, F.M., Zaleski, S. and Chavent, G., 1995. Multiscale seismic waveform inversion, Geophysics, **60**(5), 1457–1473.

Chase, D., Koren, Z., 2015. 5D Local Angle Domain Gathers as an ideal representation for directivity driven imaging (ID: 25878), 77th EAGE Conference and Exhibition 2015 in Madrid, Spain.

Cheng, X., J. Mao, Z. Feng, D. Vigh, K. Glaccum, and Y. Yu, 2025, Advancing seismic imaging: From kinematics to dynamics with elastic full-waveform inversion: The Leading Edge, 44, no. 5, 362–370, doi: 10.1190/tle44050362.1.

Dafni, R. & Symes, W.W., 2016. Scattering and dip-angle decomposition based on subsurface-offset extended wave-equation migration, Geophysics, **81**(3), S119–S138.

Dafni, R. & Symes, W.W., 2018. Kinematics of reflections in subsurface-offset and angle-domain image gathers, Geophysical Journal International, **213**(2), 1212–1230.

de Hoop, M. V., and N. Bleistein, 1997, Generalized Radon transform inversions for reflectivity in anisotropic elastic media: Inverse Problems, 13, 669–690, doi: 10.1088/0266-5611/13/3/009.

Dekel, G., Chase, D., Koren, Z., 2016, Q compensation imaging in the local angle domain, SEG International Exposition and 86th Annual Meeting in Dallas.

Dekel, G., Chase, D., Levy, R., and Koren, Z., 2017, A Novel Q Compensation Filter Implementation in the Migrated Local Angle Domain, 79th EAGE Conference and Exhibition 2017 Paris.

De Ribet, B., Yelin, G., Serfati, Y., Chase, D., Kelvin, R., and Koren, Z., 2018, High resolution diffraction imaging for reliable interpretation of fracture systems, First Break, Issue: Vol: 35, February 201**8**, pp. 43-47.

Engquist, B., and B. D. Froese, 2014, Application of the Wasserstein metric to seismic signals: Communications in Mathematical Sciences, 12, 979–988.

Eskozha, B., Aimagambetov, M., Kondratenko, A., Sementsov, V., Pankratov, V., Kuanysheva, A., Inozemtsev, A, Soloviev, V., and Koren Z., 2017, Applying the full-azimuth angle domain imaging to study carbonate reefs at great depth, special topic: petroleum geology, First Break, March 2017.

Forgues, E., and G. Lambaré, 1997, Parameterization study for acoustic and elastic ray + Born inversion: Journal of Seismic Exploration, 6, 253–278.

Gholami, Y., R. Brossier, S. Operto, A. Ribodetti, and J. Virieux, 2013, Which parameterization is suitable for acoustic vertical transverse isotropic full waveform inversion? Part 1: Sensitivity and trade-off analysis: Geophysics, 78(2), R81–R105, doi: 10.1190/geo2012-0204.1.

Ghosh, S., M. Jain, and R. Singh, 2023, An integrated approach of fracture characterization using seismic azimuthal anisotropy analysis and diffraction imaging: A case study from Padra field, Cambay Basin, India: 14th Biennial International Conference and Exposition, SPG.

He, Y., H. Xing, Y. Huang, and B. Wang, 2021, Inversion-based imaging: FWI beyond velocity: 82nd Conference and Exhibition, EAGE, Extended Abstracts.

Hosgood, K., Masako Robb, and Zvi Koren, 2015, Local angle domain migration for image improvements and its application to HTI fracture characterizations. Proceedings of the 12th SEGJ International Symposium, Tokyo, Japan, 18–20 November 2015: pp. 38–41. doi: 10.1190/segj122015-011.

Huang, G., R. Nammour, and W. W. Symes, 2018, Volume source-based extended waveform inversion: Geophysics, 83, no. 5, R369–R387.

Inozemtsev, A., Nicolaevich, S., Viktorovich, I., Galkin, A., Vasilyevich, R., and Koren, Z., 2013. Applying full-azimuth angle domain pre-stack migration and AVAZ inversion to study fractures in carbonate reservoirs in the Russian Middle Volga region, First Break, vol. 31, Feb. 2013.

Inozemtsev, A., Koren, Z., and Zapprikaspiygeofizika, G., 2015. Noise Suppression and Multiple Attenuation using Full-Azimuth Angle Domain Imaging: Case Studies, First Break, June 2015.

Inozemtsev, A., Koren, Z., and Galkin, A., 2017, Applying full-azimuth depth imaging in the local angle domain to delineate hard-to-recover hydrocarbon reserves, First Break, 35, December 2017.

Inozemtsev, I., Koren, Z., and Galkin, A., 2019, Applying full azimuth depth processing in the local angle domain for frequency absorption versus azimuth (FAVAZ), First Break, 37, January 2019.

Itan, L., Serfaty, Y., Levi, R., and Koren, Z., 2018, Seismic characterizations of geological features via PCA and Deep Learning in the Local Angle Domain, EAGE International Exposition and Annual Meeting in Copenhagen, Copenhagen, Denmark.

Kalinicheva, T., M. Warner, and F. Mancini, 2020, Full-bandwidth FWI: 90th Annual International Meeting, SEG, Expanded Abstracts, 651–655, doi: 10.1190/segam2020-3425522.1.

Kletenik-Edelman, O., Zaslavsky, A., Shustak, M., Levy, R., and Koren, Z., 2025. Local Directional Projection (LDP) specularity weights for enhancing and characterizing seismic reflectivity images, Extended Abstract, IMAGE 2025 - International Meeting for Applied Geoscience & Energy, Houston.

Koren, Z., X. Sheng, and D. Kosloff, 2002, Target-oriented common reflection angle migration: 72nd Annual International Meeting, SEG, Expanded Abstracts, 1196–1199.

Koren, Z., I. Ravve, A. Bartana, and D. Kosloff, 2007, Local angle domain in seismic imaging: 69th Conference and Exhibition, EAGE, Extended Abstracts, P287.

Koren, Z., I. Ravve, E. Ragoza, A. Bartana, and D. Kosloff, 2008, Full-azimuth angle domain imaging: 78th Annual International Meeting, SEG, Expanded Abstracts, 2221–2225.

Koren, Z., 2009. Multiple suppression in angle domain time and depth migration. U.S. Patent No. 7,584,056, granted September 1, 2009.

Koren, Z., I. Ravve, and R. Levy, 2010, Specular-diffraction imaging from directional angle decomposition: 72nd Conference and Exhibition, EAGE, Extended Abstracts, G045.

Koren, Z. & Ravve, I., 2011. Full-azimuth subsurface angle domain wavefield decomposition and imaging: Part 1 - Directional and reflection image gathers. Geophysics, **76**(1), S1–S13.

Koren, Z., Ravve, I., 2012, Patent US 8120991B2, System and method for full azimuth angle domain imaging in reduced dimensional coordinate systems, Feb 21, 2012

Koren, Z., and I. Ravve, 2021, Eigenrays in 3D heterogeneous anisotropic media, Part I: Kinematics: Geophysical Prospecting, 69, 3–27, doi: 10.1111/1365-2478.13052.

Korkidi, L., Litvak, R., Ayach, C., Levy, R., and Koren, Z., 2018, Azimuthally anisotropic effective parameters from full-azimuth reflection angle gathers, SEG Technical Program Expanded Abstracts 2018 https://doi.org/10.1190/segam2018-2995956.1 .

Kozlov, E. A., Z. Koren, D. N. Tverdohlebov, and A. N. Badeikin, 2009, Corner reflectors- a new concept of imaging vertical boundaries: 71st Conference and Exhibition, EAGE, Extended Abstracts, Z039.

Kumar, D., and R. Ali, 2024, Integration of FWI-derived reflectivity in seismic interpretation: The Leading Edge, 43, no. 12, 836–842, doi: 10.1190/tle43120836.1.

Lailly, P., 1983, The seismic inverse problem as a sequence of before stack migrations. Conference on Inverse Scattering: Theory and Application, SIAM, 206–220.

Landa, E., and S. Fomel, 2026, Time-reversal full wavefield inversion: From waveform misfit to wavefield focus: Geophysical Prospecting, 74, no. 7, e70237, doi: 10.1111/1365-2478.70237.

Luo, S., and P. Sava, 2011, A deconvolution-based objective function for wave-equation inversion: 81st Annual International Meeting, SEG, Expanded Abstracts, 2788–2792.

McLeman, J., T. Burgess, M. Sinha, G. Hampson, and T. Thompson, 2021, Reflection FWI with an augmented wave equation and quasi-Newton adaptive gradient scheme: First International Meeting for Applied Geoscience & Energy, SEG, Expanded Abstracts, 667–671.

McLeman, J., K. Dancer, and T. Burgess, 2022, Next-generation resolution through multiparameter FWI imaging: Second International Meeting for Applied Geoscience & Energy, SEG/AAPG, Expanded Abstracts, 922–926, doi: 10.1190/image2022-3745707.1.

Métivier, L., R. Brossier, Q. Mérigot, E. Oudet, and J. Virieux, 2016, Measuring the misfit between seismograms using an optimal transport distance: Application to full-waveform inversion: Geophysical Journal International, 205, 345–377.

Nolan, C., and W. Symes, 1996, Imaging in complex velocities with general acquisition geometry: TRIP, the Rice Inversion Project, Rice University, technical report.

Olneva, T., D. Semin, A. Inozemtsev, I. Bogatyrev, K. Ezhov, E. Kharyba, and Z. Koren, 2019, Improved seismic images through full-azimuth depth migration: updating the seismic geological model of an oil field in the pre-neogene base of the Pannonian Basin, First Break, 37, October 2019.

Operto, S., Y. Gholami, V. Prieux, A. Ribodetti, R. Brossier, L. Métivier, and J. Virieux, 2013, A guided tour of multiparameter full-waveform inversion with multicomponent data: From theory to practice: The Leading Edge, 32, no. 9, 1040–1054, doi: 10.1190/tle32091040.1.

Podolak, M., H. Kowalski, P. Godlewski, W. Kobusinski, J. Makarewicz, A. Nowicka, Z. Mikolajewski, D. Chase, R. Dafni, A. Canning and Z. Koren, 2014, Imaging and Characterization of a Shale Reservoir in Onshore Poland Using Full-azimuth Seismic Depth Imaging, 76th EAGE Conference and Exhibition 2014.

Pratt, R.G., 1999. Seismic waveform inversion in the frequency domain. Part 1: Theory and verification in a physical-scale model, Geophysics, 64(3), 888–901.

Prieux, V., R. Brossier, S. Operto, and J. Virieux, 2013, Multiparameter full waveform inversion of multicomponent ocean-bottom-cable data from the Valhall field. Part 1: Imaging compressional wave speed, density and attenuation: Geophysical Journal International, 194(3), 1640–1664, doi: 10.1093/gji/ggt177.

Ravve, I., and Z. Koren, 2011, Full-azimuth subsurface angle domain wavefield decomposition and imaging: Part 2 - Local angle domain: Geophysics, 76, no. 2, S51–S64.

Ravve, I., and Z. Koren, 2021, Eigenrays in 3D heterogeneous anisotropic media, Part II: Dynamics: Geophysical Prospecting, 69, 28–52, doi: 10.1111/1365-2478.13053.

Rickett, J. & Sava, P., 2002. Offset and angle-domain common-image-point gathers for shot-profile migration, Geophysics, 67(3), 883–889.

Rosales, D. A., S. Fomel, B. L. Biondi, and P. C. Sava, 2008, Wave-equation angle-domain common-image gathers for converted waves: Geophysics, 73(1), S17–S26, doi: 10.1190/1.2821193.

Rousseau, V., L. Nicoletis, J. Svay-Lucas, and H. Rakotoarisoa, 2000, 3D true amplitude migration by regularisation in angle domain: 62nd Conference and Exhibition, EAGE, Extended Abstracts, B-13.

Sava, P., and B. Biondi, 2004, Wave-equation migration velocity analysis. Part I: Theory: Geophysical Prospecting, 52(6), 593–606, doi: 10.1111/j.1365-2478.2004.00447.x.

Sava, P. & Fomel, S., 2003. Angle-domain common-image gathers by wavefield continuation methods, Geophysics, **68**(3), 1065–1074.

Sava, P., and I. Vasconcelos, 2011, Extended imaging conditions for wave-equation migration: Geophysical Prospecting, 59, 35–55.

Shen, P., and W. W. Symes, 2008, Automatic velocity analysis via shot profile migration: Geophysics, 73, no. 5, VE49–VE59.

Shustak, M., Sharabi, I., Kletenik, O. Menyoli, E., Imala, I., Levy, R. and Koren, Z., 2022, Anisotropic Full Azimuth Velocity Model Building Using Joint Reflection-Refraction Tomography, 92nd Annual

Meeting of Society of Exploration Geophysicists, Expanded Abstracts.

Shustak, M., Kletenik, O., Levy, R., and Koren, Z., 2024. Least-squares Migration utilizing high-resolution Point Spread Functions generated from Local Angle Domain illumination vectors, Extended Abstract, IMAGE 2024 - International Meeting for Applied Geoscience & Energy, Houston.

Shustak, M., Levy, R., Mandelman E., and Koren Z., 2026. Reflection/Diffraction Separation of Local Directional Projection Signals Using Implicit Neural Representations, Extended Abstract, IMAGE 2026, International Meeting for Applied Geoscience & Energy, Houston

Singh, S., Dopkin, D., and Koren, Z., 2025. High resolution diffraction imaging applied to a karst reservoir in a South Texas Field, Extended Abstract, IMAGE 2025 - International Meeting for Applied Geoscience & Energy, Houston.

Sirgue, L. & Pratt, R.G., 2004. Efficient waveform inversion and imaging: A strategy for selecting temporal frequencies, Geophysics, **69**(1), 231–248.

Sollid, A., and B. Ursin, 2003, Scattering-angle migration of ocean-bottom seismic data in weakly anisotropic media: Geophysics, 68, 641–655, doi: 10.1190/1.1567234.

Soubaras, R., 2003, Angle gathers for shot-record migration by local harmonic decomposition: 73rd Annual International Meeting, SEG, Expanded Abstracts, 889–892.

Stewart, R. R., J. E. Gaiser, R. J. Brown, and D. C. Lawton, 2002, Converted-wave seismic exploration: Methods: Geophysics, 67(5), 1348–1363, doi: 10.1190/1.1512781.

Stewart, R. R., J. E. Gaiser, R. J. Brown, and D. C. Lawton, 2003, Converted-wave seismic exploration: Applications: Geophysics, 68(1), 40–57, doi: 10.1190/1.1543193.

Symes, W.W., 2008. Migration velocity analysis and waveform inversion, Geophysical Prospecting, **56**(6), 765–790.

Symes, W. W., 2020, Wavefield reconstruction inversion: An example: Inverse Problems, 36, no. 10, 105010, doi: 10.1088/1361-6420/abaf66.

Symes, W. W., and J. J. Carazzone, 1991, Velocity inversion by differential semblance optimization: Geophysics, 56, 654–663.

Tarantola, A., 1984. Inversion of seismic reflection data in the acoustic approximation, Geophysics, **49**(8), 1259–1266.

Tarantola, A., 1986. A strategy for nonlinear elastic inversion of seismic reflection data, Geophysics, **51**(10), 1893–1903.

ten Kroode, A. P. E., D. J. Smit, and A. R. Verdel, 1994, Linearized inverse scattering in the presence of caustics: SPIE — The International Society for Optical Engineering, Expanded Abstracts, 28–42.

Thomsen, L., 1986, Weak elastic anisotropy: Geophysics, 51(10), 1954–1966, doi: 10.1190/1.1442051.

Thomsen, L., 1999, Converted-wave reflection seismology over inhomogeneous, anisotropic media: Geophysics, 64(3), 678–690, doi: 10.1190/1.1444577.

Tsvankin, I., 1997, Anisotropic parameters and P-wave velocity for orthorhombic media: Geophysics, 62(4), 1292–1309, doi: 10.1190/1.1444231.

van Leeuwen, T., and F. J. Herrmann, 2013, Mitigating local minima in full-waveform inversion by expanding the search space: Geophysical Journal International, 195, 661–667.

van Leeuwen, T., and W. A. Mulder, 2010, A correlation-based misfit criterion for wave-equation traveltime tomography: Geophysical Journal International, 182, 1383–1394.

Vasco, D. W., and K. T. Nihei, 2019, A trajectory mechanics approach for the study of wave propagation in an anisotropic elastic medium: Geophysical Journal International, 219, no. 3, 1885–1899, doi: 10.1093/gji/ggz406.

Virieux, J. and Operto, S., 2009. An overview of full-waveform inversion in exploration geophysics, Geophysics, **74**(6), WCC1–WCC26.

Wang, C., D. Yingst, P. Farmer, and J. Leveille, 2016, Full-waveform inversion with the reconstructed wavefield method: 86th Annual International Meeting, SEG, Expanded Abstracts, 1237–1241.

Warner, M., and L. Guasch, 2016, Adaptive waveform inversion: Theory: Geophysics, 81, no. 6, R429–R445.

Wu, R.-S., and K. Aki, 1985, Scattering characteristics of elastic waves by an elastic heterogeneity: Geophysics, 50(4), 582–595, doi: 10.1190/1.1441934.

Wu, R.-S., and L. Chen, 2006, Directional illumination analysis using beamlet decomposition and propagation: Geophysics, 71, no. 4, S147–S159, doi: 10.1190/1.2204963.

Xu, S., H. Chauris, G. Lambaré, and M. Noble, 2001, Common-angle migration: A strategy for imaging complex media: Geophysics, 66, 1877–1894, doi: 10.1190/1.1487131.

Yan, J., and P. Sava, 2008, Isotropic angle-domain elastic reverse-time migration: Geophysics, 73(6), S229–S239, doi: 10.1190/1.2981241.

Zhang, Z., Z. Wu, Z. Wei, J. Mei, R. Huang, and P. Wang, 2020, FWI Imaging: Full-wavefield imaging through full-waveform inversion: 90th Annual International Meeting, SEG, Expanded Abstracts, 656–660, doi: 10.1190/segam2020-3427858.1.

## APPENDIX A. Sensitivity, Resolution, and Conditioning in LAD-FWI

The sensitivity structure analyzed in this appendix is a property of the general phase-space framework; the explicit beam-based assembly of the Fréchet and Hessian operators, however, presumes the Born-linearized ray/beam engine of Section 9.

A central hypothesis of the proposed LAD-FWI framework is that velocity-type and reflectivity-type perturbations occupy different, although partially overlapping, regions of phase space; the analytic benchmark of **Appendix F** verifies the associated sensitivity separation, resolution gain, and conditioning improvement exactly in a controlled setting, while this appendix develops the general framework. To understand this distinction, it is necessary to separate the concepts of parameter sensitivity, parameter resolution, and inverse-problem conditioning:

- Sensitivity describes how strongly the objective function responds to a specific model perturbation.
- Resolution describes how accurately a specific model perturbation can be distinguished from other perturbations.
- Conditioning describes how stably the inverse problem can recover the desired specific model update.

### A1. Linearized Phase-Space Observation Equation

Under the Born approximation, the LAD image perturbation may be expressed as,

$$\delta I(\mathbf{x},\mathbf{k}_\nu,\mathbf{k}_\gamma) = \frac{\partial I}{\partial \mathbf{m}_i}\delta\mathbf{m}_i = \mathbf{J}_i\,\delta\mathbf{m}_i \quad , \tag{A.1}$$

where $\mathbf{J}_i$ is the LAD Jacobian operator w.r.t $\delta\mathbf{m}_i$. For the simplified parameterization $\mathbf{m} = \{V, \mathbf{R}\}$, where $V$ denotes velocity-related parameters and $\mathbf{R}$ denotes vector reflectivity-related parameters,

$$\delta I = \mathbf{J}_V\ \delta V + \mathbf{J}_R\,\delta\mathbf{R}\ . \tag{A.2}$$

Note: A distinction should be drawn between the Jacobian vectors and Hessian matrices associated with the LAD image $I$ - the derivatives of the image with respect to the model parameters - and the gradient and Hessian of the scalar objective function $\Phi$ of the LAD-FWI of equation (8.5) constructed from it. Throughout this appendix, to simplify the notation, all Jacobians and Hessians refer to the image $I$; the corresponding derivatives of the objective follow from these by the chain rule.

The corresponding Fréchet derivatives, $\mathbf{J}_V = \frac{\partial I}{\partial V}$ and $\mathbf{J}_R = \frac{\partial I}{\partial \mathbf{R}}$, describe the local sensitivity of the LAD image to velocity and reflectivity perturbations. Consistent with the impedance parameterization adopted in Section 7, $\mathbf{J}_R^{(I)}$ and $\mathbf{H}_{RR}$ denote reflectivity-domain quantities; the corresponding impedance-domain Jacobian and Hessian follow directly through the relative-gradient operator $\mathbf{G}$ as $\mathbf{J}_Z = \mathbf{J}_R\,\mathbf{G}$ and $\mathbf{H}_{ZZ} = \mathbf{G}^T\,\mathbf{H}_{RR}\,\mathbf{G}$ (equations 7.1b, 7.1c).

### A2. Sensitivity Analysis

The Jacobian defines parameter sensitivity. Large Jacobian amplitudes imply that small model perturbations generate large image perturbations.

The LAD analysis developed in the main text suggests that the dominant velocity sensitivity follows approximately,

$$\mathbf{J}_V(\mathbf{v},\boldsymbol{\gamma}) \propto \sin^2\left(\frac{\gamma_1}{2}\right) \quad \text{and} \quad \mathbf{J}_R(\mathbf{v},\boldsymbol{\gamma}) \propto W_{\text{spec}}(\mathbf{v})\cos^2\left(\frac{\gamma_1}{2}\right) \quad . \tag{A.3}$$

Consequently:

- Velocity sensitivity is strongest for $\gamma \rightarrow 180°$, and is primarily associated with wide-angle propagation and low-wavenumber model updates.
- Reflectivity sensitivity is strongest for $\gamma \rightarrow 0°$, and is associated with high-resolution impedance contrasts concentrated around specular directions.
- These observations describe where information exists within phase space, but they do not yet indicate whether the corresponding model parameters can be recovered independently.

The distinct angular supports of the parameter classes are summarized by the asymptotic radiation patterns of **Figure A1**.

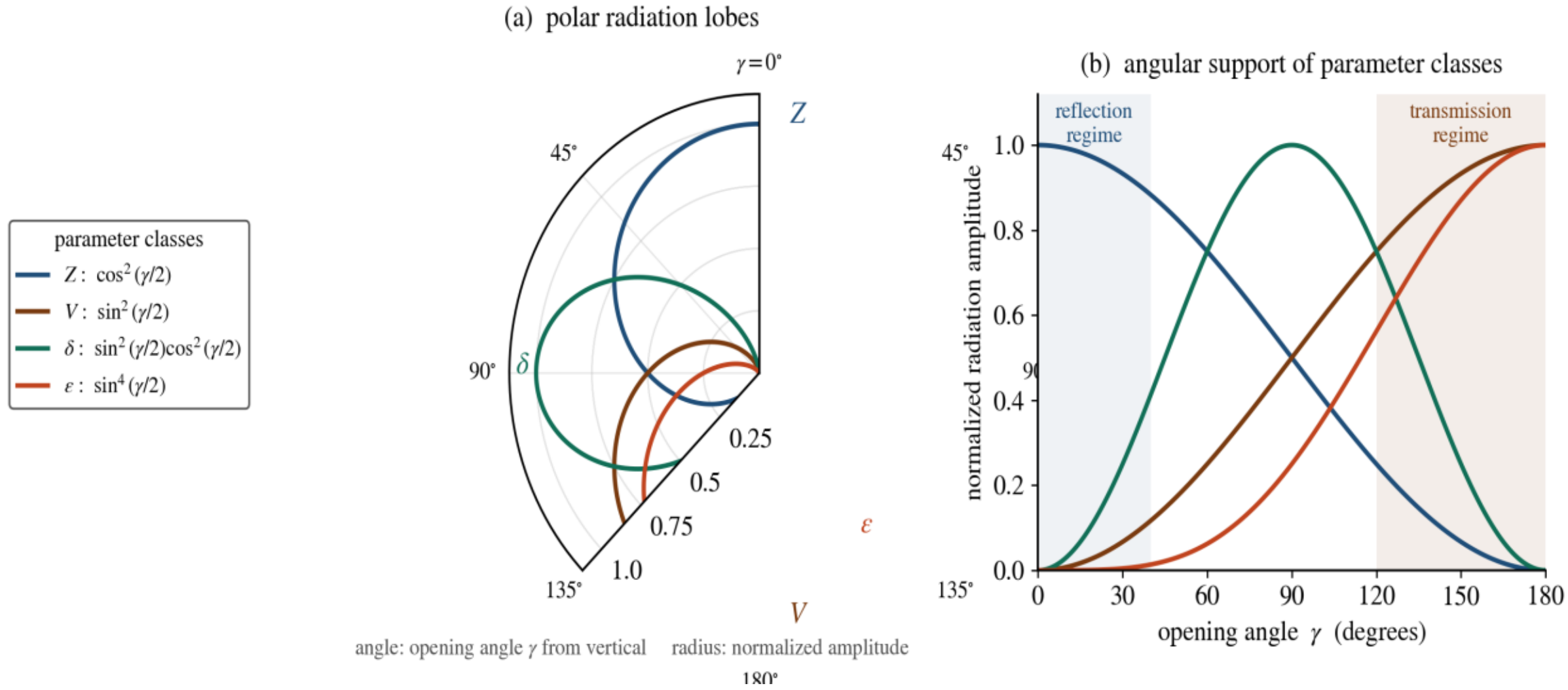

**Figure A1.** Asymptotic Born radiation patterns of the principal parameter classes versus the opening angle γ, for a locally specular P-P configuration in a VTI medium. (**a**) Polar representation of the scattering response: the opening angle γ is measured from the upward vertical (γ = 0°, backscatter / specular-reflection direction; γ = 180°, forward-scattering / transmission direction), the radial coordinate is the normalized radiation amplitude, and the lobes are mirrored about the vertical axis; the key at the left identifies the parameter classes and their angular dependences. (**b**) The same patterns as functions of the opening angle. The impedance (reflectivity) response, $\propto \cos^2(\gamma/2)$, dominates the small-angle reflection regime, whereas the velocity response, $\propto \sin^2(\gamma/2)$, dominates the wide-angle transmission regime, consistent with equation (A.3). The Thomsen parameters occupy intermediate and wide-angle bands, $\delta \propto \sin^2(\gamma/2)\cos^2(\gamma/2)$ and $\varepsilon \propto \sin^4(\gamma/2)$; the angular overlap of the ε and V patterns at wide opening angles visualizes the origin of velocity–anisotropy crosstalk. These distinct angular supports underlie the LAD separation of **Figure 4** and the mode-dependent wavenumber bounds of **Figure D2**.

**Figure 4** demonstrates the velocity and the reflectivity sensitivity regions and their overlap region in phase domain.

### A3. Resolution Analysis

Parameter resolution is governed by the Hessian operator. For a least-squares objective function,

$$J = \frac{1}{2} \parallel \delta I \parallel^2 , \tag{A.4}$$

and the Gauss-Newton Hessian becomes,

$$\mathbf{H} = \mathbf{J}^T\mathbf{J} = \begin{bmatrix} \mathbf{H}_{VV} & \mathbf{H}_{VR} \\ \mathbf{H}_{RV} & \mathbf{H}_{RR} \end{bmatrix}. \tag{A.5}$$

Velocity resolution: $\mathbf{H}_{VV} = \mathbf{J}_V^T\,\mathbf{J}_V$ : This term measures how strongly the data constrain velocity perturbations.
Reflectivity resolution: $H_{RR} = \mathbf{J}_R^T\,\mathbf{J}_R$ : This term measures the resolution attainable for reflectivity updates.

Parameter coupling: $\mathbf{H}_{VR} = \mathbf{J}_V^T\,\mathbf{J}_R$ : This term measures the similarity between velocity and reflectivity sensitivities. Large off-diagonal terms imply strong crosstalk and poor parameter separation.

### A4. LAD Separation of Velocity and Reflectivity

The overlap between velocity and reflectivity sensitivity regions may be quantified by,

$$C_{VR} = \int \mathbf{J}_V(\mathbf{v},\boldsymbol{\gamma})\mathbf{J}_R(\mathbf{v},\boldsymbol{\gamma})\,d\mathbf{v}d\boldsymbol{\gamma} \quad . \tag{A.6}$$

Substituting the asymptotic sensitivity relations A3 yields,

$$C_{VR} \propto \int \sin^2\left(\frac{\gamma_1}{2}\right)\cos^2\left(\frac{\gamma_1}{2}\right)\mathbf{W}_{\text{spec}}(\mathbf{v})\,d\mathbf{v}\,d\boldsymbol{\gamma}. \tag{A.7}$$

Since $\mathbf{W}_{\text{spec}}$ occupies only a limited directional region, the overlap remains substantially smaller than the diagonal energies,

$$E_V = \int \mathbf{J}_V^2\, d\mathbf{v}d\boldsymbol{\gamma}, \text{ and } E_R = \int \mathbf{J}_R^2\, d\mathbf{v}d\boldsymbol{\gamma}. \tag{A.8}$$

Therefore,

$$C_{VR} \ll E_V, E_R. \tag{A.9}$$

This implies that the corresponding off-diagonal Hessian terms are reduced in phase space.

**A5. Hessian Conditioning**

The Hessian may therefore be decomposed as,

$$\mathbf{H} = \mathbf{H}_0 + \epsilon\mathbf{H}_c, \tag{A.10}$$

where,

$$\mathbf{H}_0 = \begin{bmatrix} \mathbf{H}_{VV} & 0 \\ 0 & \mathbf{H}_{RR} \end{bmatrix} \tag{A.11}$$

and,

$$\eta = \frac{\| \mathbf{H}_{VR} \|}{\sqrt{\| \mathbf{H}_{VV} \| \| \mathbf{H}_{RR} \|}} \quad , \tag{A.12}$$

measures parameter coupling. When $\eta \ll 1$, the inverse problem approaches a block-diagonal structure, and velocity and reflectivity updates become approximately separable.

**Figure A2** illustrates this structural contrast schematically: the disjoint LAD support of the velocity and reflectivity sensitivities suppresses the coupling block and drives the Hessian toward the block-diagonal form of equation (A.11).

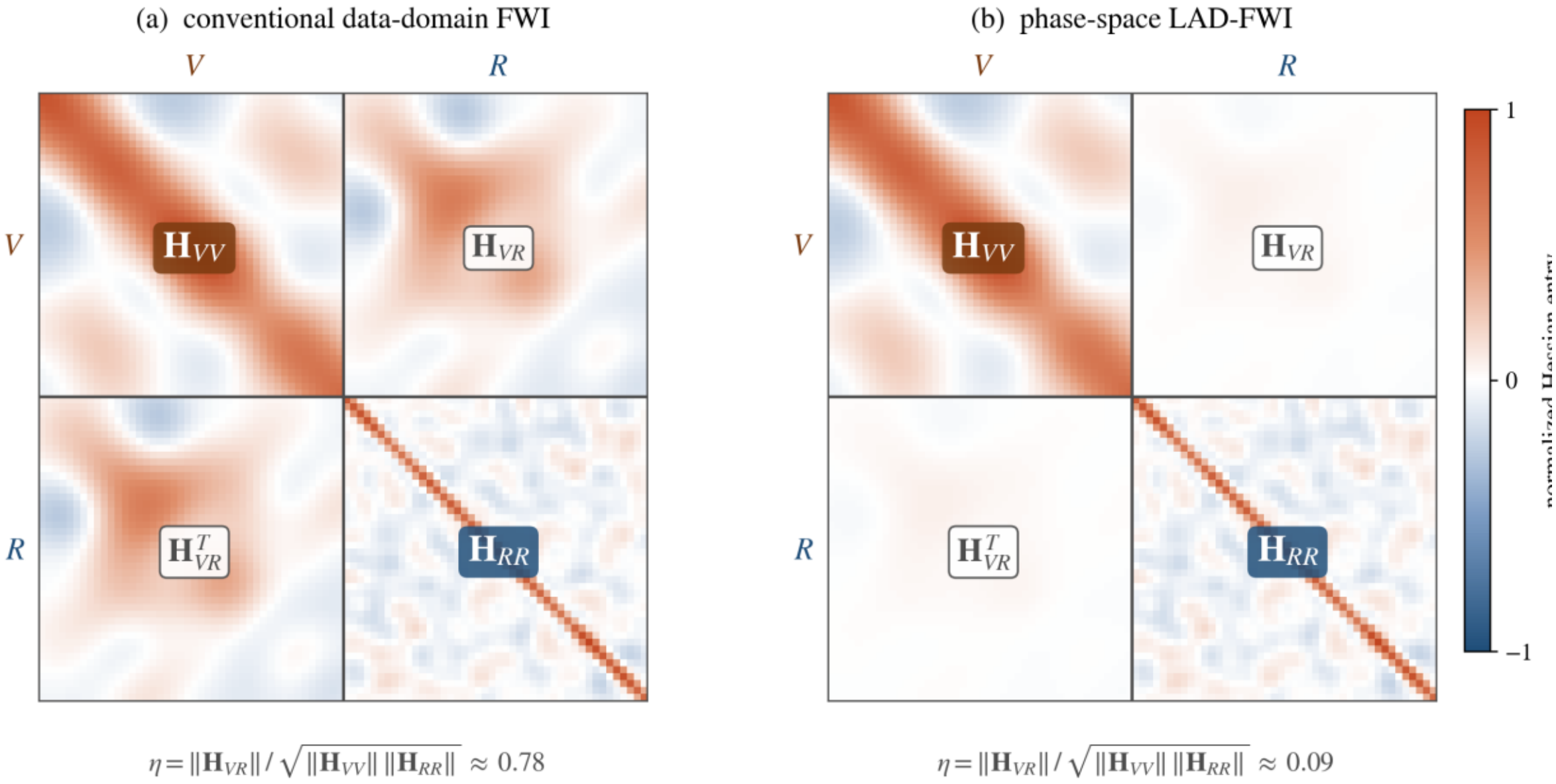

**Figure A2.** Schematic structure of the Gauss-Newton Hessian over the velocity (V) and reflectivity (**R**) parameter classes. (a) In conventional data-domain FWI, entangled kinematic and dynamic sensitivities produce a strong off-diagonal coupling block $H_{VR}$, so velocity and reflectivity updates contaminate one another. (b) In LAD-FWI, the disjoint wavenumber and opening-angle support of the two sensitivity classes (**Figure 4**) suppresses $H_{VR}$ and drives the Hessian toward the block-diagonal form of equations (A.10) - (A.11). The coupling measure η of equation (A.12), reported beneath each panel, quantifies the contrast. The matrices are schematic, drawn to emphasize the correlation structure of each block: broad and smooth for the long-wavelength velocity block $H_{VV}$, narrowly banded for the localized reflectivity point-spread functions of $H_{RR}$.

### A6. Resolution Matrix

The model-resolution matrix may be written as,

$$\boldsymbol{\mathcal{R}} = \mathbf{H}^{-1}\mathbf{J}^T\mathbf{W}\mathbf{J}. \tag{A.13}$$

Diagonal elements $\mathcal{R}_{ii}$ measure the recoverability of parameter $i$. Off-diagonal terms $\mathcal{R}_{ij}$ measure leakage of parameter $j$ into parameter $i$. Reducing the overlap term $\mathbf{H}_{VR}$, therefore improves not only conditioning but also parameter resolution and interpretability.

In summary, the Jacobian determines where information exists in phase space, whereas the Hessian determines how well that information can be separated and recovered.

Within the proposed LAD framework:

- Velocity and reflectivity exhibit different sensitivity distributions in $(\boldsymbol{\nu}, \boldsymbol{\gamma})$ space.
- This reduces the overlap of their Jacobian kernels.
- The off-diagonal Hessian terms decrease.
- The Hessian approaches a block-diagonal structure.
- Parameter resolution improves and crosstalk decreases.
- The condition number is expected to be reduced relative to conventional FWI; in the analytic benchmark of **Appendix F** the reduction is exact, the velocity–depth condition number falling from 269 in the narrow-aperture data domain to cond = 1 for the LAD flatness residual.

Consequently, LAD imaging $\overbrace{D(\mathbf{x}_s, \mathbf{x}_r, t)}^{5D} \rightarrow \overbrace{I(\mathbf{x}, \mathbf{v}, \boldsymbol{\gamma})}^{7D}$ does not create information beyond that contained in the recorded seismic data. Rather, it reorganizes and concentrates that information within a higher dimensionality LAD representation that enhances its effective utilization by improving parameter separability, reducing velocity–reflectivity coupling, strengthening Hessian conditioning, and increasing the recoverability of physically meaningful model updates.

**A7. Numerical Computation of the Hessian**

For realistic three-dimensional LAD-FWI problems, the full Hessian matrix is prohibitively large and cannot be explicitly assembled or stored. Let $N_m$ denote the number of model unknowns. The Hessian contains $N_m^2$ elements and quickly exceeds practical memory limits. Consequently, practical implementations rely on matrix-free representations in which only Hessian actions on vectors are evaluated.

The Gauss-Newton Hessian is $\mathbf{H} = \mathbf{J}^T\mathbf{W}\mathbf{J}$, where $\mathbf{J} = \frac{\partial I}{\partial \mathbf{m}}$ is the LAD image Jacobian and $\mathbf{W}$ is a weighting operator associated with the objective function. The Hessian-vector product for an arbitrary perturbation $\delta\mathbf{m}$ is $\mathbf{H}\delta\mathbf{m} = \mathbf{J}^T\mathbf{W}\mathbf{J}\delta\mathbf{m}$. The computation proceeds in three steps:

**Step 1. Forward Jacobian Projection**
Propagate the model perturbation through the Jacobian $\delta I = \mathbf{J}\delta\boldsymbol{m}$. This produces a perturbation of the LAD image volume. Physically:

- velocity perturbations generate opening-angle residuals,
- reflectivity perturbations generate image amplitude residuals,
- anisotropic perturbations generate directional and azimuthal residuals.

**Step 2. Weighting**
Apply the objective-function weighting $\delta I_w = W\delta I$. Examples include:

- semblance weighting,
- angle-dependent weighting,
- illumination compensation,
- diffraction weighting.

**Step 3. Adjoint Projection**
Back-project the weighted image perturbation $\mathbf{H}\delta\mathbf{m} = \mathbf{J}^T\delta I_w$. This operation is mathematically identical to a second migration or adjoint-state propagation. Thus, a Hessian-vector product requires:

- one linearized forward propagation,
- one adjoint propagation.

No Hessian matrix is ever explicitly constructed. A practical approximation consists of retaining only the diagonal terms: $\mathbf{H} \approx \mathrm{diag}(\mathbf{H})$.
The diagonal elements may be estimated as $\mathbf{H}_{ii} = \sum_k \mathbf{J}_{ki}^2$.
Physically, this measures:

- illumination,
- sensitivity,
- local parameter resolution.

For LAD-FWI, $\mathbf{H}_{ii} = \sum_{\nu,\gamma,f} \mathbf{J}_i^2\,(\boldsymbol{\nu}, \boldsymbol{\gamma}, f)$. This quantity may be interpreted as the total LAD illumination of parameter $i$. The corresponding preconditioner is $\mathbf{P} = \mathrm{diag}(\mathbf{H})^{-1}$. Such diagonal Hessian scaling is often sufficient to compensate for uneven illumination and geometrical spreading.

A more complete characterization of resolution may be obtained by computing selected rows or columns of the Hessian. Consider a unit perturbation $e_i = (0, \ldots, 1, \ldots, 0)^T$. The corresponding point-spread function becomes $PSF_i = He_i$. The PSF describes how a model perturbation located at point $i$spreads through the inversion. In LAD coordinates, PSF = PSF($\boldsymbol{x}, \boldsymbol{\nu}, \boldsymbol{\gamma}$). The localization of the PSF directly measures parameter resolution:

- narrow PSF → high resolution,
- broad PSF → poor resolution.

- **Lanczos Eigenvalue Analysis:** To evaluate conditioning, only a small number of dominant eigenvalues are required. Using a matrix-free Hessian-vector product, the Lanczos iteration can be applied: $\mathbf{Hq}_i = \lambda_i \mathbf{q_i}$. Only repeated evaluations of $\mathbf{HV} = \mathbf{J}^T\mathbf{WJV}$ are required. Typically, the largest 50–200 eigenvalues are sufficient to estimate cond $= \frac{\lambda_{\max}}{\lambda_{\min}}$ and characterize the effective condition number. This provides a practical method for validating the central hypothesis of LAD-FWI.

- **Beam-Based Computation of the Hessian**

The matrix-free strategy described above is dictated by grid-based propagation engines, for which each Hessian-vector product entails additional wave-equation solves. Within the EigenRay/EigenWave framework of Section 9, however, the Jacobian rows are semi-analytic per-beam quantities, and the Gauss-Newton Hessian becomes directly computable: its entries are integrals of overlapping finite-frequency kernels, nonzero only where the Fresnel volumes of the contributing beams intersect, so that the Hessian is sparse, banded, and diagonally dominant.

In particular, the velocity block, defined on the coarse long-wavelength parameterization, can be assembled explicitly - in analogy with ray-based tomographic normal equations, but with finite-frequency kernels and full LAD weighting - enabling direct regularized solves, explicit resolution matrices (equation A13), and uncertainty estimates. The reflectivity/impedance block is dominated by localized point-spread functions and is well approximated by a diagonal or narrowly banded operator computed from per-beam illumination, while the coupling block and the coupling measures of equations (A.12) and (C.2) become directly computable and can be monitored throughout the inversion.

This contrasts with conventional data-domain FWI, where explicit Hessian information is unavailable and quasi-Newton (L-BFGS) approximations are the practical default. The accuracy of the beam-based Hessian is that of the underlying asymptotic kernels - exact for the beam-approximated physics and dependent on adequate path coverage, including multipathing branches - and it approximates the Gauss-Newton rather than the full-Newton operator; both limitations are mitigated by the finite-frequency EigenWave kernels and by the rich two-point coverage of the eigenray solutions.

## APPENDIX B. Size and Sampling of the LAD Projection Data

The proposed local directional projection (LDP) data, $I(\mathbf{x_p}, \boldsymbol{\nu}, \boldsymbol{\gamma}, \tau)$, are represented in an eight-dimensional (8D) phase-space comprising three spatial coordinates $\mathbf{x}$, two pairs of spherical angular parameters $\boldsymbol{\nu} = \{\nu_1, \nu_2\}$ and $\boldsymbol{\gamma} = \{\gamma_1, \gamma_2\}$ (4D in total), and a small time-lag dimension $\tau$. The discretization of this multidimensional domain should be designed to follow the local frequency bandwidth and dominant wavenumber content of the wavefield, thereby adapting the sampling density to the expected spatial and angular variations of the data.

### B1. Spatial Sampling

The spatial grid can be defined with high lateral and vertical resolution in the shallow subsurface, followed by a monotonic decrease in resolution with depth. This strategy reflects both the reduction in dominant frequency content and the increase in Fresnel-zone dimensions with propagation distance. The vertical sampling interval can be parameterized according to

$$dz_i = v_{\min}(z_i)\, \Delta T_{\text{ver}}, \tag{B.1}$$

where $v_{\min}(z_i)$ denotes the minimum velocity within the $i$-th depth slice and,

$$\Delta T_{\text{ver}} \approx \frac{1}{2f(z_i)} \quad , \tag{B.2}$$

with $f(z_i)$ representing the local dominant frequency. This criterion provides a practical balance between resolution requirements and computational cost while preserving the dominant kinematic and dynamic characteristics of the wavefield.

**B2. Angular Sampling**

The discretization of the - $\boldsymbol{\gamma}$ and $\boldsymbol{\nu}$-axes is determined according to their respective parameter sensitivities (**Appendix A**) and the intended inversion objectives.

**B3. Sampling of the $\boldsymbol{\gamma}$-Axis**

The $\boldsymbol{\gamma}$-axis, representing opening angles and azimuths, is parameterized using a uniform spherical-spiral discretization (Koren and Ravve, 2012). Because both kinematic attributes (e.g. residual moveout) and dynamic attributes (e.g. AVA(Z) responses) vary relatively smoothly along this axis, a moderate angular sampling is generally sufficient. In practical implementations, approximately 90 spherical bins are used to describe opening angles spanning the range $0°–180°$.

The availability of large opening angles decreases with depth because of finite acquisition offset and aperture. Consequently, direct arrivals, transmitted waves, diving waves, and head waves that exhibit opening angles approaching $180°$at a given subsurface scattering point are naturally restricted to limited penetration depths. Increasing the maximum source-receiver offset increases the depth range over which these large-opening-angle components can be observed.

**B4. Sampling of the $\boldsymbol{\nu}$-Axis**

The $\boldsymbol{\nu}$-axis describes directional dip and azimuth and is likewise discretized using a uniform spherical-spiral parameterization. In contrast to the $\boldsymbol{\gamma}$-axis, wavefield amplitudes and phases exhibit significantly stronger variations along the $\boldsymbol{\nu}$-direction. The amplitude difference between specular and strongly non-specular events may exceed two orders of magnitude, requiring substantially finer angular sampling.

To adequately capture these rapid variations, approximately 1000 spherical bins are used over the dip range $0°–90°$. The inward- and outward-propagating wavefields are discretized independently, ensuring appropriate directional resolution for both illumination and scattering analyses.

The availability of steep dip illumination largely depends on the lateral extension of the surface aperture. The steep dip illumination naturally decreases with the increasing depth.

**B5. Time-Lag Sampling**

The projected signals are sampled along the time-lag axis using approximately eleven samples centered around zero lag. The sampling interval is chosen according to

$$dt \approx \frac{1}{2f} \quad , \tag{B.3}$$

where $f$denotes the dominant local frequency. This sampling is sufficient to preserve the phase and amplitude behavior required for subsequent inversion and attribute analysis.

**B6. Resolution Requirements for Velocity and Scattering Parameters**

As demonstrated in **Appendix A**, the high-sensitivity components of the velocity model and those of the scattering field occupy predominantly different regions of phase space. Consequently, the corresponding inversion objectives can be effectively decoupled through appropriate spatial, angular, and frequency-domain filtering.

**Figure B1** illustrates the resulting dual-grid model parameterization for a structurally realistic setting, demonstrating that a fine, depth-graded Cartesian reflectivity grid (blue dots) and a sparse, structure-conformal velocity grid (brown dots) accommodate faulted and pierced stratigraphy without modification.

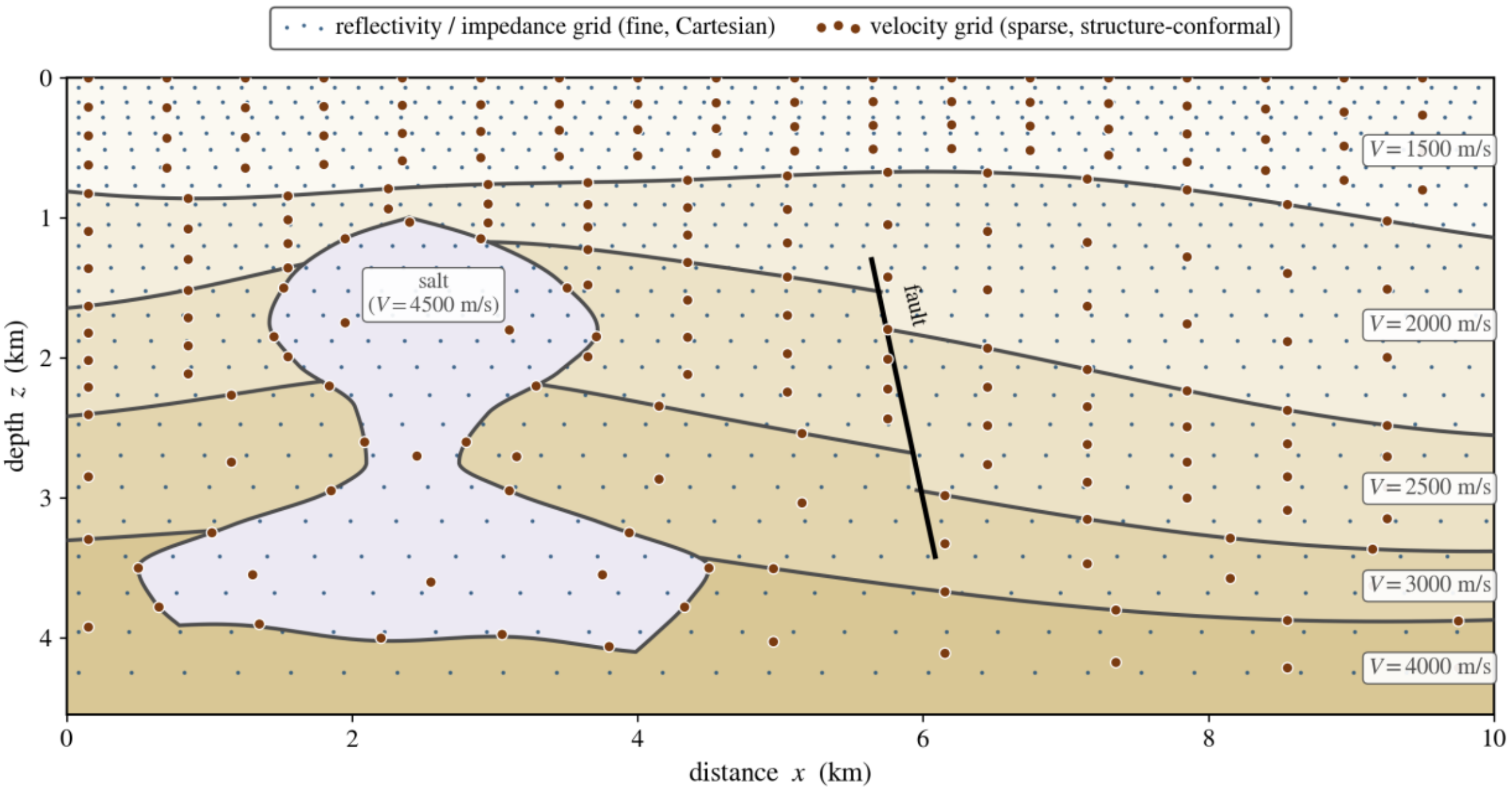

**Figure B1.** Dual-grid model parameterization consistent with the depth-graded sampling strategy of equations (B.1) - (B.2), illustrated for a structurally complex five-layer model. The reflectivity/impedance field is sampled on a fine Cartesian grid (blue) whose vertical and lateral spacings increase monotonically with depth, following the growth of the dominant wavelength; this grid is unaffected by the structural framework and continues uninterrupted through the salt body and into the sediments beneath its base. The long-wavelength velocity model is sampled on a much sparser, structure-conformal grid (brown): nodes are placed along the interpreted horizons, intermediate levels within each layer are obtained by linear interpolation between the bounding horizons, node columns follow the faulted horizons and are offset across the fault, nodes track the closed salt outline, and a few coarse nodes sample the (more or less) homogeneous salt interior. The contrast in grid density reflects the resolution requirements of the two parameter classes (**Appendix A**), consistent with the block-structured Hessian of **Figure A2**.

## APPENDIX C. Adaptive Frequency and Wavenumber Selection in LAD-FWI

Conventional multiscale FWI introduces frequencies sequentially from low to high values, $f_{\min} \rightarrow f_{\max}$, under the assumption that low frequencies recover large-scale velocity structure while higher frequencies provide increasing spatial resolution.

However, temporal frequency is only an indirect measure of subsurface resolution. The recoverable model wavenumber depends not only on frequency but also on scattering geometry. Consequently, observations recorded at the same frequency may contribute very different information to velocity and reflectivity estimation.

The LAD representation provides direct access to the LAD coordinates $(\mathbf{x_p}, \mathbf{k_v}, \mathbf{k_\gamma}, f)$, allowing inversion scheduling to be formulated in terms of parameter sensitivity, model resolution, and Hessian conditioning rather than frequency alone.

**C1. Frequency and Model Resolution**

In conventional FWI, the maximum recoverable model wavenumber increases approximately with frequency, $k_m \propto f$. More generally, $k_m \approx 2k \sin\left(\frac{\gamma}{2}\right)$, where $k$ is the wavefield wavenumber and $\gamma$ is the opening angle. Thus, model resolution is controlled jointly by frequency and scattering geometry. The LAD framework explicitly accounts for both through the coordinates $(\mathbf{k_v}, \mathbf{k_\gamma})$.

**C2. Sensitivity-Based Selection**

Let $\mathbf{J}_V(\mathbf{k_v}, \mathbf{k_\gamma}, f)$ and $\mathbf{J}_R(\mathbf{k_v}, \mathbf{k_\gamma}, f)$ denote the Jacobians for velocity and reflectivity parameters. The corresponding sensitivity energies are given below. As in Section 7, the reflectivity Jacobian $\mathbf{J}_R$ maps to the impedance domain through $\mathbf{J}_Z = \mathbf{J}_R\,\mathbf{G}$, so the coupling coefficient below may equivalently be expressed in the velocity–impedance blocks.

$$E_v(f) = \int \mathbf{J}_V^2\left(\mathbf{k_v}, \mathbf{k_\gamma}, f\right) d\mathbf{k_v}\, d\boldsymbol{k_\gamma}, \quad (C.1a)$$
$$E_r(f) = \int \mathbf{J}_R^2\left(\mathbf{k_v}, \mathbf{k_\gamma}, f\right) d\mathbf{k_v}\, d\mathbf{k_\gamma}. \quad (C.1b)$$

These quantities measure the amount of velocity-sensitive and reflectivity-sensitive information available at each frequency.
Frequencies may therefore be selected according to their contribution to the desired parameter update rather than their numerical value

**C3. Hessian-Based Selection**

For each frequency, $\mathbf{H}(f) = \mathbf{J}^T(f)\mathbf{J}(f)$. The velocity–reflectivity coupling coefficient is,

$$\rho_{VR}(f) = \frac{\mathbf{J}_V^T\mathbf{J}_R}{\sqrt{\left(\mathbf{J}_V^T\mathbf{J}_V\right)\left(\mathbf{J}_R^T\mathbf{J}_R\right)}}, \qquad 0 \le \rho_{VR} \le 1. \quad (C.2)$$

Large values indicate parameter crosstalk, whereas small values indicate increasing parameter separability. New frequencies are introduced only when they improve resolution or reduce parameter coupling.

**C4. Resolution Gain**

The Hessian eigenvalue spectrum $\mathbf{H}(f)\mathbf{q}_i = \lambda_i \mathbf{q}_i$ provides a measure of recoverable information. Define $R(f) = \sum_i \lambda_i$ , as the cumulative resolution measure. The contribution of a candidate frequency is $\Delta R(f) = R(f) - R(f - \Delta f)$. Frequencies giving negligible resolution gain, $\Delta R(f) \approx 0$, may be omitted from the inversion schedule.

**C5. Wavenumber Illumination**

The recoverable model-wavenumber illumination can be estimated as,

$$W(\mathbf{k}_m) = \sum_f |\,\mathbf{J}(\mathbf{k}_m, f)\,|^2 \, . \quad (C.3)$$

The inversion strategy is then:

1. Estimate the current illumination spectrum.
2. Identify unresolved model wavenumbers.
3. Select frequencies and LAD regions that illuminate them.
4. Update the model using only the most informative data subsets.

This transforms frequency continuation into a resolution-driven process.

**Figure C1** depicts this continuation directly in model-wavenumber space: successive stages expand the aperture-limited coverage sector from the low-wavenumber transmission regime toward the reflection-dominated band, and converted modes extend its outer bound.

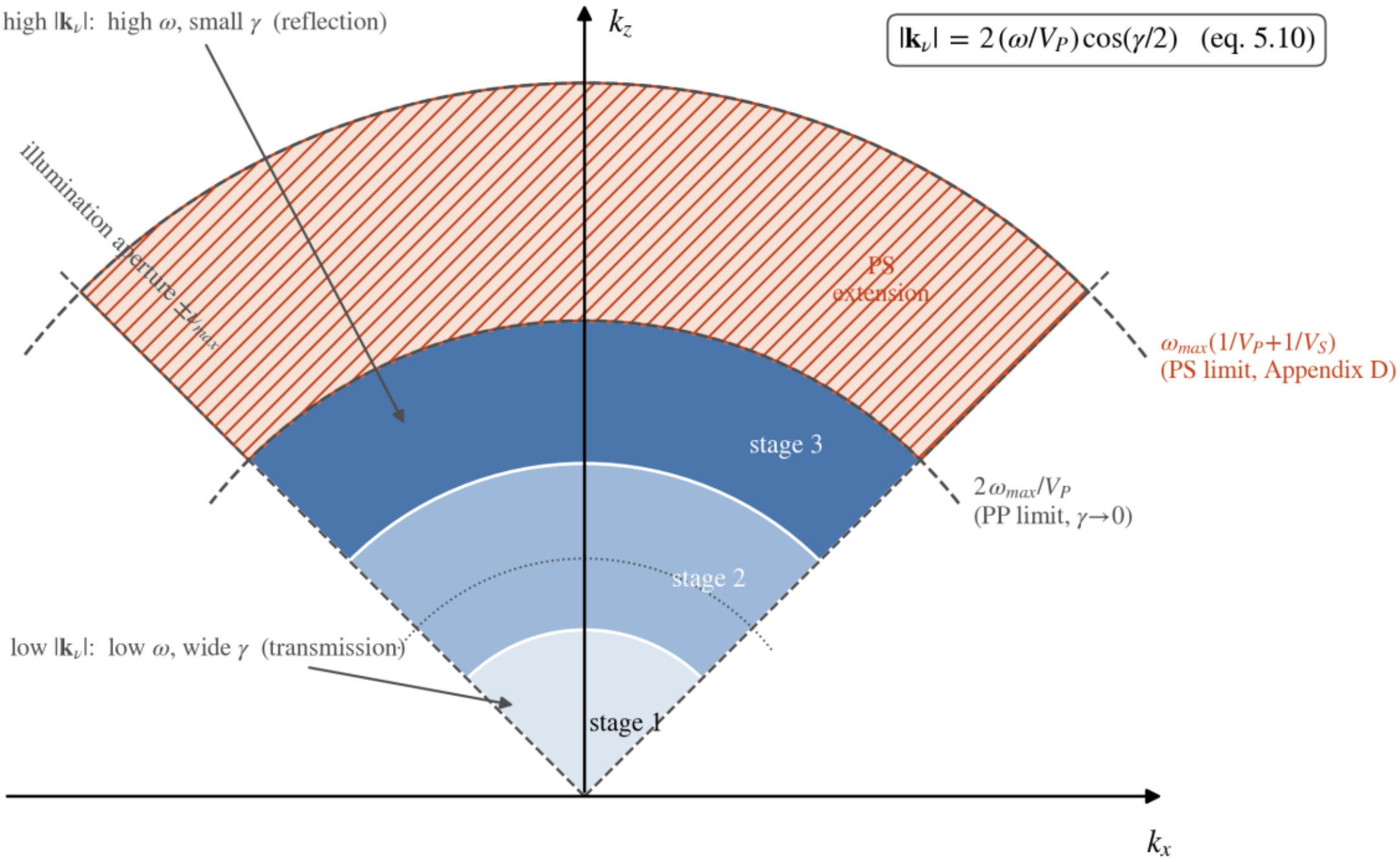

**Figure C1.** Model-wavenumber coverage and phase-space continuation. Each recorded frequency ω and opening angle **γ** contribute a model wavenumber of magnitude $|\mathbf{k}_\nu| = 2(\omega/V_P)\cos(\gamma/2)$ along the illumination direction **ν** (equation 5.10), so the accessible region of model-wavenumber space is a sector bounded by the illumination aperture $\pm\nu_{max}$. Early inversion stages (light shading) assemble the low-wavenumber velocity backbone from low frequencies and wide opening angles; later stages (darker shading) extend the coverage toward the reflection-dominated high-wavenumber band as higher frequencies and smaller opening angles are admitted by the resolution-driven selection of equations (C.1) - (C.3). Converted PS data (hatched) extend the recoverable band beyond the pure-mode limit $2\omega_{max}/V_P$ toward $\omega_{max}(1/V_P + 1/V_S)$ (**Appendix D, Figure D2**). Continuation is thus scheduled directly in ($|\mathbf{k}_\nu|$, ν) space rather than through temporal frequency alone (cf. **Figure 11**).

### C6. Adaptive Parameter Updating

**Velocity Updating:** Velocity estimation is associated primarily with large opening angles and low model wavenumbers,

$$E_V = \int_{\gamma>\gamma_c} \mathbf{J}_V^2 \; d\boldsymbol{\nu}\, d\boldsymbol{\gamma}. \tag{C.4}$$

During velocity model building, frequencies maximizing $E_v$ are preferentially selected.

**Reflectivity Updating:** Reflectivity estimation is associated primarily with small opening angles and high model wavenumbers.

$$E_R = \int_{\gamma<\gamma_c} \mathbf{J}_R^2 \, d\mathbf{\nu} \, d\boldsymbol{\gamma}. \tag{C.5}$$

At later stages, frequency selection is driven by reflectivity resolution rather than frequency itself, emphasizing fine-scale structural information.

**C7. Phase-Space Continuation Strategy**

The inversion progresses through increasingly resolved regions of phase space:

**Stage 1**: Macro-Velocity Building: $\gamma \to 180°, \ |\, \mathbf{k}_\nu \,| \to 0$

- transmitted and diving waves;
- low-wavenumber velocity updates;
- maximum inversion stability.

**Stage 2**: Model Refinement: $25° < \gamma < 60°$

- reflection-based residual moveout analysis;
- anisotropic parameter estimation;
- directional illumination refinement.

**Stage 3**: Reflectivity Reconstruction: $\gamma \to 0°$

- impedance inversion;
- high-wavenumber reflectivity;
- structural sharpening.

**Stage 4**: Diffraction Characterization

- fractures and small faults;
- localized heterogeneities;
- sub-wavelength features.

**Stage 5**: Full LAD Optimization

Joint inversion of all LAD components using the complete LAD representation.

**C8. Summary**

Conventional FWI uses temporal frequency as a proxy for model resolution. In contrast, LAD-FWI directly evaluates parameter sensitivity, model resolution, and parameter coupling in phase space. Consequently, inversion progression is controlled by illumination, sensitivity, resolution, conditioning and parameter separability, rather than frequency alone.

The resulting framework replaces classical frequency continuation with a resolution-driven continuation strategy, naturally aligned with the phase-space formulation of LAD-FWI and its progression toward increasingly resolved spatial, directional, and scattering wavenumbers.

**APPENDIX D. Converted-Wave Imaging and Inversion in Phase-Space**

Converted waves - most commonly downgoing compressional energy scattered into upgoing shear energy - recorded by multicomponent acquisition carry complementary information on shear properties, lithology, fluids, and fractures. Their processing in the acquisition-data domain, however, has long been burdened by asymmetric kinematics, model-dependent conversion-point binning, polarity reversals, and ill-posed PP-PS event registration (Thomsen, 1999; Stewart et al., 2002, 2003).

The underlying reason is that mode conversion is a local scattering phenomenon: it is governed entirely by the incident and scattered slowness directions and by the local perturbation at the image point - quantities that are projected only indirectly, and mode-asymmetrically, onto acquisition coordinates. The LAD phase space, whose coordinates are precisely these in-situ quantities, is therefore the natural domain for converted-wave imaging, inversion, and characterization. This appendix outlines the extension of the LAD-FWI framework to multicomponent, multi-mode data.

### D1. Mode Conversion as a Phase-Space Scattering Process

Under the Born approximation in an anisotropic elastic background, an incident wave of mode $c \in$ {qP, qS1, qS2} interacting with a local perturbation of the density-normalized stiffness tensor and of the density at image point $\mathbf{x_p}$ radiates scattered energy into all modes $c'$, with conversion strength governed by the incident slowness $\boldsymbol{p}_{in}^{(c)}$, the scattered slowness $\boldsymbol{p}_{sc}^{(c')}$, and the perturbation type (Wu and Aki, 1985). The converted-wave sensitivity kernels are therefore functions of exactly the LAD coordinates - position, propagation direction, opening angle and azimuth, and temporal support - augmented by the mode pair ($c$, $c'$). The LAD image data of the main text accordingly generalize to mode-indexed image data $I^{(cc')}(\mathbf{x_p}, \boldsymbol{\nu}, \gamma, \tau)$, with the pure-mode data recovered as the special cases $c = c'$.

### D2. Converted-Wave Specularity and Kinematics

For unequal incident and scattered phase velocities, the specular reflector normal aligns with the direction of the slowness sum $\boldsymbol{p}_{in} + \boldsymbol{p}_{sc}$ (e.g., Rosales et al., 2008; **Figure D1**). Since the LAD construction defines the migration dip and the opening angle through the slowness vectors, the converted-wave specular condition is exact, including in general anisotropic media, and the dip/opening-angle decomposition - together with the LDP specularity weights - carries over to mode-converted energy with no structural change.

Because the incident and scattered wave legs are propagated independently, each in its own mode, the asymmetric converted-wave kinematics are represented exactly, with no need for an asymptotic conversion-point approximation. Moreover, the magnitude of the recoverable directional wavenumber at the image point for a mode pair (c, c′) is bounded by

$$|\mathbf{k}_{\nu}^{(cc')}| \leq \omega(\frac{1}{V_c} + \frac{1}{V_{c'}}) \quad , \tag{D.1}$$

with equality at zero opening angle. Since the shear slowness magnitude exceeds the compressional one, converted modes extend the recoverable wavenumber coverage beyond that of pure compressional data at comparable frequencies - particularly where $V_P/V_S$ is large - thereby enriching the sensitivity and resolution analyses of **Appendices A and B** (**Figure D2**).

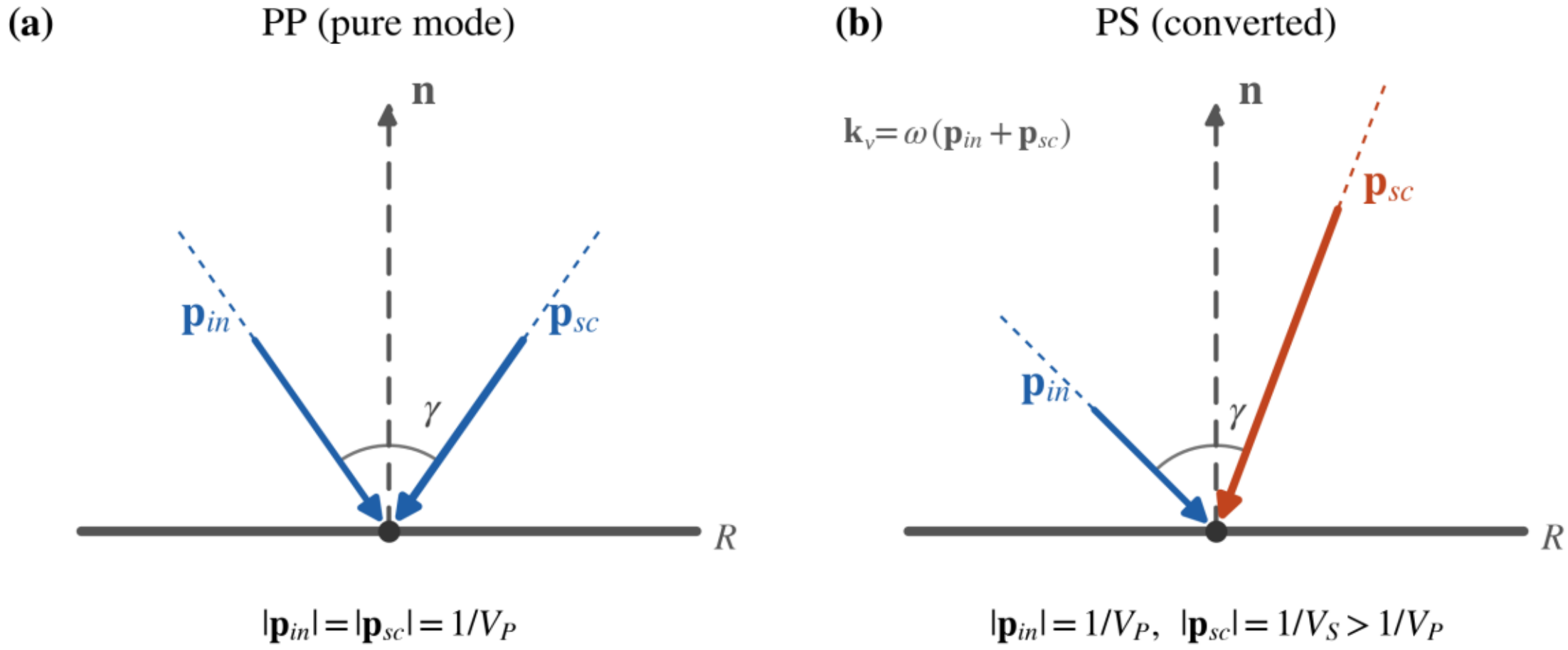

**Figure D1.** Specular scattering geometry at an image point for (a) pure-mode PP and (b) converted PS reflection, with the incident and scattered slowness vectors $\mathbf{p}_{in}(c)$ and $\mathbf{p}_{sc}(c')$ drawn pointing toward the scattering point. For PP, the equal slowness magnitudes render the reflector normal $\mathbf{n_r}$ the bisector of the two specular legs. For PS, the shear-leg slowness magnitude $1/V_S$ exceeds the compressional $1/V_P$; the specular legs are asymmetric, the normal is aligned with the direction of the slowness sum, and the illumination directional wavenumber vector $\mathbf{k_\nu} = \omega(\mathbf{p}_{in} + \mathbf{p}_{sc})$ accesses higher model wavenumbers than in the pure-mode case at the same frequency.

### D3. Polarity and Amplitude Behavior

The PS scattering response is antisymmetric in the opening angle and vanishes at normal incidence. Within the LAD image data this sign structure is explicit: polarity is resolved per directional and opening-angle bin prior to any partial summation, converting the destructive-stacking hazard of acquisition-domain processing into a well-defined angle-dependent observable that constrains the shear reflectivity vector $\mathbf{R}_S$ and the density term. The odd symmetry further provides a useful internal consistency check on the recovered reflector orientation.

### D4. Mode-Decomposed LDP Signals from Multicomponent Data

For multicomponent data, the elastic wavefield at the image point is decomposed into its qP, qS1, and qS2 constituents by projection onto the polarization eigenvectors of the local Christoffel matrix (cf. Yan and Sava, 2008, in the context of elastic angle-domain imaging). The LDP construction of Section 6 then applies mode by mode, yielding projected signals for each mode pair (c, c′). In azimuthally anisotropic media, the qS1/qS2 distinction retains shear-wave splitting as an in-situ observable, so that splitting-based fracture characterization is performed in the same LAD coordinates as the inversion itself.

### D5. Cross-Mode Consistency as an Inversion Objective

Because all mode-pair images are formed on a common spatial grid with a common model, structural consistency across modes becomes an explicit LAD objective: co-location of the reflectivity support and co-orientation of the reflectivity vectors inferred from the PP and PS images. The residual of this objective is directly and quasi-linearly sensitive to the $V_P/V_S$ (equivalently $Z_P/Z_S$) field, replacing data-domain PP-PS traveltime registration - a notoriously ill-posed preprocessing step - with a well-posed model-domain constraint that participates in the adjoint-state gradient like any other term of the generalized objective function of Section 8.

### D6. Joint Multi-Mode Hessian Structure

The LAD Hessian generalizes to mode-indexed blocks. The PS radiation patterns constrain the shear impedance and the density over opening-angle bands in which the PP patterns are weak or degenerate, while the PP data dominate the constraint on the compressional velocity and impedance. The joint (PP + PS) multiparameter Hessian therefore exhibits increased diagonal dominance relative to either mode alone (**Figure D2**) - the cross-mode analogue of the velocity-impedance separation that motivates the main text - improving the conditioning of the $V_P/V_S$, shear-impedance, and anisotropic-parameter estimation, and tightening the a posteriori derivation of the density discussed in Section 7. The sensitivity, resolution, and conditioning analyses of **Appendix A** apply per mode pair, and the adaptive selection strategy of **Appendix C** extends naturally to scheduling over wave modes as well as over frequencies and LAD regions.

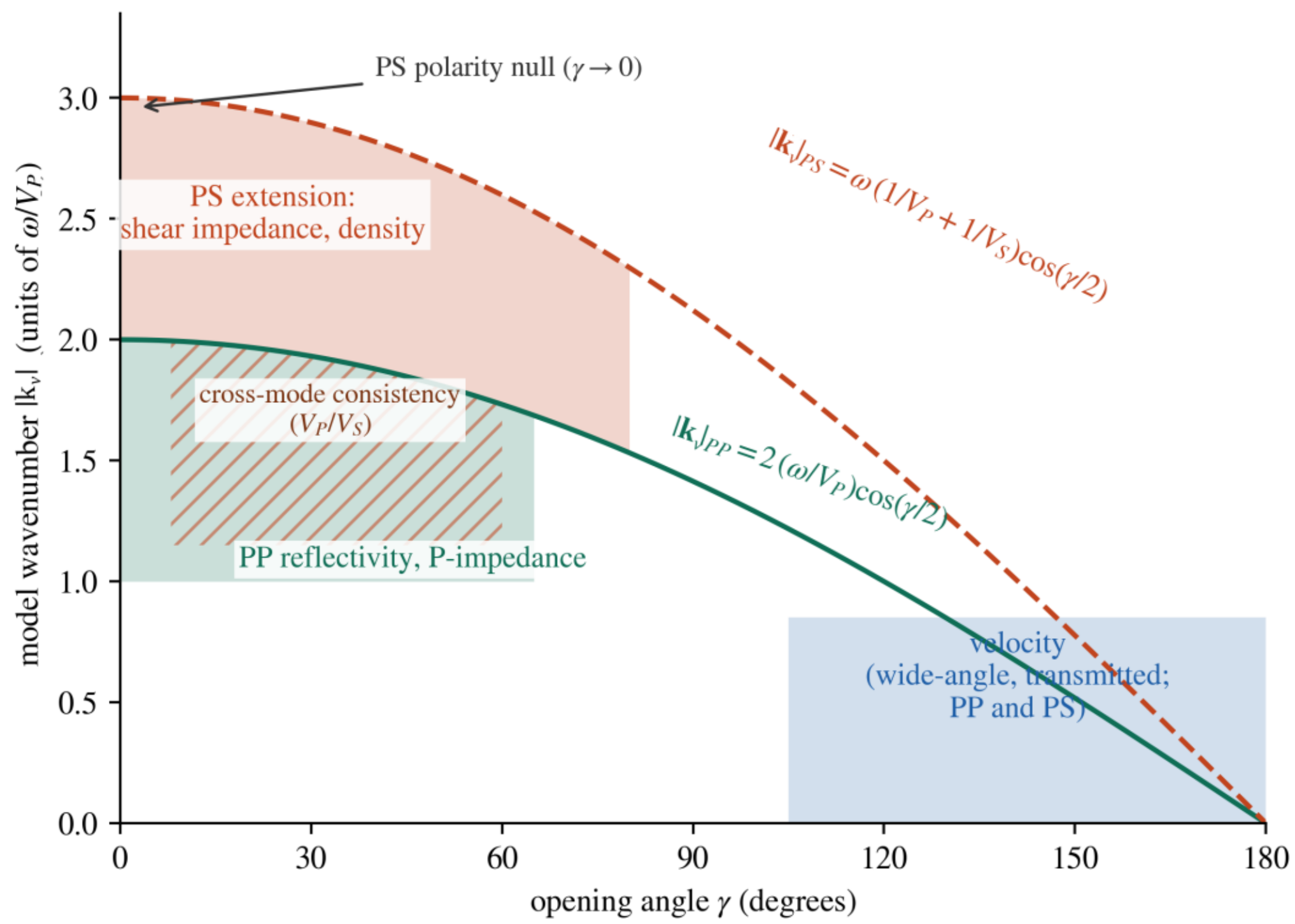

**Figure D2.** Schematic phase-space support of PP and PS parameter sensitivities in the plane of opening angle versus model wavenumber (in units of $\omega/V_P$, drawn for $V_P/V_S = 2$). Velocity sensitivity concentrates at wide opening angles and low wavenumbers. PP reflectivity and P-impedance sensitivity occupy the small-angle band bounded by $|\mathbf{k}|_{PP}$; PS data extend the recoverable band toward $|\mathbf{k}|_{PS}$ and constrain the shear impedance and the density. The hatched region marks the cross-mode overlap supporting the $V_P/V_S$ consistency objective; the PS response vanishes at normal incidence (polarity null).

## APPENDIX E. Discrimination of Multiply Scattered Energy in Phase Space

As discussed in Section 10, the ray/beam imaging operator $\mathfrak{M}(m)$ maps coherent arrivals under the single-scattering assumption. Consequently, multiply scattered energy is mapped into the LAD image as coherent residual energy that lies outside the range of the primary-scattering operator. This appendix develops that observation into a practical discrimination framework.

Each of the five criteria described below defines a soft weighting function $w_i(\mathbf{x}, \boldsymbol{\nu}, \boldsymbol{\gamma}) \in [0, 1]$. The composite weight,

$$w = \prod_i w_i \tag{E.1}$$

is applied to the residuals entering the objective functions of Section 8 and is also used to gate the adjoint back-projection. Energy identified as inconsistent with primary scattering therefore contributes neither to the objective function nor to the resulting model updates. The proposed strategy assumes the Born-linearized ray/beam implementation of Section 9, whose explicit path-level sensitivities and local Hessian information are exploited by the following criteria:

### E1. Directional-Consistency Semblance

The inherent rejection capability of the local slant-stack (LSS) operators (Koren, 2009) can be reformulated as a continuous confidence measure. For each central primary-ray pair, the LSS semblance $S_{\mathrm{LSS}}(\mathbf{p})$ is evaluated over a range of trial surface-slowness pairs within the corresponding Fresnel-zone apertures. The directional-consistency weight is then defined as,

$$w_1 = \left[\frac{S_{\mathrm{LSS}}(\mathbf{p}_{\mathrm{mod}})}{\max_{\mathbf{p}} S_{\mathrm{LSS}}(\mathbf{p})}\right]^{q_1}, \tag{E.2}$$

where $\mathbf{p}_{\mathrm{mod}}$ denotes the modeled slowness pair associated with the primary ray.

Primary events attain their maximum semblance near $\mathbf{p}_{\mathrm{mod}}$, yielding $w_1 \approx 1$. Multiples that share a similar traveltime but arrive with different free-surface directivity achieve their maximum response at different slownesses and therefore receive reduced weights ($w_1 \ll 1$). Importantly, the ratio formulation does not penalize diffractions. Their slant-stack response is typically broadened rather than shifted, such that the semblance evaluated at the modeled slope remains close to the maximum value. Field-scale demonstrations of this mechanism are reported by Inozemtsev et al. (2015).

**E2. Range-Space Explainability**

A defining property of a primary event is that it can, in principle, be explained by an admissible model perturbation. Let **r** denote the residual vector within a local LAD neighborhood over the $(\nu', \nu)$ bins, and let,

$$\mathbf{J} = [\mathbf{J}_{V'}\, \mathbf{J}_Z] \tag{E.3}$$

represent the corresponding local Fréchet block introduced in **Appendix A**. The explainable fraction is defined as,

$$q = \frac{\|\mathbf{Pr}\|^2}{\|\mathbf{r}\|^2}, \tag{E.4}$$

where,

$$\mathbf{P} = \mathbf{J}\left(\mathbf{J}^{\mathrm{T}}\mathbf{J} + \varepsilon\mathbf{I}\right)^{-1}\mathbf{J}^{\mathrm{T}}, \tag{E.5}$$

is the regularized orthogonal projector onto the range of $\mathbf{J}$. The associated weight is,

$$w_2 = q^{q_2}. \tag{E.6}$$

Velocity errors generate flattenable residual moveout, and impedance errors generate $\gamma$-coherent amplitude residuals; both belong to the range of $J$. Surface multiples, in contrast, commonly produce residual moveout patterns that cannot be reproduced by any admissible perturbation $(\delta V, \delta Z)$, leaving a substantial component orthogonal to the range of the linearized operator.

In conventional data-domain FWI, such projections are generally impractical to evaluate. Here, however, $\mathbf{J}^{\mathrm{T}}\mathbf{J}$ corresponds to the beam-based Gauss-Newton Hessian block assembled explicitly in **Appendix A**, making range-space discrimination available as a natural by-product of the inversion machinery.

This criterion complements the annihilator interpretation of Term 1 in Section 8. Whereas the flattening annihilator removes behavior that a correct model should not produce, the orthogonal projector $\mathbf{I} - \mathbf{P}$ removes behavior that no admissible model perturbation can produce. In weakly illuminated regions, the projection becomes less reliable and the quantity $q$should therefore be illumination-normalized, consistent with the treatment used in the semblance objective.

**E3. LDP Wavelength Consistency**

Each Local Directional Projection (LDP) packet must satisfy the local dispersion relation

$$|\,\mathbf{k_v}\,| = \frac{2\omega\cos(\gamma_1/2)}{v_{\mathrm{phs}}(\mathbf{x})} \quad , \qquad \text{(E.7)}$$

at its imaged location (Section 6).

A multiple accumulates its traveltime along slower shallow paths but is frequently mapped to a deeper and faster location. As a result, the packet exhibits a spatial wavelength that is inconsistent with its imaged position. In other words, the measured spectral content is not compatible with the local velocity implied by the imaging geometry.

Because the LDP representation preserves the temporal waveform, the instantaneous wavenumber $k_{\mathrm{meas}}$ can be estimated directly from each packet. The resulting weight is defined as

$$w_3 = \exp\left[-\frac{(k_{\mathrm{meas}} - k_{\mathrm{pred}})^2}{2\sigma_k^2}\right], \qquad \text{(E.8)}$$

where $k_{\mathrm{pred}}$ is the wavenumber magnitude predicted by the local dispersion relation.

This criterion is unique to the LDP representation, which simultaneously preserves both the local waveform characteristics and the velocity information required for the consistency test.

**E4. Predictive LAD Masking**

Once strong shallow generators such as the water-bottom reflector or top-salt interface have been identified, mirror-ray pairs containing a single free-surface reflection can be traced using the same propagation engine. Predicted first-order surface multiples and peg-leg events are then mapped through the primary imaging operator into their expected LAD coordinates $\left(\mathbf{x}_{\mathrm{p}}, \mathbf{v}, \boldsymbol{\gamma}, \tau\right)$.

The resulting occupancy map defines a soft mask $w_4$, representing the phase-space analogue of surface-related multiple prediction (SRMP). In contrast to conventional adaptive subtraction, however, the predicted energy is used to construct a weighting function rather than to remove data directly. Within the EigenRay framework, the additional bounce-path tracing is computationally inexpensive.

The same mirror-ray formulation also enables an optional extension in which first-order surface multiples are treated as bounce-path primaries and incorporated into the inversion. In this interpretation, energy that would otherwise be suppressed may contribute additional small-opening-angle illumination of the shallow subsurface.

Internal multiples generally evade this predictive criterion and must instead be addressed through the complementary mechanisms provided by criteria E2, E3, and E5.

**E5. Iteration-Persistence Reweighting**

The persistence measure $P(\mathbf{x}, \boldsymbol{v}, \boldsymbol{\gamma})$ quantifies the normalized residual energy that remains in a LAD bin over multiple inversion iterations relative to the overall reduction of the objective function. Residuals associated with correctable velocity or impedance errors decay as the model converges and therefore exhibit low persistence, whereas residuals generated by multiples remain largely unchanged and exhibit high persistence.

A simple implementation for $P$ could be,

$$P_n(\mathbf{x}, \boldsymbol{\nu}, \boldsymbol{\gamma}) = \frac{1}{K}\sum_{k=n-K+1}^{n} \frac{|r_k(\mathbf{x}, \boldsymbol{\nu}, \boldsymbol{\gamma})|}{\|r_k\|}, \tag{E.9}$$

where $r_k$ is the local residual at iteration $k$. A more useful version measures lack of decay:

$$P_n = \frac{\sum_{k=n-K+1}^{n} |r_k|}{|r_{n-K+1}| + \varepsilon}. \tag{E.10}$$

Then,

- $P \ll 1$: residual is decaying rapidly → likely primary-related.
- $P \approx 1$: residual is essentially unchanged → candidate multiple.
- $P > 1$: residual is growing → likely unstable or inconsistent.

Conceptually, $P \propto 1 - \text{corr}\,(|, r_n$ | misfit reduction$)$, so $P$ measures how poorly the local residual follows the global convergence trend. If the global objective decreases by 80% while a particular LAD bin decreases by only 5%, that bin receives a large persistence value and is progressively down-weighted through,

$$w_5 = \exp(-\beta P). \tag{E.11}$$

Bins that retain significant residual energy over many iterations receive progressively smaller weights, whereas bins that respond to model updates remain largely unaffected. This criterion operationalizes the flattening-consistency concept introduced in Section 10.

This criterion is especially valuable. The other four criteria are **kinematic** or **physical**:

- E1: directional consistency,
- E2: explainability by the Jacobian,
- E3: wavelength consistency,
- E4: predicted-multiple occupancy.

E5 is fundamentally different because it is **dynamic**. It does not ask whether an event looks like a multiple; instead, it asks: *Has this event behaved like something the inversion can fit?* Anything that systematically survives many iterations despite substantial model improvement becomes suspicious.

**E6. Composite Discrimination Strategy**

The five criteria are complementary and operate on distinct aspects of the imaging and inversion process. Criterion E1 acts before stacking through free-surface directional consistency. Criteria E2 and E3 operate in the image domain through kinematic and spectral consistency, respectively. Criterion E4 exploits explicit forward prediction of multiple events, while Criterion E5 exploits their distinctive persistence during inversion.

Consequently, multiply scattered energy must evade all five discrimination mechanisms to influence a model update, whereas genuine primary-scattering energy is largely unaffected by any of them. A quantitative evaluation of the composite weighting strategy on synthetic and field data is left for future work.

**APPENDIX F. Synthetic Examples: Analytic and Numerical Benchmarks**

The purpose of this appendix is to validate the central hypothesis of LAD-FWI under progressively more realistic conditions. The hypothesis is that velocity-type and reflectivity-type perturbations occupy predominantly different regions of phase space and therefore produce reduced coupling in the LAD representation relative to conventional data-domain formulations. The analytical examples provide exact tests of this hypothesis, while the numerical example demonstrates the persistence of the same behavior in a geologically realistic setting. The benchmarks investigate three related questions:

1. Do velocity and reflectivity sensitivities exhibit distinct support in LAD space?
2. Does the LAD objective produce a broader basin of attraction than waveform-residual objectives?
3. Does sensitivity separation lead to improved Hessian conditioning and parameter resolution?

The examples are organized in increasing complexity. Example 1 examines a point diffractor embedded in a constant-velocity-gradient medium, where all ray quantities are available analytically. Example 2 introduces a dipping reflector and allows direct evaluation of sensitivity, conditioning, and resolution. Example 3 extends the analysis to a complex structural model representative of realistic exploration settings.

**Note**: The benchmarks below are not intended to quantify expected field-data performance. Their purpose is to validate the mathematical mechanisms underlying LAD-FWI in environments where exact solutions are available.

### F1. EXAMPLE 1: Constant-Velocity-Gradient Model

This example provides an exact analytical benchmark. Because ray paths, traveltimes, LAD coordinates, and model perturbations can all be written in closed form, it allows the key (kinematic-based) claims of LAD-FWI to be tested without numerical approximations. A point scaterrer is placed at 2 km depth in a constant-gradient velocity model,

$$V(z) = V_0 + gz \tag{F.1}$$

in which every ray is a circular arc whose center lies on the level $z = -V_0/g$, and the two-point traveltime between any points A and B a distance R apart is given in closed form by,

$$t = \frac{1}{g}\,\mathrm{arccosh}\left(1 + \frac{g^2R^2}{2V_AV_B}\right), \tag{F.2}$$

while the ray parameter p, determined by the circular arc through the two endpoints, obeys Snell's invariant

$$\sin\theta(z) = p\,V(z). \tag{F.3}$$

The acquisition geometry covers a wide range of illumination angles, generating nearly complete coverage of the LAD coordinates $(\nu, \gamma)$. For a surface source and any image point, the incident and scattered slowness vectors at the image point - and therefore the directional angle $\nu$, the opening angle $\gamma$, and the directional wavenumber $k_\nu$ of equation (5.10) - follow analytically from equations (F.2) - (F.3), with no numerical ray tracing or wavefield simulation.

The benchmark places a point diffractor at 2 km depth in a medium with $V_0$ = 1.8 km/s and g = 0.6 $s^{-1}$ (local velocity 3.0 km/s; two-way vertical time 1.703 s), recorded by surface sources and receivers at offsets up to ±8 km. The resulting illumination is exceptionally broad: symmetric source-receiver pairs sweep the opening angle over the full range 0°–180°, with horizontally arriving rays at ±4.0 km offset, while turning rays beyond this offset illuminate the diffractor from below, reaching one-way propagation angles of 121° from the vertical at ±8 km.

**Figure F1** shows the analytic ray fan and the attained $(\nu, \gamma)$ coverage; the attained wavenumber coverage follows directly from $|\mathbf{k}_\nu| = 2(\omega/V)\cos(\gamma/2)$, providing an exact instance of the coverage sector of **Figure C1**.

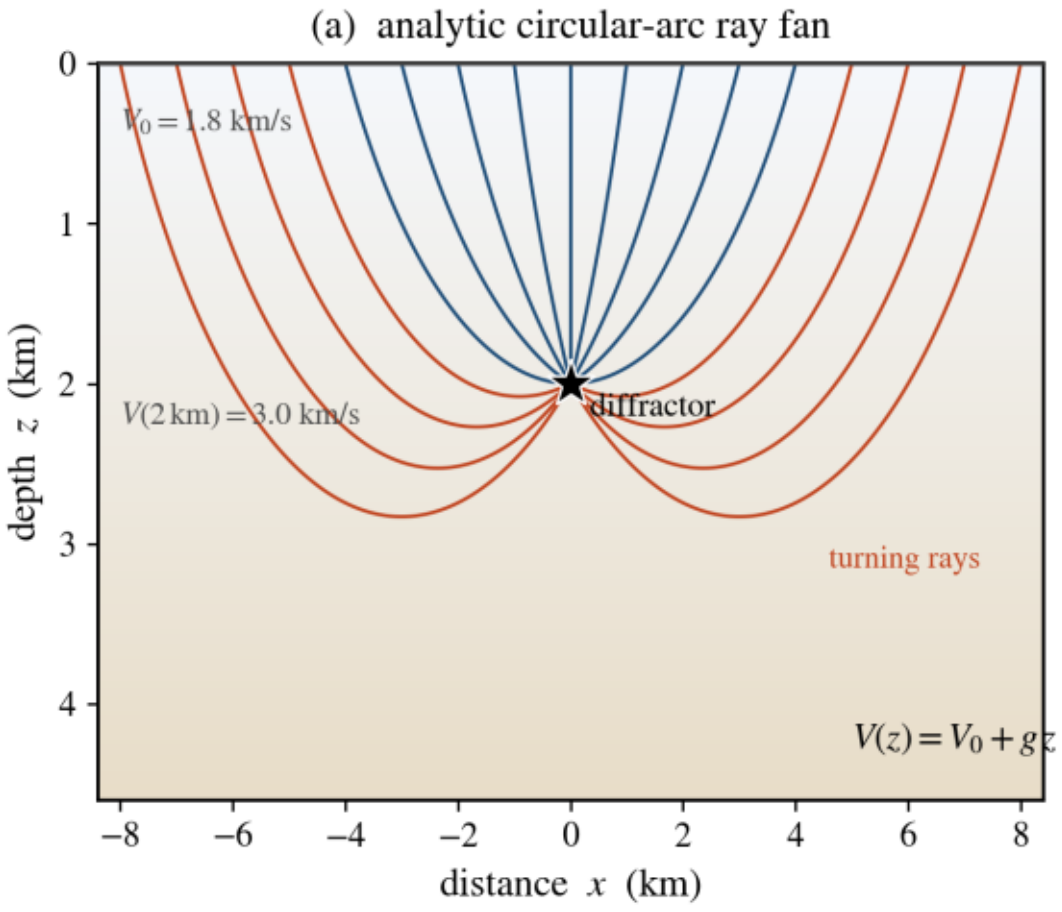

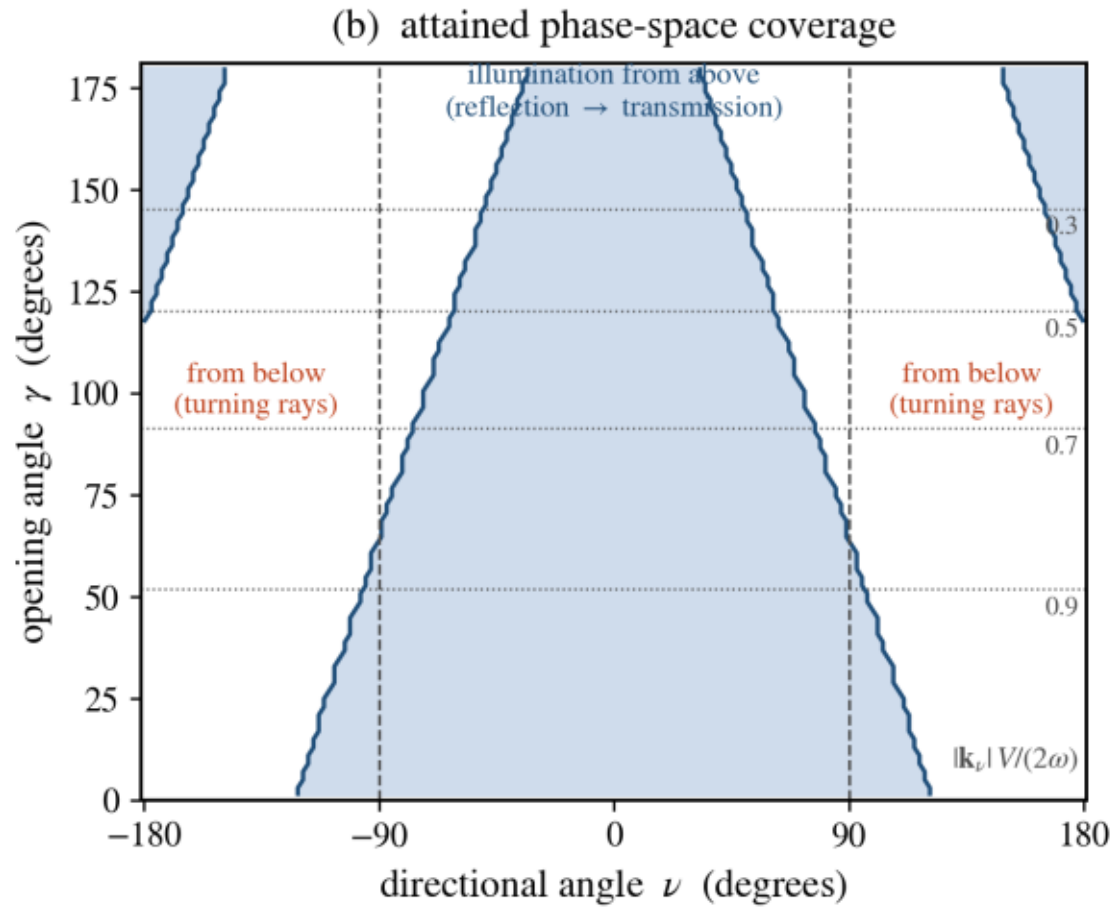

**Figure F1.** Analytic constant-gradient benchmark: acquisition and LAD coverage. (**a**) Circular-arc rays traced analytically from surface positions (offsets to ±8 km) to a point diffractor at 2 km depth in V(z) = $V_0$ + gz ($V_0$ = 1.8 km/s, g = 0.6 s$^{-1}$); rays that bottom below the diffractor (orange) illuminate it from below. (**b**) The attained region of the ($\nu$, $\gamma$) plane (shaded): symmetric pairs sweep the full opening-angle range, with horizontally arriving rays at ±4 km offset, and turning-ray combinations extend the directional coverage to illumination from below ($|\nu| > 90°$). Dotted horizontal lines mark levels of the normalized directional-wavenumber magnitude $|\mathbf{k}_\nu| V/(2\omega) = \cos(\gamma/2)$ of equation (5.10).

The property that makes this medium ideally suited to testing LAD-FWI is its exact behavior under a uniform velocity scaling. Since $V_0$ and $g$ scale together, the argument of the arccosh in equation (F.2) is invariant, and therefore,

$$V(z) \to (1+\varepsilon)\,V(z) \quad \Longrightarrow \quad t(\nu,\gamma) \to \frac{t(\nu,\gamma)}{1+\varepsilon}, \tag{F.4}$$

every ray path is unchanged while every traveltime scales by $1/(1+\varepsilon)$. Migrating data recorded in the true medium with the scaled model therefore displaces each image contribution along its fixed ray pair by an exactly computable depth shift; to first order,

$$\delta z(\nu,\gamma;\varepsilon) = \frac{\varepsilon\, t(\nu,\gamma)\, V(z_I)}{2\cos(\gamma/2)\cos\nu} + O(\varepsilon^2). \tag{F.5}$$

So, the residual moveout is expressed directly in the LAD coordinates of the method, with the vertical slowness sum $2\cos(\gamma/2)\cos\nu / V$ - the vertical component of the directional wavenumber - appearing in the denominator. The linearization is accurate to a few meters over moderate scaling (at $\gamma = 67°$, the exact shift for $\varepsilon = 5\%$ is 172 m against 171 m from equation (F.5)) and degrades only as $\gamma \to 180°$, where the isochron becomes vertical and the depth sensitivity diverges - itself a direct expression of the wide-angle concentration of velocity sensitivity.

**Figure F2** collects the two principal demonstrations. Panel (**a**) shows the exact residual moveout of the diffractor image in the opening-angle gather: the shift grows rapidly with opening angle, realizing the $\sin^2(\gamma/2)$ velocity kernel of equation (A.3) with exact rather than asymptotic kinematics. Panel (**b**) contrasts the inversion objectives over the same wide aperture: the data-domain waveform misfit develops cycle-skipped secondary minima already at $\varepsilon \approx \pm3\%$, with depths reaching 40% of the misfit range, whereas the flatness objective $\Phi_1$ of equation (8.7), stacked over a 10–20 Hz band, retains a single basin: its largest secondary undulation anywhere within ±15% amounts to about one percent of the misfit range.

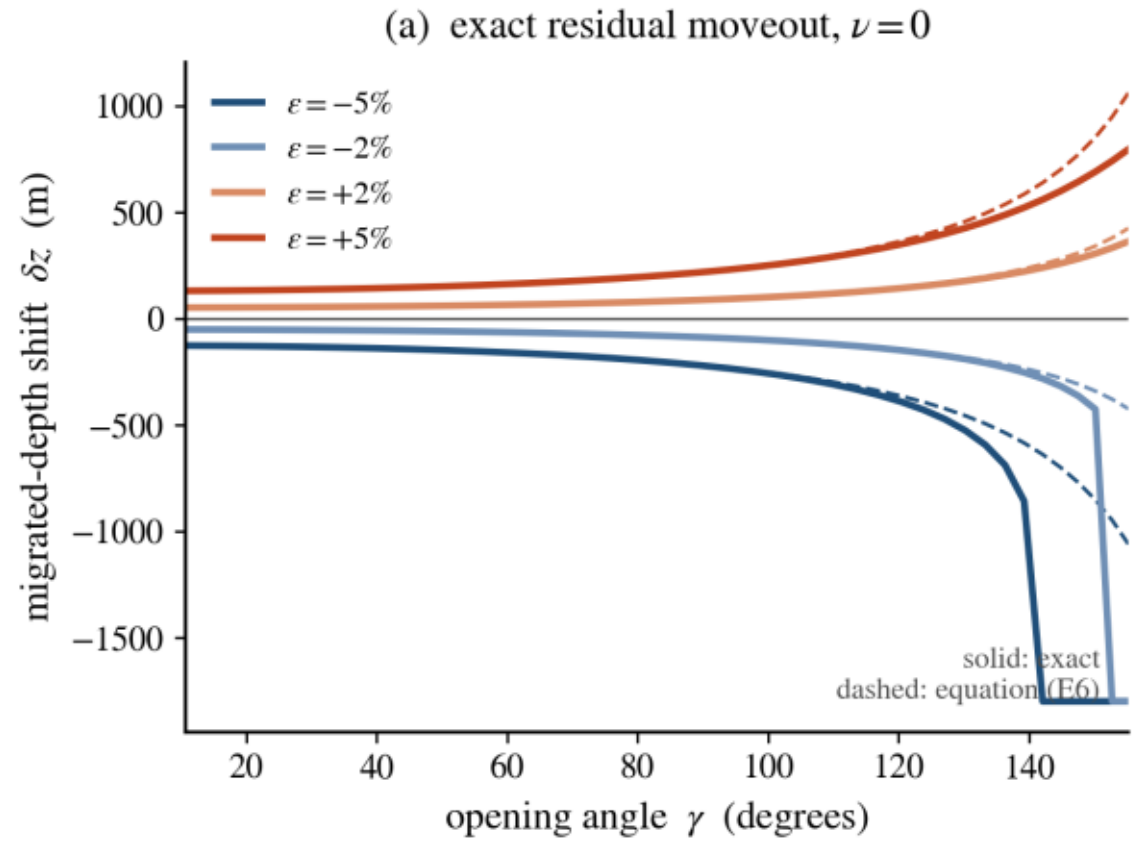

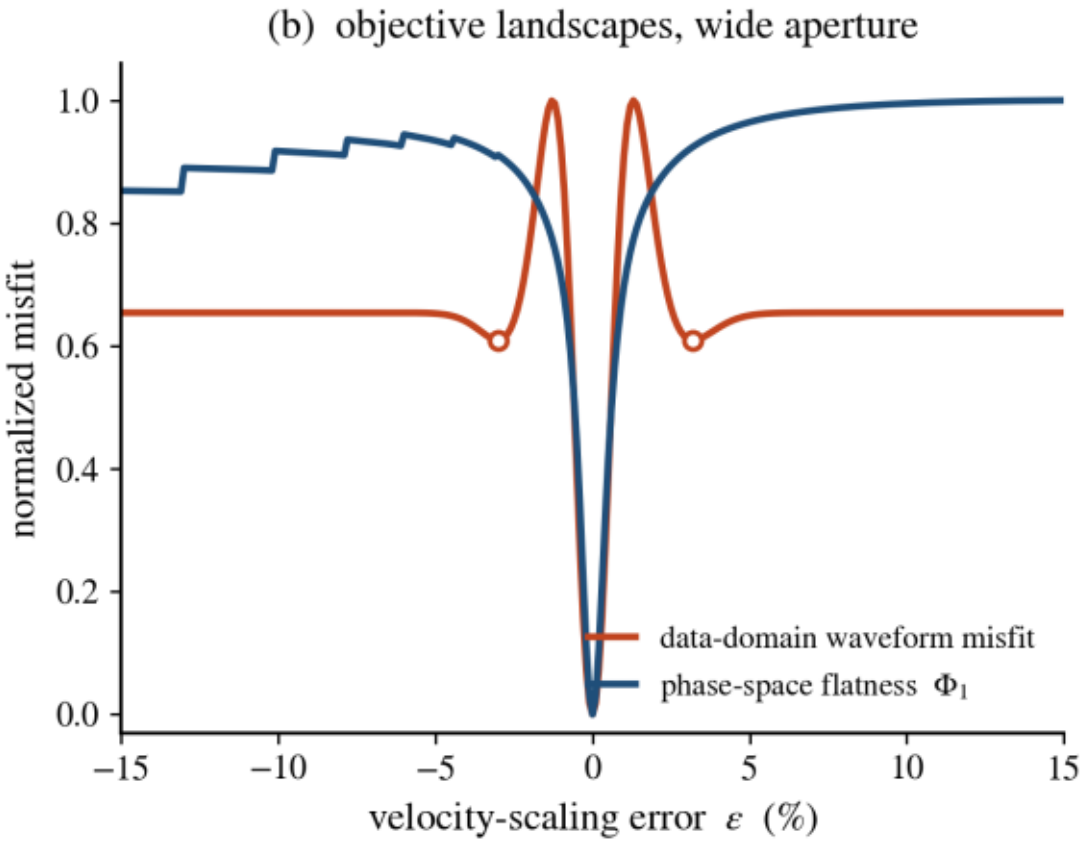

**Figure F2.** Exact LAD response of the benchmark under uniform velocity scaling $V \rightarrow (1+\varepsilon)V$. (a) Residual moveout $\delta z$ of the diffractor image in the opening-angle gather at $\nu = 0$: exact solutions of the analytic traveltime equation (solid) and the first-order expression of equation (F.5) (dashed) for $\varepsilon = \pm 2\%$ and $\pm 5\%$. (b) Objective landscapes over the same aperture: the data-domain waveform misfit (orange; Ricker wavelet, $f_0$ = 15 Hz) develops cycle-skipped secondary minima at $\varepsilon \approx \pm 3\%$ (open circles), whereas the flatness objective $\Phi_1$ of equation (8.7), evaluated on the opening-angle gather and stacked over a 10–20 Hz band, retains a single basin of attraction across the full ±15% range.

The aperture dependence of the velocity-reflectivity coupling can likewise be evaluated in closed form. Applying the asymptotic kernels of equation (A.3) over the attained opening-angle range, the coupling measure $\eta$ of equation (A.12) falls from 0.72 for a narrow reflection aperture ($\gamma \leq 40°$) to 0.33 for the full aperture - a reduction obtained from the angular supports alone, before the additional wavenumber-band separation of **Appendix C** is applied. Finally, the imaging operator in this setting is literally a mapping of the observed samples onto analytic isochrons with LAD binning - the architectural principle of Section 8 realized with no synthetic forward-modeling step - and the same framework extends at negligible cost to converted waves by assigning each leg its own gradient medium (**Appendix D**).

**Objective-Function Behavior:** The conventional waveform-misfit objective exhibits the well-known cycle-skipping phenomenon. Multiple local minima appear as soon as the kinematic error exceeds approximately half a cycle. By contrast, the LAD flattening objective depends on image consistency across opening angle rather than waveform phase alignment. The resulting objective remains smooth and quasi-convex across the entire tested velocity-error range. Consequently, LAD attributes provide a substantially broader basin of attraction for long-wavelength velocity updates.

### F2. EXAMPLE 2: A Tilted Reflector: Sensitivity, Resolution, and Conditioning

The same machinery extends to a planar reflector dipping at 15° through the point (0, 2 km), with the specular reflection point for each source–receiver pair obtained by a ray bending method - Fermat minimization of the closed-form two-leg traveltime (F.2). This configuration furnishes an exact test of the specularity principle at the heart of the LAD construction: at every genuine reflection point the computed directional angle equals the reflector dip, $\nu = 15°$, to the accuracy of the search (better than $10^{-3}$ degrees) for all half-offsets up to the critical aperture of ≈ 3.7 km, beyond which the specular branch disappears into the turning-ray regime. Because the medium is heterogeneous and the reflector tilted, this is a non-trivial verification that the LAD angle system delivers the migration dip exactly, with no small-angle or homogeneity approximation.

The tilted event also provides a clean quantitative comparison of the three **Appendix A** diagnostics.

**Sensitivity**: at the specular direction $\nu$ = dip, the exact velocity sensitivity of equation (F.5) grows with opening angle while the reflectivity response decays as $\cos^2(\gamma/2)$ (**Figure F3b**) - the disjoint angular supports of equation (A.3) realized on a dipping event.

**Conditioning**: the classical velocity-depth trade-off is measured by the normalized correlation $\rho$ between the traveltime sensitivities to a uniform velocity scaling ($\partial t/\partial\varepsilon = -t$, exact) and to a normal shift of the reflector ($\partial t/\partial n = -2\cos(\gamma/2)/V$). In the data domain the associated condition number cond $= (1+|\rho|)/(1-|\rho|)$ is 269 for a narrow reflection aperture ($|\rho| = 0.993$) and 8 for the full pre-critical aperture, whereas the LAD flatness residual of Term (1) is invariant to a bulk depth shift by construction, so its coupling to the reflector position vanishes identically and cond = 1.

**Resolution**: with the columns normalized and one percent damping, the resolvable fraction of each parameter (the diagonal of the resolution operator, equation A13) is 0.71 in the narrow-aperture data domain, recovering to 0.98 at full aperture, against 0.99 for the decoupled LAD blocks (**Figure F3c**). The benchmark thus reproduces, with exact arithmetic, the qualitative structure asserted in **Appendix A** and visualized in **Figure A2**.

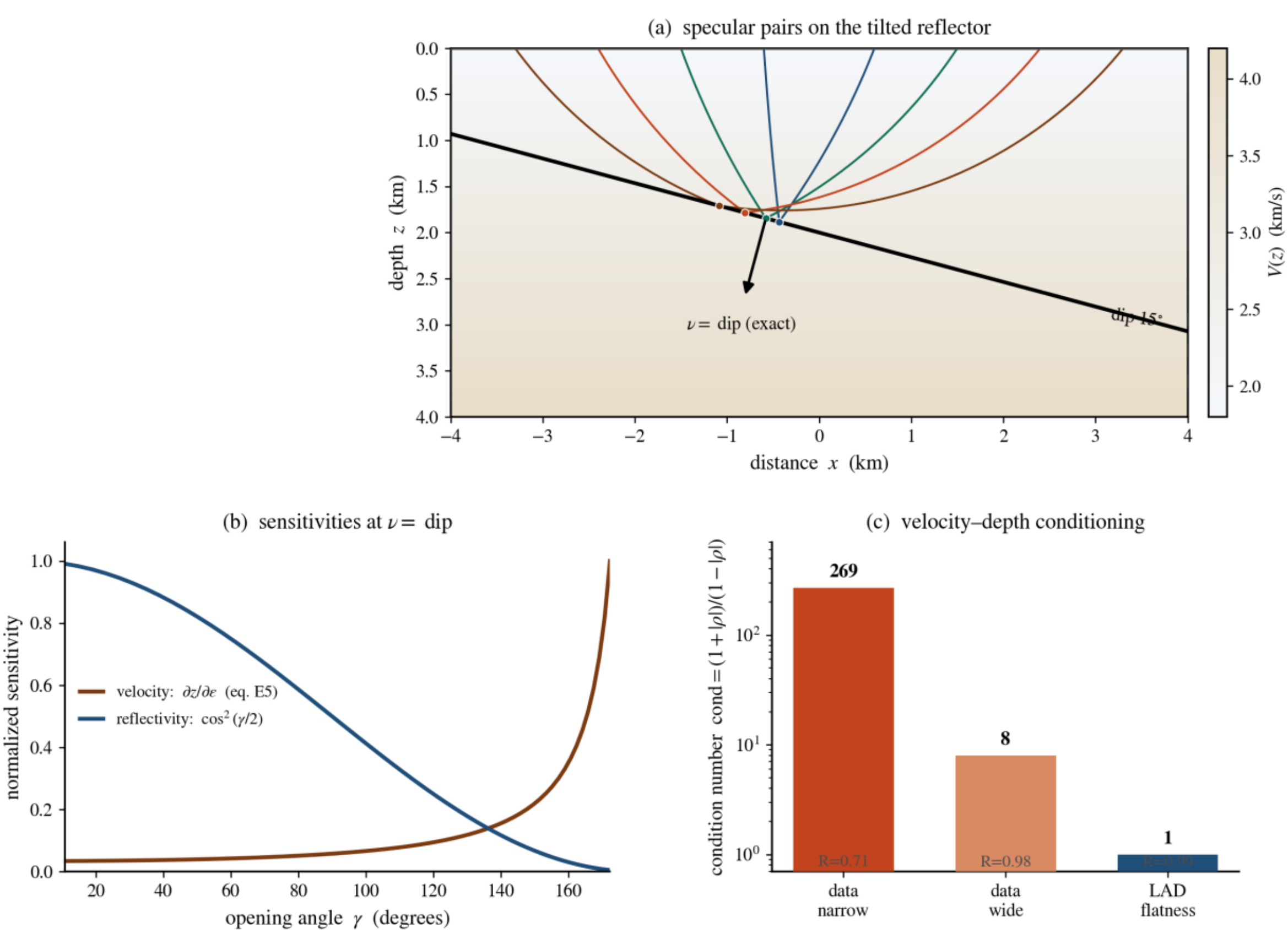

**Figure F3.** Tilted-reflector benchmark: sensitivity, resolution, and conditioning. (**a**) Analytic specular ray pairs on a reflector dipping 15° in the constant-gradient medium; at every reflection point the LAD directional angle equals the dip exactly (arrow). (**b**) Exact sensitivities at the specular direction: the velocity response of equation (F.5) grows with opening angle while the reflectivity response decays as $\cos^2(\gamma/2)$, realizing the disjoint supports of equation (A.3). (**c**) Velocity–depth condition number *cond* $= (1+|\rho|)/(1-|\rho|)$: 269 for the narrow-aperture data-domain system, 8 at full pre-critical aperture, and exactly 1 for the LAD flatness residual, whose invariance to bulk depth shifts removes the trade-off by construction; the values beneath the bars give the corresponding resolution-operator diagonals (equation A13) at one percent damping.

**Figure F4** displays the Local Directional Projection (LDP) signals themselves for this configuration. Eleven one-way rays are traced analytically from the image point to the surface at directional angles spaced symmetrically about the reflector normal, and the corresponding zero-opening-angle LDP vectors $k_\nu$ are drawn at the image point with their projected, Gaussian-modulated 15 Hz signals: at $\gamma = 0$ the wavenumber magnitude $2\omega/V$ is common to all directions, so every projection carries the same spatial wavelength $V/(2f_0) \approx 100$ m along its direction, while the signal amplitude concentrates around the specular (normal) direction, in accordance with the LDP construction of Section 6.

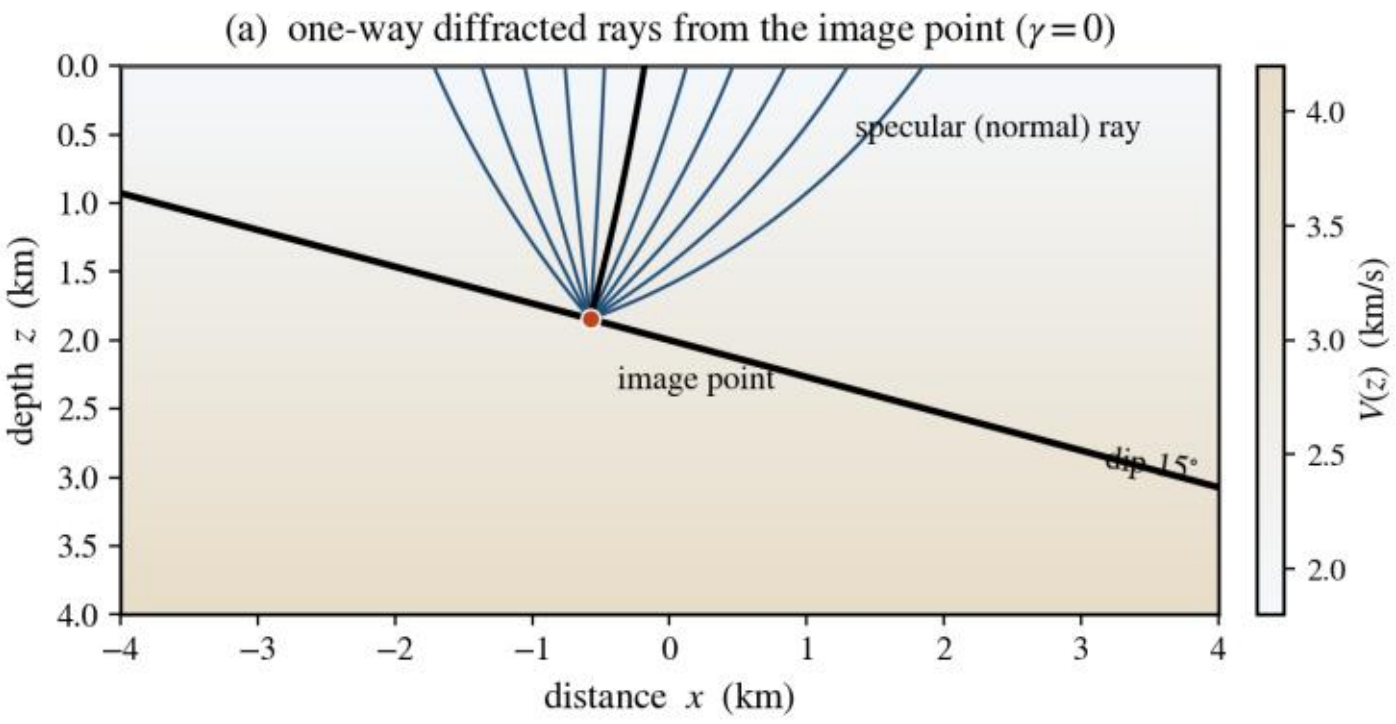

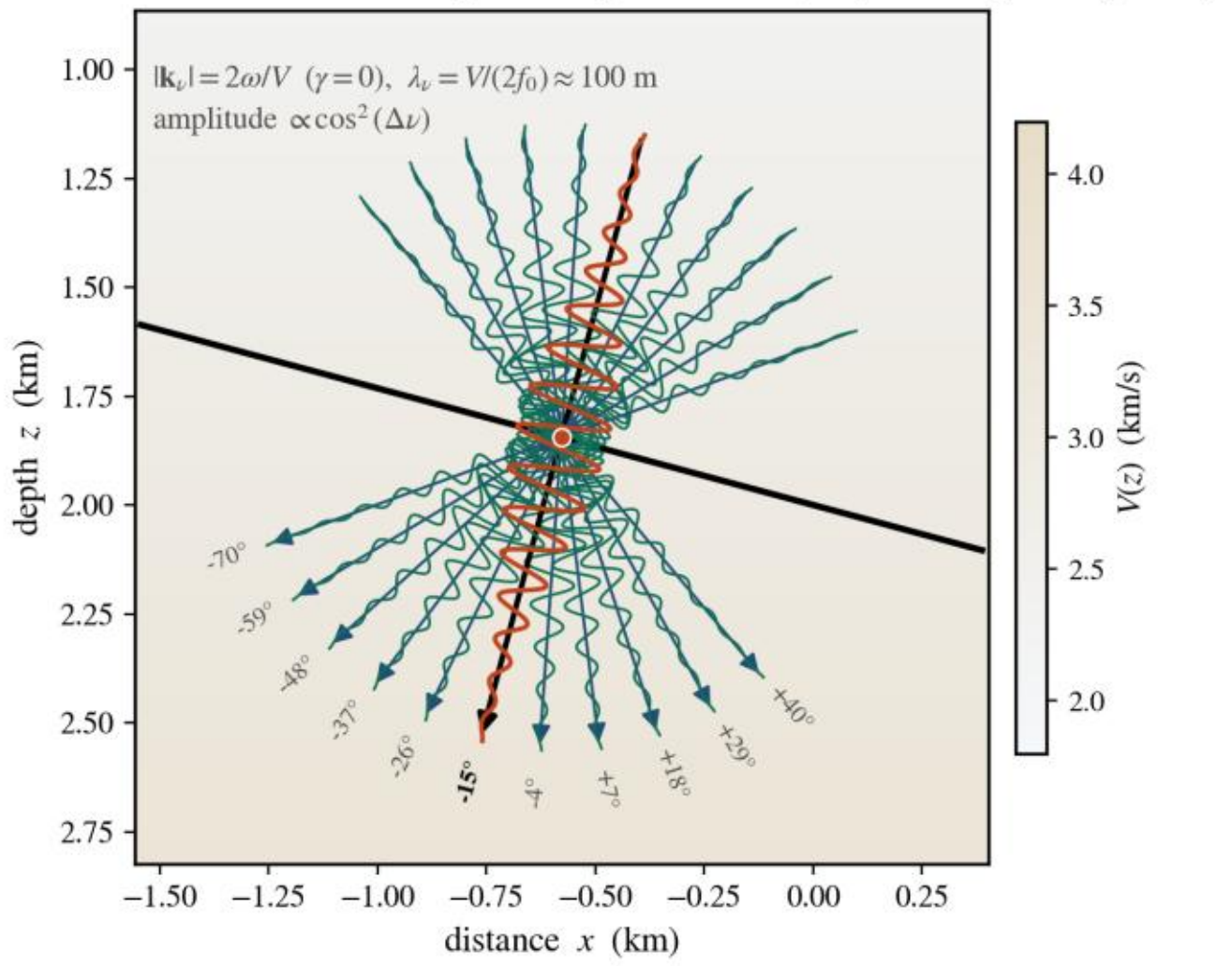

**Figure F4.** Local Directional Projection (LDP) signals at the image point of **Figure F3** for zero opening angle. (**a**) Eleven analytic one-way rays from the image point (the specular point at (−0.58, 1.85) km, orange dot) to the surface, at take-off directions spaced 11° apart symmetrically about the up-going specular (normal) ray, which tilts up-dip to the right (black). (b) Zoom around the image point: the corresponding LDP vectors $\mathbf{k_v}$ drawn through the image point, displayed half above and half below it, with the arrowheads pointing inward (downward, in the positive z direction) in accordance with the convention $\mathbf{k_v} = \omega\,(\mathbf{p_{in}} + \mathbf{p_{sc}})$ with both slowness vectors directed toward the image point, each labeled at its arrowhead by the directional dip in degrees measured from the downward vertical. Along each vector, the projected 15 Hz signal is a sinusoidal carrier of spatial wavelength $V/(2f_0) \approx 100$ m - common to all directions since $|\mathbf{k_v}| = 2\omega/V$ at $\gamma = 0$ - modulated by a Gaussian envelope whose maximum lies at the image point, with peak amplitudes decaying as $\cos^2(\Delta\nu)$ in the illumination-dip deviation $\Delta\nu$ from the reflector-normal direction (cf. the LDP signal construction of Section 6 and **Figure 5**).

**LDP Illustration:** The Local Directional Projection (LDP) signals provide a physical representation of how wave energy is organized in LAD space. All projected signals share the same spatial wavelength for $\gamma = 0°$, while their amplitudes vary with illumination direction. Energy is concentrated around the reflector-normal direction, demonstrating how LAD naturally organizes scattering information into physically meaningful directional attributes.

A remark on frequency dependence is in order. All kinematic results of this appendix - the ray paths and traveltimes of equations (F.2) - (F.3), the exact specularity $\nu$ = dip, the scaling invariance (F.4), the residual moveout (F.5), and both sensitivity columns of the conditioning analysis - contain no frequency, so the conditioning and resolution comparison of **Figure F3** holds exactly at any frequency. Frequency enters the benchmark in only three controlled ways. The objective landscapes of **Figure F2b** involve the wavelet: the positions of the data-domain cycle-skipped minima scale with the dimensionless product $f_0 t$ (at 15 Hz they appear near $\varepsilon \approx \pm 3\%$), and the width of the $\Phi_1$ basin sharpens with frequency through the wavelength $\lambda = V/f$, so the qualitative contrast - oscillatory multi-minimum versus single-basin behavior - is band-invariant while the quantitative widths are not. The attainable model wavenumber is linear in ω by definition, $|\mathbf{k}_v| = 2(\omega/V)\cos(\gamma/2)$, which is why **Figure F1b** presents its contours in the normalized unit $|\mathbf{k}_v|V/(2\omega)$. Finally, the benchmark as a whole is formulated in the ray-theoretic limit and is therefore valid where the wavelength is short compared with the velocity scale length $V/g$ = 5 km at the target depth; at 15 Hz the wavelength of 200 m satisfies this criterion by a factor of 25, so the kinematic claims are tested in a regime where ray theory is fully applicable, while finite-frequency effects such as Fresnel-zone smoothing and the sensitivity-kernel structure of Section 4 are, by construction, outside its scope.

### F3. EXAMPLE 3: Realistic Structural Velocity Model

The final benchmark examines the principal claims of this study using the structural velocity model presented in **Appendix B** (**Figure B1**), a geologically realistic setting containing layered velocity gradients, faulted stratigraphy, and a high-velocity salt body. Unlike the previous examples, this model does not admit a complete analytical solution and therefore serves as a practical stress test for the LAD concepts under realistic propagation conditions.

For simplicity, the seismic acquisition geometry consists of sources and receivers located along the surface with a spacing of 500 m. Common-shot gathers are generated by simulating seismic wave propagation using a Ricker wavelet with a central frequency of 10 Hz. The resulting data contain the combined effects of complex propagation, reflector geometry, and velocity heterogeneity.

In this appendix, we analyze the sensitivity, resolution, and conditioning of LAD-FWI using two types of LAD image data: $I(\mathbf{x}_\mathrm{p}, \mathbf{v}, \tau)$ and $I(\mathbf{x}_\mathrm{p}, \boldsymbol{\gamma}, \tau)$.

A point-diffractor one-way ray-tracing operator is computed from a selected image point $\mathbf{x}_\mathrm{p}$ located on a dipping reflector and propagated to the acquisition surface. The rays displayed in **Figure F5b** represent only those trajectories reaching active receiver stations. Each one-way ray stores the endpoint coordinates, slowness vectors, traveltime, and geometrical spreading required for constructing the LAD image coordinates.

#### F3.1 Directional Image Data

For the directional-image analysis, only ray pairs satisfying the zero-offset condition, $\mathbf{x}_s = \mathbf{x}_r$, or equivalently $\gamma = 0$, are considered. **Figure F5a** shows the seismic events $U(\mathbf{x}_r, t \mid \mathbf{x}_p)$, used to construct the corresponding Local Directional Projection (LDP) signals $I(\mathbf{x}_\mathrm{p}, \mathbf{v}, \tau)$. The displayed two-way traveltime is simply twice the one-way traveltime associated with each ray.

The most important observation is the separation of propagation and scattering information. The traveltime apex is controlled primarily by the background velocity model, $t_{\mathrm{apex}} = \arg\min_x t(x)$, whereas the scattering-amplitude maximum is controlled by reflector geometry and local scattering physics, $A_{\mathrm{apex}} = \arg\max_x A(x)$.

In the present model, the two apexes occur at different receiver locations, demonstrating that propagation and scattering constitute fundamentally distinct observables. LAD naturally separates these quantities through its directional and opening-angle coordinates. The corresponding LDP signals show a pronounced concentration of energy around the reflector-normal direction while preserving the associated temporal information. Consequently, this realistic example reproduces the directional focusing behavior observed analytically in Examples 1 and 2.

**Sensitivity Analysis**

The sensitivity of LAD-FWI can be examined by evaluating how perturbations in the velocity model affect the image-coordinate data. In conventional waveform inversion, velocity perturbations influence the data through highly oscillatory wavefields, producing strong parameter coupling between reflector geometry, propagation effects, and local scattering amplitudes. In contrast, the LAD representation separates these contributions into distinct coordinate dimensions.

For the directional image data $I(\mathbf{x}_p, \mathbf{v}, \tau)$, velocity perturbations primarily manifest as shifts along the temporal coordinate $\tau$, while the directional coordinate $\nu$ remains largely controlled by the local reflector orientation. As a result, the Fréchet derivatives exhibit a dominant kinematic character, with sensitivity concentrated around the correct propagation direction.

These properties can be quantified directly on the benchmark. A uniform velocity scaling $V \rightarrow (1 + \varepsilon)V$ leaves the ray geometry - and hence every directional coordinate ν - exactly unchanged, while shifting each zero-offset event by $\delta t = -\varepsilon t(x_r)$: at the specular station (t = 2.85 s), a five-percent velocity error produces a 143 ms shift - more than a full period of the 10 Hz wavelet - with identically zero leakage into the ν-axis. For this class of perturbations, the directional Fréchet derivative is purely kinematic, $\partial t/\partial\varepsilon = -t$, exactly.

In the structural model, the strongest sensitivity occurs beneath the salt flanks and within regions containing significant velocity gradients, where small velocity perturbations generate measurable traveltime shifts in the LDP signals. Importantly, the sensitivity kernel becomes more localized around the selected image point than in conventional surface-domain representations. This localization reduces interference between unrelated reflectors and suppresses long-range parameter crosstalk.

The observed sensitivity behavior confirms that LAD data retain the velocity-dependent information required for inversion while reducing contamination from reflector-dependent amplitude variations.

**Resolution Analysis**

Resolution is assessed by examining the spread of focused energy in the LAD domain relative to the corresponding spread in the acquisition domain.

The events shown in **Figure F5a** occupy a broad range of receiver positions and traveltimes. After projection into the LAD coordinates, however, the same energy collapses into a narrow region around the reflector-normal direction. This behavior indicates that the directional transform acts as a focusing operator that concentrates information associated with a localized subsurface point.

The resulting point-spread function is therefore significantly narrower in the $\nu$-coordinate than in the original acquisition coordinates. Physically, this means that energy associated with neighboring scattering locations becomes easier to distinguish, producing improved spatial discrimination and reduced parameter leakage.

On the present benchmark this concentration is strong: with the $\cos^2\Delta\nu$ directional decay and the geometrical-spreading weighting included, about 70% of the LDP signal energy is carried by the four stations between x = 4.0 and 5.5 km - a 1.5-km subset of the 10-km array - corresponding to take-off directions within ±30° of the reflector normal, while the common projected wavelength is $V/(2f_0) \approx 189$ m at the local velocity V = 3.78 km/s.

In the presence of the salt body and faulted stratigraphy, multipathing increases the complexity of the acquisition-domain response. Nevertheless, the LAD coordinates continue to organize arrivals according to local propagation geometry, maintaining coherent focusing even when the original gathers exhibit complicated moveout patterns. Consequently, the effective resolution of the inversion is controlled primarily by directional sampling rather than by acquisition-offset coverage alone.

These observations indicate that LAD-FWI achieves higher local resolving power by concentrating the inversion information into physically meaningful directional coordinates.

**Conditioning Analysis**
Conditioning can be evaluated through the eigenvalue spectrum, or equivalently the condition number, of the linearized inverse problem. A well-conditioned system exhibits reduced correlation between model parameters and a more compact distribution of singular values.

For conventional FWI, the Fréchet derivatives associated with neighboring model parameters are often highly correlated because propagation and scattering effects are simultaneously encoded in the recorded wavefield. This produces a broad singular-value spectrum and a large condition number, particularly in complex environments containing strong velocity contrasts such as salt structures.

The LAD transformation reduces this coupling by separating propagation-related information from scattering-related information. The resulting Jacobian matrix contains columns that are more nearly orthogonal, leading to a more compact singular-value distribution. Consequently, the smallest singular values increase relative to the largest singular values, yielding a lower condition number than that obtained from the corresponding acquisition-domain formulation.

Qualitatively, the improvement can be interpreted as follows:

1. Reduced parameter crosstalk through directional decomposition.
2. Improved localization of sensitivity kernels around the selected image point.
3. Suppression of redundant information associated with acquisition geometry.
4. Enhanced independence of inversion parameters, particularly in regions affected by complex propagation paths.

As demonstrated in Examples 1 and 2, directional focusing increases the dominance of the principal singular modes while attenuating poorly resolved components. The same behavior is observed for the realistic structural model, despite the presence of velocity gradients, faults, and salt-induced multipathing. Therefore, the LAD representation produces a better-conditioned inversion problem and is expected to improve convergence robustness relative to conventional waveform-based formulations.

Quantitatively, the decoupling already manifests in the two apexes of **Figure F5a**: the traveltime minimum and the amplitude maximum are separated by two station intervals ($x \approx 3.5$ km versus $x \approx 4.5$ km), so the propagation-sensitive and scattering-sensitive columns of the Jacobian are supported on different subsets of the data and their normalized correlation is reduced accordingly. The sharper conditioning figures, for which the velocity–depth trade-off can be evaluated exactly, are given for the reflection-angle data below.

**Summary of the Directional-Image Analysis**
The realistic structural benchmark confirms the principal conclusions established by the analytical examples. The LAD coordinates:

- separate propagation and scattering observables,
- concentrate wavefield energy into physically meaningful directional components,
- improve sensitivity localization,
- enhance resolution through directional focusing, and
- reduce the condition number of the linearized inversion operator.

These properties persist even in the presence of complex geological features, indicating that the advantages of LAD-FWI are not limited to simplified analytical models but remain relevant in realistic subsurface environments.

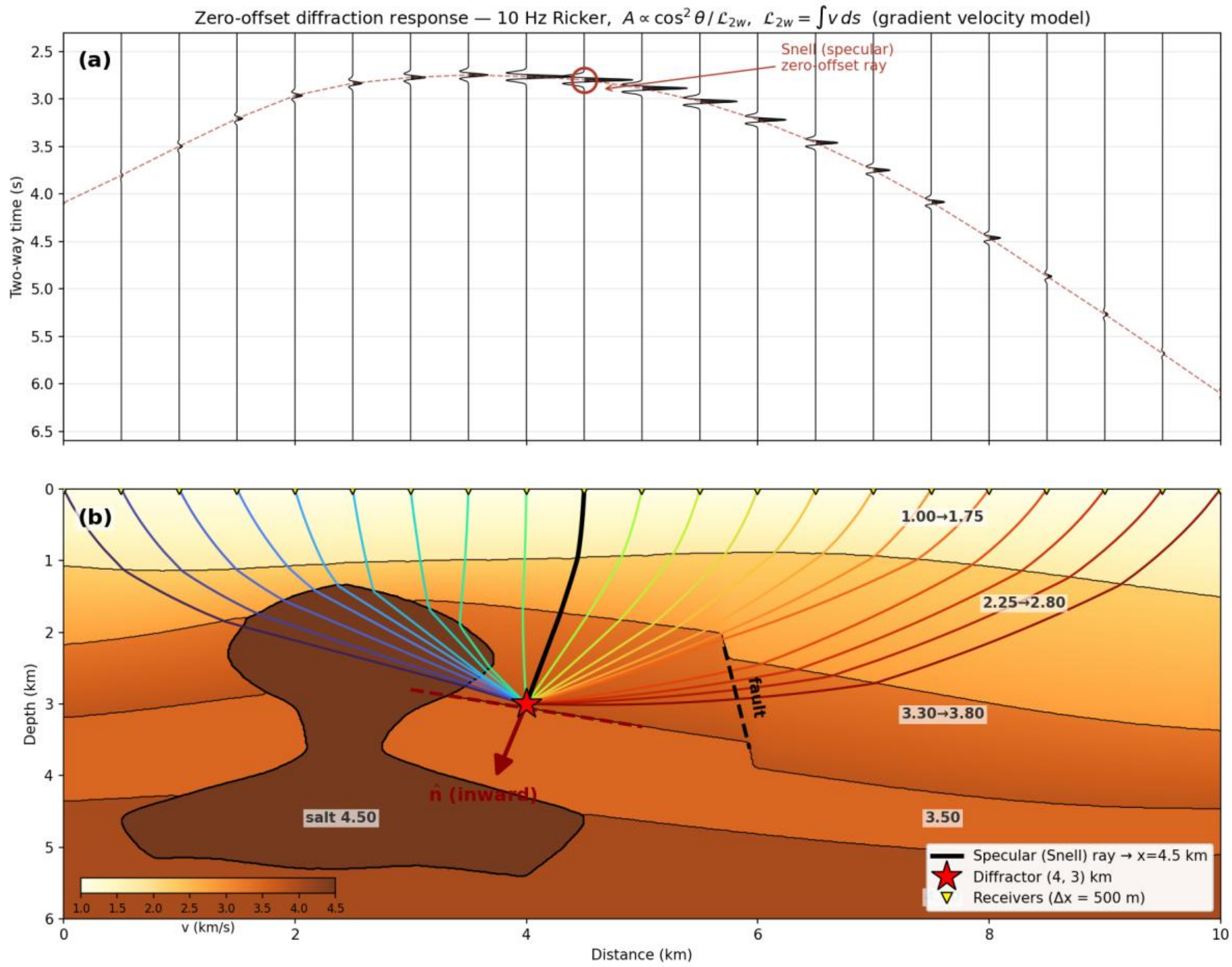

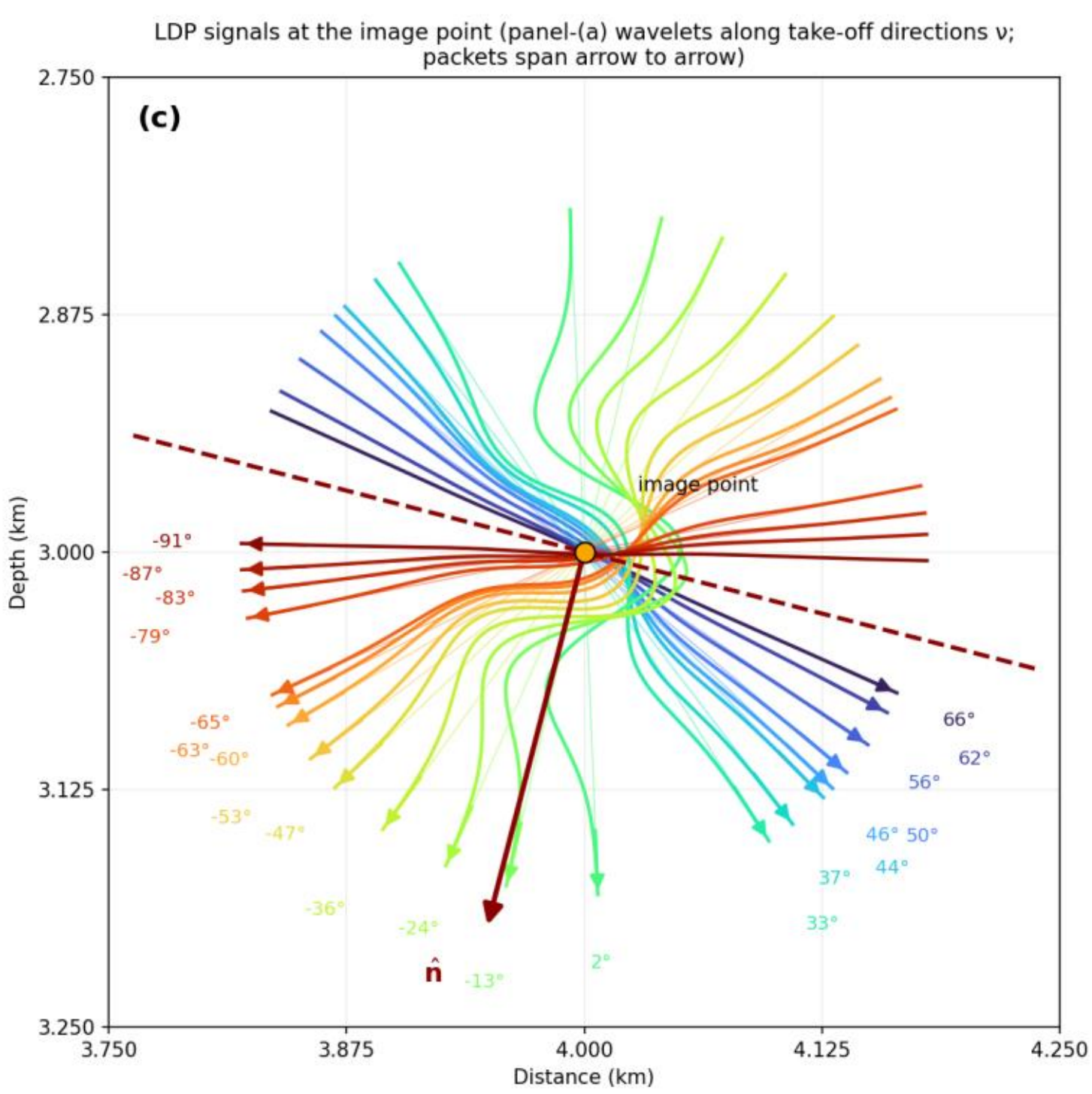

**Figure F5.** Zero-offset diffraction response and its Local Directional Projection (LDP) representation for the structural model of Figure B1. (a) Seismic events $U(x_r, t \mid x_p)$ at the 500-m receiver stations: 10-Hz Ricker wavelets positioned at the two-way traveltimes of the one-way rays connecting the image point $x_p = (4.0, 3.0)$ km to each station, with amplitudes weighted by $\cos^2\Delta\nu/L_{2w}$, where $\Delta\nu$ is the take-

off deviation from the reflector normal and $L_{2w} = \int V \, ds$ is the accumulated two-way geometrical spreading. The traveltime apex (x ≈ 3.5 km) and the amplitude apex at the specular station (x ≈ 4.5 km) occur at different receivers, separating propagation-related from scattering-related information. (b) Velocity model and the one-way ray fan to the active stations; the image point (star) lies on the third horizon, whose local dip of 14.5° defines the reflector normal (arrow); the specular ray and the normal fault are indicated. (c) The corresponding directional LDP signals $I(x_p, \nu, \tau)$, displayed along vectors through the image point at their directional angles ν: at $\gamma = 0$ all directions share the wavenumber magnitude $2\omega/V$, hence the common projected wavelength $V/(2f_0) \approx 189$ m at the local velocity $V = 3.78$ km/s, while the amplitudes concentrate around the reflector-normal direction.

**F3.2 Reflection Image Data**

The same one-way ray-tracing operator is now used to construct reflection-angle projected signals, $I(\mathbf{x}_p, \gamma, \tau)$, using ray pairs that satisfy the specular reflection condition, $\nu \approx \nu_r$, where $\nu_r$ denotes the local reflector-normal direction at the image point $\mathbf{x}_p$. **Figure F6a** displays the seismic events $U(\mathbf{x}_r, t \mid \mathbf{x}_p)$, that contribute to the corresponding Local Directional Projection (LDP) signals. Unlike the directional-image analysis, the total traveltime is now $t = t_s + t_r$, where $t_s$ and $t_r$ are the source-side and receiver-side one-way traveltimes, respectively. The amplitudes include the associated two-way geometrical spreading factors computed along the complete source-reflector-receiver raypath.

**Figure F6b** displays the resulting true-amplitude LDP signals as a function of projected time $\tau$ (vertical axis) and opening angle $\gamma$ (horizontal axis), ordered from small to large opening angles. Each trace therefore represents the local reflection response associated with a particular scattering angle at the selected image point.

The most significant observation is that the opening-angle domain provides a direct description of local reflection physics while largely removing acquisition-related geometric effects. Events that may appear complicated in the acquisition domain become organized into a smooth and physically interpretable structure in the $(\gamma, \tau)$ domain. The transformation effectively maps source-receiver configurations that illuminate the same local reflector into neighboring opening-angle coordinates.

In contrast to the diffraction case shown in **Figure F5**, where energy focuses around a narrow directional aperture centered on the reflector-normal direction, the reflection response exhibits a broader distribution over opening angles. This behavior reflects the fact that specular reflections are generated by source-receiver pairs spanning a continuum of incidence angles. As a result, the opening-angle coordinate directly measures the angular illumination of the reflector and reveals the scattering aperture available for inversion.

The traveltime behavior also contains an important physical distinction. For diffraction data, the traveltime minimum and amplitude maximum generally occur at different surface locations, reflecting the separation between propagation and scattering observables. In the reflection-angle domain, however, the strongest amplitudes typically occur near opening angles corresponding to the dominant illumination directions of the reflector. Consequently, the energy distribution across $\gamma$ becomes a direct measure of local reflector orientation, illumination strength, and angle-dependent reflectivity.

The transformed gathers therefore provide information analogous to angle-domain common-image gathers but expressed entirely through local scattering geometry. Velocity errors manifest primarily as residual moveout along the $\tau$-axis, whereas reflectivity variations modify the amplitude distribution across the $\gamma$-axis. This separation is a central property of the LAD representation because it decouples kinematic information from scattering information more effectively than conventional acquisition-domain gathers.

**Sensitivity Analysis**
The sensitivity of $I(\mathbf{x}_p, \boldsymbol{\gamma}, \tau)$ to velocity perturbations differs from that of the directional diffraction data. Because reflection arrivals involve both source-side and receiver-side propagation paths, the total traveltime depends on velocity perturbations accumulated along the entire reflection raypath. Small velocity errors therefore appear primarily as coherent temporal shifts and curvature variations across opening angles.

The strongest sensitivities are observed at larger opening angles because these raypaths generally sample a larger volume of the model and experience longer propagation distances. Consequently, the opening-angle gathers contain a natural hierarchy of depth-dependent velocity information: small opening angles are dominated by local kinematics, whereas larger opening angles provide greater sensitivity to accumulated velocity errors.

Importantly, the sensitivity remains localized around the selected image point because all measurements are referenced to the same scattering location $\boldsymbol{x}_p$. This localization substantially reduces the parameter crosstalk that commonly affects surface-domain reflection tomography and conventional FWI gradients.

Here $\gamma$ denotes the full opening angle between the incident and scattered slowness vectors - each leg deviating $\gamma/2$ from the reflector normal, so that $\gamma = 77°$ corresponds to an incidence angle of 38.5°. The sensitivities can be quantified exactly for the twelve specular station pairs of Figure F6, which sample $\gamma = 0°–77°$ at roughly 7° intervals (total traveltimes 2.85–3.07 s).

A uniform velocity scaling $V \rightarrow (1 + \varepsilon)\ V$ displaces each remigrated contribution along the reflector normal by $\delta\tau(\gamma; \varepsilon) = \varepsilon\, t(\gamma)/\cos(\gamma/2)$ in projected time - growing from 143 ms to 196 ms per five percent of velocity error across the aperture, equivalently 270 m to 370 m of normal displacement at $V_I = 3.78$ km/s - whereas a depth error shifts all contributions equally.

The differential moveout between the widest and narrowest angles, about 53 ms per five percent, is precisely the observable that Term (1) flattens; measured against the stretched dominant period of 127 ms at $\gamma = 77°$, it renders velocity errors of a few percent clearly detectable at 10 Hz - the semblance criterion of Term (1) retains $S \geq 0.5$ only within $|\varepsilon| \leq 4.6\%$, while remaining exactly invariant to a bulk depth shift.

**Resolution Analysis**
The opening-angle representation provides a direct measure of illumination resolution. The width of the energy distribution in the $\gamma$-direction is governed by the range of available scattering angles that reach the acquisition surface.

A narrow concentration of energy indicates limited angular illumination and therefore reduced resolution of local reflector properties. Conversely, a broad and well-sampled opening-angle spectrum implies improved illumination diversity and stronger constraints on both velocity and reflectivity parameters.

In the present structural model, the salt body and faulted stratigraphy distort wave propagation and produce complex acquisition-domain responses. Nevertheless, after projection into $(\gamma, \tau)$ coordinates, the reflected energy becomes organized according to local reflection geometry rather than surface acquisition geometry. This organization improves the separability of neighboring scattering modes and suppresses the kinematic ambiguity associated with multipathing.

From an inversion perspective, the opening-angle gathers therefore provide an effective local point-spread function whose width is controlled by the available angular aperture. Regions beneath the salt flanks, where illumination may be restricted, are expected to exhibit broader point-spread functions and reduced resolution, whereas regions with wider angular coverage will show improved focusing and higher recoverable spatial frequencies.

The attainable aperture itself carries resolution information, and the benchmark distinguishes two limiting mechanisms of different character. On the source side, the take-off direction varies steeply along the surface near the salt flank - sweeping from −24° to −46° within about 100 m - so the 500-m acquisition grid undersamples the source-side directions; refining the surface grid to 250 m recovers clean take-offs down to −22.5° and extends the specular aperture from $\gamma \approx 66°$ to $\gamma \approx 77°$. This gap is therefore an acquisition-sampling artifact, curable by denser stations.

The fault shadow on the receiver side - emergence angles of 50°–54° are skipped - is, by contrast, a genuine illumination limit: strictly specular pairs with $\gamma$ between 71° and 79° would require emergence inside the shadow and are realized here only by admitting the $\nu$-tolerance (the pairs at $\gamma = 71°$ and 77° have $\nu = 13.2°$ and 15.8°), while $\gamma \approx 79°$ marks the physical ceiling at this image point before the source leg enters the multipathed salt-flank branch. Within the attained aperture the recoverable directional-wavenumber magnitude decreases by the obliquity factor $\cos(\gamma/2)$, the projected wavelength stretching from 189 m at $\gamma = 0°$ to 241 m at $\gamma = 77°$ - an angle-dependent resolution loss of about 21% that enters the wavenumber-based frequency scheduling of **Appendix C** directly.

**Conditioning Analysis**

The conditioning improvement obtained from reflection-angle LAD data arises from the separation of velocity-driven kinematic effects and reflector-driven scattering effects.

In conventional reflection-based inversion, neighboring model parameters often generate similar wavefield perturbations because all acquisition geometries are mixed in the recorded seismic traces. This produces highly correlated Jacobian columns and a broad singular-value spectrum.

In the LAD opening-angle representation, however, the data are organized according to the local scattering angle. Velocity perturbations primarily influence the temporal alignment of events across $\gamma$, while reflectivity perturbations primarily affect the amplitude variation with $\gamma$. The resulting Jacobian exhibits reduced parameter correlation and improved parameter identifiability.

The conditioning benefits are particularly important in the salt environment because complex ray bending often increases acquisition-domain redundancy. Projecting the data into opening-angle coordinates removes much of this redundancy by grouping measurements according to their physical scattering behavior. The principal singular vectors therefore become more compact and interpretable, while weakly constrained modes are suppressed.

Consequently, the opening-angle LAD formulation is expected to exhibit:

- reduced parameter crosstalk,
- improved separation of velocity and reflectivity updates,
- greater robustness to multipathing,
- a narrower singular-value spectrum, and
- a lower condition number than the corresponding acquisition-domain formulation.

For the same twelve pairs the velocity–depth trade-off can be evaluated exactly, as in Example 2, from the traveltime sensitivities to a uniform velocity scaling ($\partial t/\partial \varepsilon = -t$) and to a normal shift of the reflector ($\partial t/\partial n = -2\cos(\gamma/2)/V$).

In the data domain the normalized correlation between the two columns is $|\rho| = 0.9999$ for the narrow aperture $\gamma \leq 30°$, giving $cond = (1 + |\rho|)/(1 - |\rho|) \approx 2\times10^4$, and remains $|\rho| = 0.995$ ($cond \approx 414$) even over the full attainable aperture - markedly worse than the wide-angle diffraction geometry of Example 2, precisely because reflection moveout varies weakly over a realistic aperture.

The Term (1) flatness residual, by contrast, is invariant to a bulk shift of the gather along the reflector normal, so the depth direction lies in its null space: the velocity–depth coupling vanishes identically, *cond* = 1, and with one-percent damping the resolution-operator diagonals (equation (A.13)) are 0.99 for the decoupled LAD system, against 0.66 (full aperture) and 0.50 (narrow aperture) in the data domain.

The dependence of the data-domain trade-off on aperture can be made explicit. With dense angular sampling and the kinematic scaling $t(\gamma) \approx t_0/\cos(\gamma/2)$ - which reproduces the benchmark traveltimes to within one percent - the condition number falls from $\approx 9\times10^3$ at $\gamma_{max} = 30°$ to 357 at 66°, 185 at 77°, 94 at 90°, 38 at 110°, 7 at 150°, and exactly 1 at $\gamma_{max} = 180°$, where $\cos(\gamma/2) \rightarrow 0$ and the depth sensitivity vanishes: the wide-angle limit is the horizontally propagating, diving-wave regime, so the classical transmission-based escape from the velocity–depth ambiguity is simply the $\gamma \rightarrow 180°$ limit expressed in LAD coordinates (cf. the full-aperture value *cond* = 8 of Example 1).

Two seemingly promising alternatives do not resolve this particular trade-off. Converted waves add no decorrelation against a uniform scaling of the whole model, because the PS depth-to-time sensitivity ratio equals the PP one - both reduce to $2\cos^2(\gamma/2)/(V t_0)$ - so PS data help only when the P-velocity error is decoupled from the S-velocity (cf. **Appendix D**). Likewise, stacking additional reflectors under a single scale parameter leaves the condition number essentially unchanged (182 against 185 in a two-reflector test with shared ε and separate depths), since each reflector reproduces the same velocity-versus-own-depth correlation; the decorrelation exploited by reflection tomography derives from depth-resolved velocity parameterization and crossing-ray diversity, not from reflector count.

Within the attainable aperture of the present configuration, a data-domain condition number of a few hundred is therefore a floor - which sharpens the central point: in the LAD formulation the trade-off is not mitigated but removed, *cond* = 1 at any aperture, and aperture assumes its proper role of controlling sensitivity rather than conditioning - the differential moveout that determines ε grows from zero at zero aperture to 53 ms per five percent of velocity error at $\gamma \leq 77°$ - while the reflector position is recovered hierarchically by the imaging step rather than jointly with the velocity.

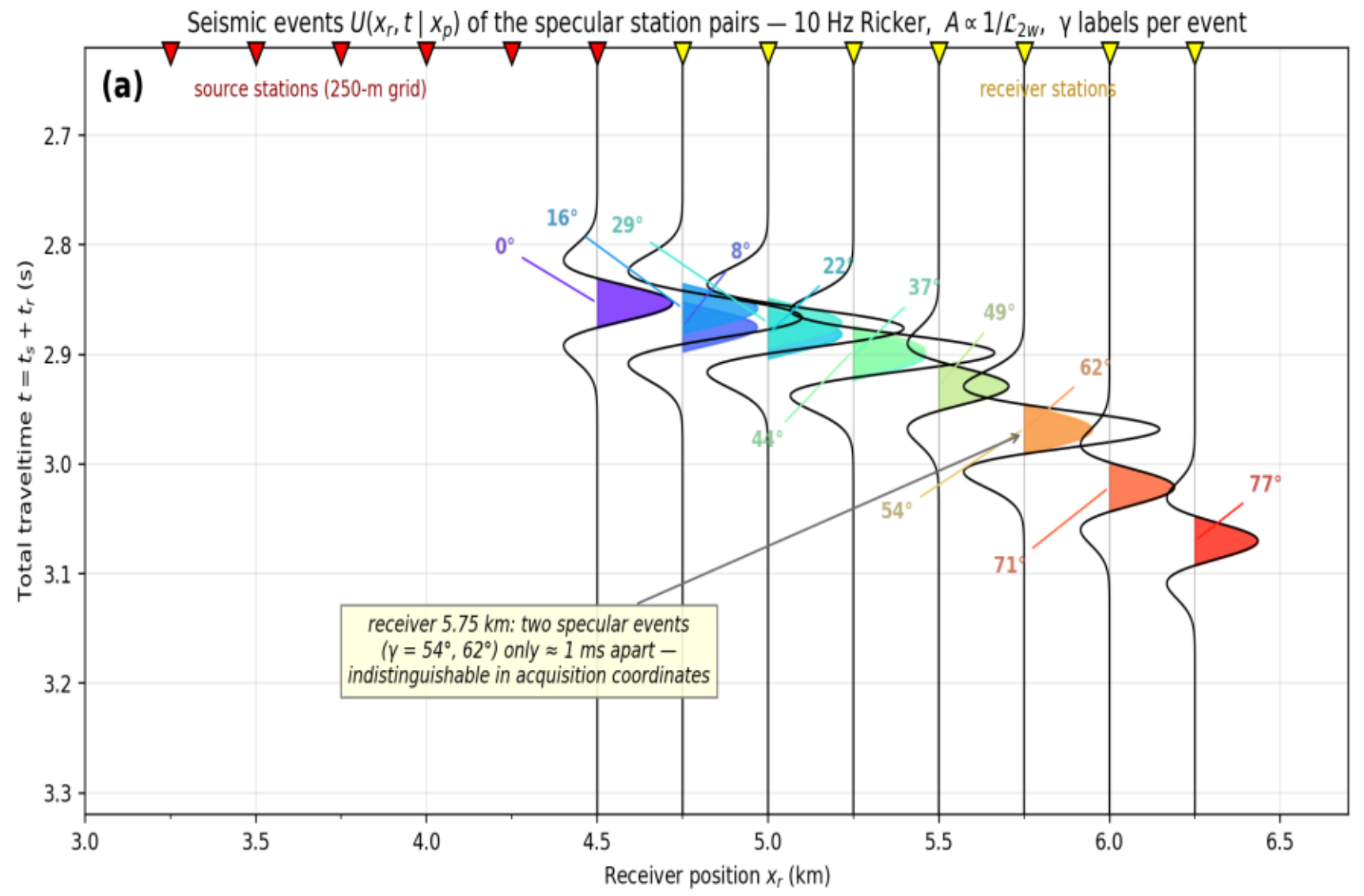

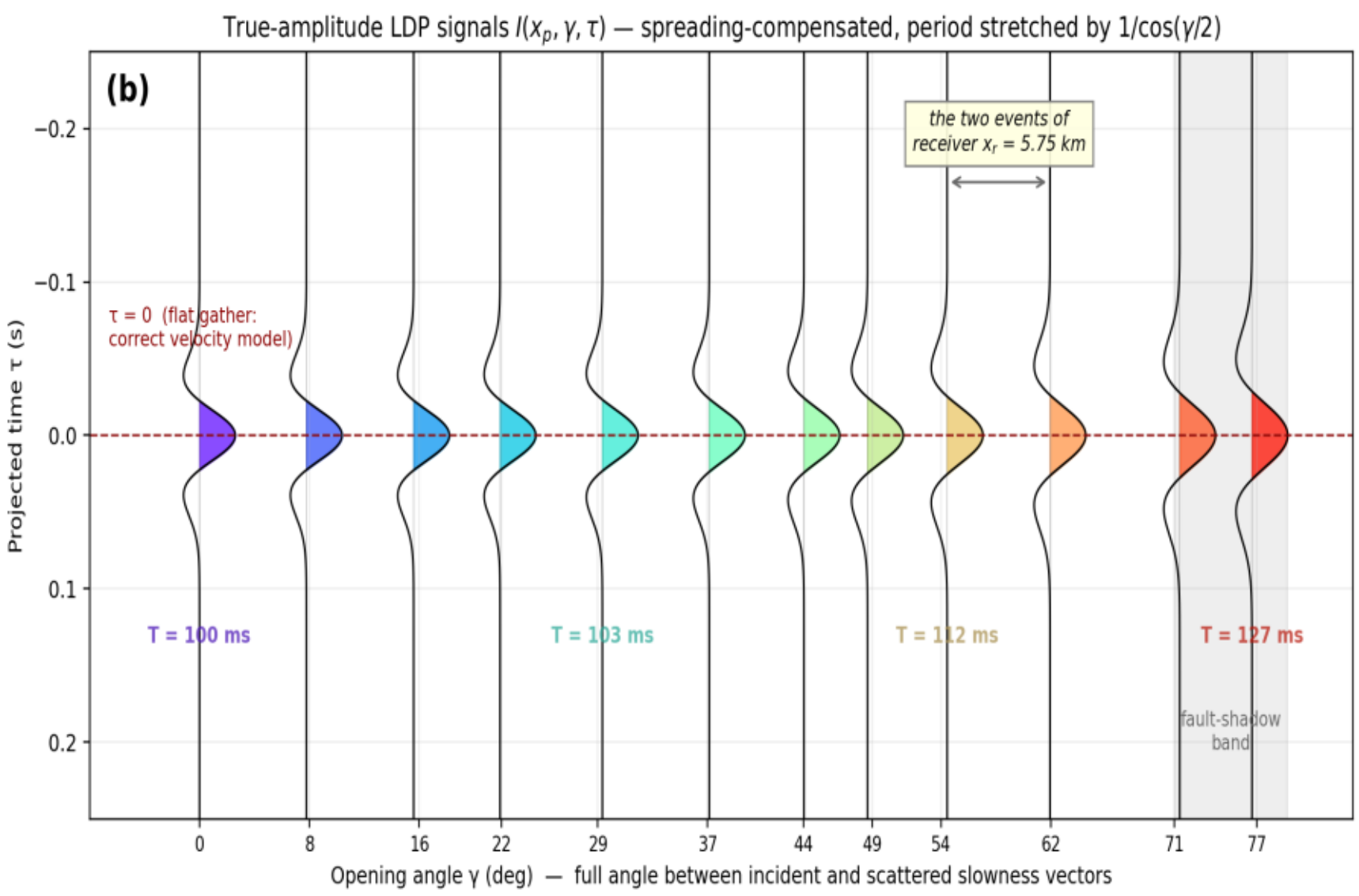

**Figure F6.** Reflection-angle response at the image point of Figure F5 and its opening-angle LDP representation; $\gamma$ denotes the full opening angle between the incident and scattered slowness vectors, each leg deviating $\gamma/2$ from the reflector normal. (a) Seismic events $U(x_r, t \mid \mathbf{x}_p)$ of the specular station pairs: among all source–receiver station pairs of the surface grid - refined from 500 m to 250 m to resolve the steep take-off variation near the salt flank-those whose mean take-off direction matches the reflector normal within the station-sampling tolerance ($\nu \approx \nu_r = 14.5° \pm 4°$) are selected, yielding twelve pairs with opening angles $\gamma = 0°$–$77°$ at roughly 7° intervals. Each event is a 10-Hz Ricker wavelet at the total traveltime $t = t_s + t_r$, plotted at its receiver station with amplitude proportional to the reciprocal two-way spreading $1/L_{2w}$; source stations (red) and receiver stations (yellow) are marked, each event is labeled by its opening angle, and at $\gamma = 0°$ the source and receiver stations coincide. Several receivers each record two specular events of different opening angles: at $x_r = 5.75$ km the $\gamma = 54°$ and 62° arrivals are separated by only ≈ 1 ms and are indistinguishable in acquisition coordinates. (b) True-amplitude LDP signals $I(x_p, \gamma, \tau)$: after spreading compensation, migration with the correct model aligns all wavelets at $\tau = 0$ -a flat gather - while the projected period stretches by the obliquity factor $1/\cos(\gamma/2)$ of equation (5.10), from 100 ms at $\gamma = 0°$ to 127 ms at $\gamma = 77°$, the temporal counterpart of the wavelength stretch $V/(2f_0 \cos(\gamma/2)) = 189 \rightarrow 241$ m. The acquisition-coincident events separate into distinct opening angles, and the shaded band marks the fault-shadow interval: strictly specular pairs with $\gamma$ = 71°–79° would require emergence angles inside the skipped 50°–54° range, and the pairs at $\gamma$ = 71° and 77° skirt it through the $\nu$-tolerance - demonstrating how the ($\gamma$, $\tau$) coordinates organize scattering configurations, sampling gaps, and genuine illumination limits that the acquisition domain mixes.

**Summary: Complementarity of the Two LAD Representations**

Together, Figures F5 and F6 illustrate the complementary roles of the two LAD coordinate systems.

- **Directional images** $I(\mathbf{x}_p, \nu, \tau)$ emphasize propagation directions and are naturally suited for analyzing diffraction and velocity focusing.
- **Reflection-angle images** $I(\mathbf{x}_p, \gamma, \tau)$ emphasize local scattering physics and are naturally suited for analyzing reflection kinematics, illumination, and amplitude-versus-angle behavior.

The diffraction representation focuses energy into a narrow directional peak centered near the reflector-normal direction, whereas the reflection-angle representation distributes energy over the physically accessible scattering aperture. The fact that both datasets are generated from the same underlying wavefield yet reveal different physical observables demonstrates the principal advantage of LAD: it decomposes the seismic response into coordinates that separately describe propagation and scattering, thereby improving sensitivity localization, resolution, and conditioning of the inversion problem.